\documentclass[10pt, twoside]{article}

\usepackage{williams_notations}
\usepackage{williams_packages}

\allowdisplaybreaks

\begin{document}

\title{Derangement model of ligand-receptor binding}

\author{Mobolaji Williams}
\affil{School of Engineering and Applied Sciences, \\Harvard University, Cambridge, MA
02138, USA\footnote{\href{williams.mobolaji@gmail.com}{williams.mobolaji@gmail.com}}}
\date{August 8, 2022}

\maketitle

\begin{abstract}
We introduce a derangement model of ligand-receptor binding that allows us to quantitatively frame the question "How can ligands seek out and bind to their optimal receptor sites in a sea of other competing ligands and suboptimal receptor sites?" To answer the question, we first derive a formula to count the number of partial generalized derangements in a list, thus extending the derangement result of Gillis and Even. We then compute the general partition function for the ligand-receptor system and derive the equilibrium expressions for the average number of bound ligands and the average number of optimally bound ligands. A visual model of squares assembling onto a grid allows us to easily identify fully optimal bound states. Equilibrium simulations of the system reveal its extremes to be one of two types, qualitatively distinguished by whether optimal ligand-receptor binding is the dominant form of binding at all temperatures and quantitatively distinguished by the relative values of two critical temperatures. One of those system types (termed "search-limited," as it was in previous work) does not exhibit kinetic traps and we thus infer that biomolecular systems where optimal ligand-receptor binding is functionally important are likely to be search-limited.  \\

\noindent 
\end{abstract}
{%
	\hypersetup{linkcolor=black}
	\setcounter{tocdepth}{2}
	\tableofcontents
}

\section{Introduction \label{sec:ligand_receptor}}
The interaction between membrane receptors and extracellular ligands is the starting point for many cell-signaling pathways \cite{su2020ligand}. Given the intricacy of these pathways, one might think that the initiating ligand-receptor interaction needs to be ``highly specific" (i.e., one ligand type only binds to one receptor type). But work over the past two decades suggests the opposite: The specificity of the resulting processes requires such a precise code that only a combinatorial one, which makes use of various combinations of a finite number of inputs, can achieve it. This was found in the case of olfactory receptors \cite{malnic1999combinatorial} where different receptors recognized different combinations of ligands. Also, polypharmacology, a recent branch of drug design referring to creating ligands that act on multiple target receptors, has been found to be necessary for treating complex diseases such as schizophrenia \cite{roth2004magic}. Others have found that having many-to-many interactions between ligands and receptors promotes an increased diversity in range of responses for signaling pathways \cite{su2020ligand}.

All of these contexts for ligand-receptor interactions allow us to conceive of a stripped down model of the extracellular medium as one where receptors of various types and copy numbers exist on a cell surface surrounded by ligands of various types and copy numbers. Due to the many-to-many interactions between ligands and receptors (also called, "multi-specific" or "promiscuous" binding), in the most general case such a system exhibits bindings featuring all combinations of receptors and ligands. Still, to provide a reference point for the affinities, we can highlight, for each ligand type, a single binding between receptor and ligand that is the strongest for that ligand. We can term such interactions as "optimal" to distinguish them from other interactions.

Such a representation of the ligand-receptor system presents us with a question: How do the various binding affinities between specific receptors and ligands affect the global binding properties (e.g., total number of bound ligands, total number of optimally bound ligands, temperature at which optimal binding occurs, etc.) of the entire ligand-receptor system? 

The binding of ligands to their optimal receptor sites presents both a combinatorial and a kinetic problem. In a game of musical chairs, a person sitting in a single chair constrains the chairs available to the remaining people and thus affects the states the remaining people can occupy. Similarly, a ligand occupying one receptor site affects the receptor sites available to other ligands and thus changes the combinatorial count of the possible number of configurations of those receptors. But for optimal ligand-receptor binding, ligands not only need to strongly bind to their correct sites; they must also find these sites. Rather than musical chairs, this aspect of the problem is more like a game of tag where the size of the environment and the number of other players determines how easily one person tags another. Similarly, the size of and the number of receptors in the ligand-receptor system affects how easily ligands can find receptors.

Thus the combinatorial subproblem for ligand-receptor binding concerns how ligands can arrange themselves so that each one attaches to its optimal receptor site, and the kinetic subproblem concerns how ligands can find these optimal receptor sites in the volume they occupy. The two subproblems together are also the prototypical definitions of a self-assembly process in which initially distanced units must come together and combine in the correct ordered configuration through a thermal system's unforced evolution towards a free-energy minimum \cite{nelson2004biologicalch8, johnston2010modelling, perlmutter2015mechanisms}. There are few analytically tractable model archetypes that can treat the respective influences of combinatorics and kinetics on such processes. This work aims to propose such an archetype. 

There are some well known approaches to modeling ligand-receptor binding. Most well known is the law of mass action which has often been applied to ligand-receptor binding since it provides a coarse-grained framework to model how affinities affect bound concentrations (see chapter five of \cite{raicu2008integrated} for a summary). However, this approach does not take into account the competition between ligands that can place additional limits on the achievement of specific types of binding. 

A combinatorial model of ligand-receptor binding was presented in chapter six of \cite{phillips2012physical}. There the authors considered a collection of identical ligands in a grid-like space that contained a single receptor. The main combinatorial task in that analysis was determining the number of ways to arrange the ligands amongst the spatial grid for both bound and unbound configurations. From the answer, the authors computed the system partition function and the ligand-receptor binding concentration. However, the model did not consider different ligand species existing in the same environment and thus did not account for the combinatorial competition between various species in real systems. 

In the present work, combinatorics is incorporated through the finiteness of the number of different particle species in the system and the resulting finiteness of the possible number of ligand-receptor interactions. Also, by considering multiple ligand-types each of different copy numbers and with distinct "promiscuous" (or multi-specific) binding affinities to a similarly diverse set of receptor sites, the total possible set of combinatorial bindings better approaches that of a real system. The net consequence of these assumptions is to introduce combinatorial competition into the equilibrium statistical physics that define the system, making finite number effects particularly important in describing thermal properties.

\begin{figure}[t]
\centering
\includegraphics[width=.85\linewidth]{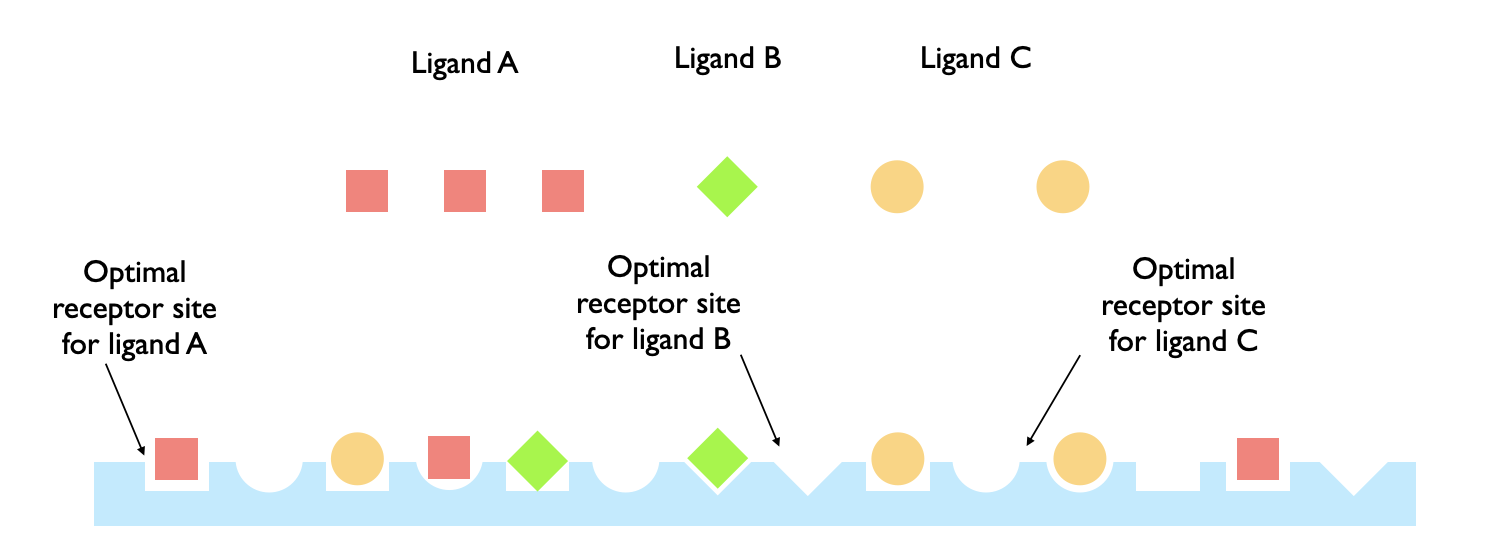}
	\caption{Example ligand-receptor system. The figure displays three different types of ligands and three corresponding types of receptors. All ligands can bind to all receptors, but each receptor has an optimal binding with a specific receptor type. The ligands not bound to any receptor are free and exist in the volume of the system.  In this work we derive the conditions under which all ligands can bind to their optimal receptor sites. }
	\label{fig:biophys_ligand_receptor}
\end{figure}

The general system we consider is shown in \reffig{biophys_ligand_receptor}. Say we have a system of ligands and receptors existing in the extracellular medium. The ligands come in multiple copies as do the receptors, and, as is consistent with the multi-specificity of some real ligand systems, we assume all ligands have the ability to bind to all receptors. However, we will also assume that each ligand type has a specific optimal binding with a particular receptor type. This latter assumption will provide us with an additional order parameter with which we can define our system.  

There are some basic questions that a corresponding model for this system should be able to answer: How does the average number of bound ligands of various types depend on the system's binding affinities? How does the average number of \textit{optimally} bound ligands of various types depend on the system's binding affinities?  What thermal conditions define a system in which all ligands are optimally bound? For what properties of the affinities are such conditions even feasible? Are there ways we can categorize these systems so as to determine \textit{a priori} from affinity properties what the expected binding behavior should be? 

We answer these questions in the subsequent sections. In Section \ref{sec:part_gen_derang}, we introduce the main combinatorial problem that underlies the binding of multiple ligand species to multiple receptor species. In Section \ref{sec:assmb_part}, we use the solution to this problem to compute the partition function for the system and show that the result generalizes a special case derived in \cite{williams2019self}. In Section \ref{sec:cases}, we consider the large particle-number limit of the partition function for two limiting cases and the general case. The limiting cases help us build the intuition relevant to understanding the conditions that define fully optimal binding in the general case. In each case, we derive expressions for the average number of bound ligands of each species and the average number of optimally bound ligands, as long as the relevant quantity is not trivially constrained by the case itself. In Section \ref{sec:simulation}, we introduce an image on a grid to give a visual handle on the various limiting cases of the system, and we simulate the grid images to affirm that the analytical results accurately predict the average grid state at various temperatures. In Section \ref{sec:system_type}, we note that the temperature curves for the average number of bound ligands and the average number of optimally bound ligands have distinct limiting behaviors contingent on how model parameters vary with one another. We explore these distinct limiting behaviors through simulations and argue that one limiting behavior is associated with kinetic traps. In Section \ref{sec:implications}, we return to the system that motivated our analyses and discuss the biophysical implications of the results. In Section \ref{sec:extend}, we conclude by considering ways to extend the general model.


\section{Partial Generalized Derangements \label{sec:part_gen_derang}}

Our ultimate goal is to model the equilibrium thermodynamics of systems of the kind shown in \reffig{biophys_ligand_receptor}. Achieving this amounts to computing a partition function and then using the partition function to find order parameters, but first we need to solve the combinatorial problem at the heart of this system.  

We recall that the derangement of a list is a rearrangement of that list such no element is in its original position. The formula for the number of derangements of a list with $N$ unique elements was first obtained by Pierre Mortmort and Nicholas Bernoulli in the the early 18th century \cite{de1713essay}: 
\begin{equation}
d_{N} = \sum_{j=0}^{N} \binom{N}{j} (-1)^{j} (N-j)!.
\label{eq:derang_simp}
\end{equation}

More than 200 years later, Gillis and Even derived the generalization to this result for the case where elements occur with multiple copy number \cite{even1976derangements}. They showed that the number of ways to completely derange an ordered list with $n_1$ elements of type 1, $n_2$ elements of type $2$, $\ldots$, and $n_D$ elements of type $D$ (where $D \leq N$) is 
\begin{equation}
G_{\boldsymbol{n}} = \int^{\infty}_{0} dx \, e^{-x} \prod_{k=1}^{D} (-1)^{n_k} L_{n_k}(x), 
\label{eq:gillis_even}
\end{equation}
where $\boldsymbol{n} \equiv (n_1, n_2, \ldots, n_D)$ and $L_{n}(x)$ is the Laguerre polynomial defined as 
\begin{equation}
L_{n}(x) = \sum_{j=0}^{n} \binom{n}{k} \frac{(-1)^j}{j!} x^{j}.
\label{eq:laguerre}
\end{equation}

We will call Gillis and Even’s result the "generalized derangement result." For this work, we want to obtain a further generalization to the generalized derangement result to the case where not necessarily all elements of an initial list are included in a rearrangement. Finding this generalization would allow us to model a system in which ligands can exist both on and off receptor sites.

The primary problem we need to solve is as follows: 

\begin{quote}
We have $r_1$ elements of type $1$, $r_2$ elements of type $2$, $\ldots$, and $r_D$ elements of type $D$, all of which are arranged in an initial list. All elements are then removed from the list. What is the number of ways that we can choose and arrange $k_1\leq r_1$ elements of type $1$, $k_2 \leq r_2$ elements of type $r_2$, $\ldots$, and $k_D\leq r_D$ elements of type $D$ such that none of the elements has the same position as it has in the original list?
\end{quote}

We call the answer to this question the "partial generalized derangement result," given that we are considering derangements of partial collections of the total set of elements with repeats. The resulting quantity will be denoted $B_{\bs{r}, \bs{k}}$ where $\bs{r} = (r_1, r_2, \ldots, r_D)$ and $\bs{k} = (k_1, k_2, \ldots, k_D)$, and we will obtain an explicit expression for it by reasoning according to the principal of inclusion and exclusion. 

To apply the principal of inclusion and exclusion in the desired case, it is helpful to first review it in the simpler case of \rfw{derang_simp}. With the summation index $j$ denoting the number of elements that are fixed in their original positions, the factor $\binom{N}{j}$ is the number of ways to choose $j$ fixed elements out of $N$ possible elements. The factor $(-1)^{j}$ is the common principle of inclusion and exclusion factor that leads sets of "correct position" elements to be alternately subtracted from and added to the first term of $N!$ which is a count of all permutations. The factor $(N-j)!$ counts the number of ways to arrange the remaining elements given that $j$ are fixed in their original  positions. The end result after summing over all $j$ is a count of only permutations that do not include any elements in their original positions. 

Thus, there are three essential factors in the summand of \rfw{derang_simp}: The counting of the number of ways to arrange elements in their original position; the principle of inclusion and exclusion factor $(-1)$ for each such original-position element; and the factor that counts the number of ways to arrange the remaining elements. 

We can define analogous factors for $B_{\bs{r}, \bs{k}}$ and use them to write a summation expression for the quantity. The result is
\begin{align}
B_{\bs{r}, \bs{k}} & = \sum_{j_1 =0}^{k_1} \cdots \sum_{j_D =0}^{k_D} \binom{r_1}{j_1} \cdots \binom{r_D}{j_D} (-1)^{j_1 + \cdots + j_D}\binom{r_1-j_1 + \cdots + r_D-j_D}{k_1-j_1 + \cdots + k_D-j_D}\mm
 & \qquad \qquad \times \frac{(k_1-j_1 + \cdots + k_D-j_D)!}{(k_1-j_1)! \cdots (k_D-j_D)!}.
\label{eq:Bnk}
\end{align} 

To understand \rfw{Bnk} we consider how each factor in the summand contributes to the final expression. The factor $\binom{r_i}{j_i}$, for $i=1, \ldots, D$, is the number of ways to fill $j_i$ out of the $r_i$ positions of type $i$ with their original elements. The factor $(-1)^{j_1+\cdots+j_D}$ is the net principal of inclusion and exclusion factor for the $j_i$ elements of type $i$ (for $i$ running from $1$ to $D$) that are in their original positions. After fixing these positions with their original elements, there are now $r_1 - j_1 + \cdots + r_D - j_D$ possible positions which we must fill with $k_1-j_1 + \cdots + k_D - j_D$ elements. The number of ways to choose which of these remaining positions to fill is represented by a binomial factor. The last factor ${(k_1-j_1 + \cdots + k_D-j_D)!}/{(k_1-j_1)! \cdots (k_D-j_D)!}$ is the number of ways to permute the $k_1-j_1 + \cdots + k_D - j_D$  elements amongst the chosen positions divided by factors to correct for the fact that elements of the same type are identical. 

To affirm correctness, we can perform some sanity checks on \rfw{Bnk} to show that this result is consistent with related ones. 

For what follows, it will be most useful to express \rfw{Bnk} as an integral expression. To do so, we introduce the generalized Laguerre polynomial:
\begin{equation}
L_{n}^{(\alpha)}(x) = \sum_{j=0}^{n} \binom{n+\alpha}{n-j} \frac{(-1)^j}{j!} x^{j}.
\label{eq:gen_laguerre}
\end{equation}
Using \rfw{gen_laguerre} and the definition of the Gamma function, we find that \rfw{Bnk} can be written as 
\begin{equation}
B_{\bs{r}, \bs{k}} = \frac{1}{(\sum_i\alpha_i)!}\int^{\infty}_{0} dx\, e^{-x}  \prod_{i=1}^D (-1)^{k_i} x^{\alpha_i} L_{k_i}^{(\alpha_i)}(x),
\label{eq:Bnkfin}
\end{equation}
where we defined $\alpha_i \equiv n_i -k_i$.

\begin{figure}[t]
\centering
\includegraphics[width=.85\linewidth]{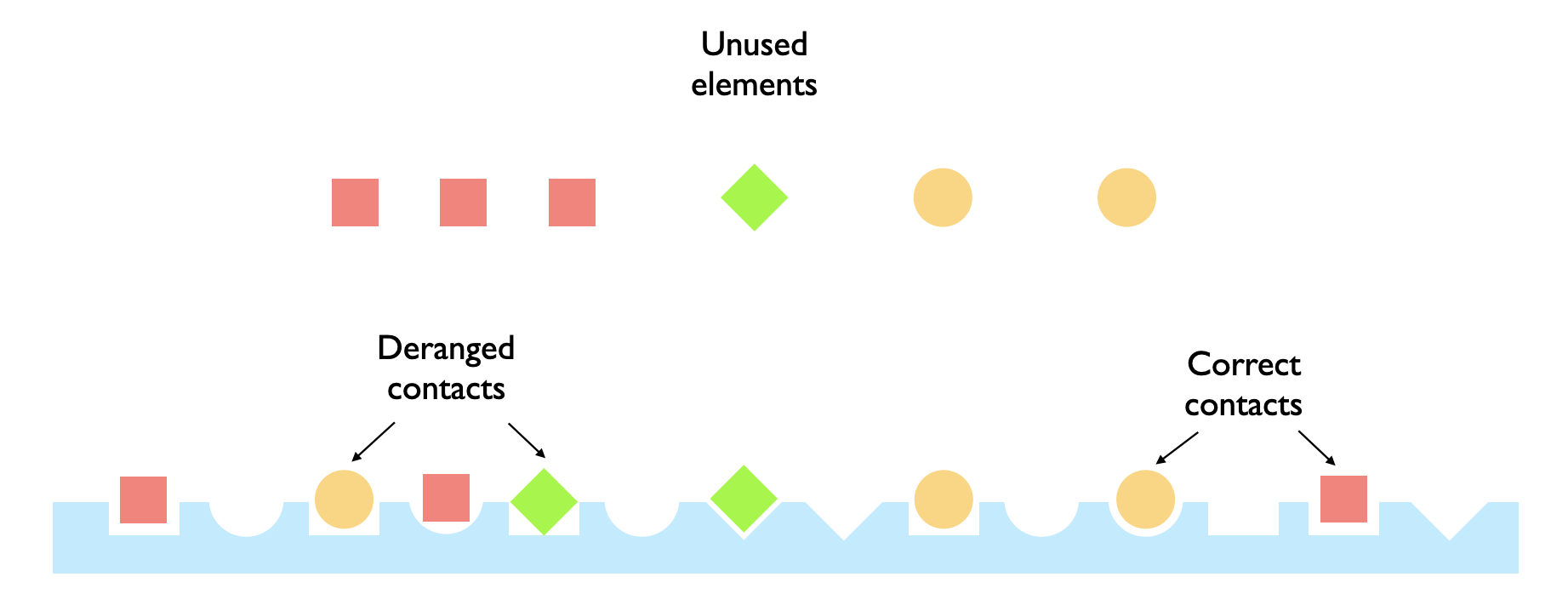}
	\caption{Microstate of partial generalized derangements. In a partial generalized derangement, elements occur in multiple copies and partially occupy deranged positions in a list. The figure shows a microstate for the system with $D=3$, $r_1 = 6$, $r_2 = 3$, and $r_3 = 5$ with squares, triangles, and circles associated with $1$, $2$, and $3$ respectively. There are $k_1 = 3$, $k_2 = 2$, and $k_3 = 3$ elements of the various types in contact with the lattice. Some of the elements on the lattice are in deranged positions and some are in correct positions. Taking $\bs{m}= (m_1, m_2, m_3)$ to define the vector counting the number of elements of each type in correct positions, we have $m_1=2$, $m_2=1$, and $m_3=1$. Given these correct positions, the microstate in this figure contributes to the count for $B_{\bs{r}-\bs{m}, \bs{k}-\bs{m}}$.}
	\label{fig:part_gen_derang}
\end{figure}

For the first sanity check, we expect that \rfw{Bnkfin} should reduce to \rfw{gillis_even} when we take $k_j = n_j$ for all $j$. Namely when we are considering the full (rather than a partial) set of elements, the partial generalized derangement result should reduce to the generalized derangement result. Imposing this equality on \rfw{Bnkfin} and noting that $L_j \equiv L^{(0)}_{j}$, we indeed find that \rfw{gillis_even} is reproduced. 

One can show (as was done in the appendix of \cite{williams2019self}) that if we have $D$ different elements each of which is associated with a particular site out of $D$ lattice sites, then the number of ways to select $K\leq D$ elements to arrange amongst the lattice sites such that none is in its associated site is 
\begin{equation}
b_{D, K} = \sum_{J=0}^D (-1)^{J}\binom{D}{J} \binom{D-J}{K-J}^2(K-J)!.
\label{eq:bNK}
\end{equation}
Thus,  we should be able to show that \rfw{Bnkfin} reduces to \rfw{bNK} under the right conditions. In particular if we take $\bs{r} = (1, 1, \ldots, 1) \equiv \bs{r}_0$ (i.e., we have $D$ unique elements, each of a single copy-number), then the vector $\bs{k}$ in \rfw{Bnkfin} can only have elements of $1$ or $0$, and thus $\bs{k}$ defines a particular subset of the total set of elements. $B_{\bs{r}_0, \bs{k}}$ then represents the number of ways to completely derange a particular collection of unique elements where the collection is defined by the vector $\bs{k}$. In order to find the total number of ways to completely derange $K$ \textit{total} elements (i.e., what is represented in \rfw{bNK}), we need to sum $B_{\bs{r}_0, \bs{k}}$ over all possible values of $\bs{k}$ such that $\sum_j k_j = K$. Thus the consistency check we must make is 
\begin{equation}
\sum_{k_1=0}^{1} \ldots \sum_{k_D=0}^{1} B_{\bs{r}_0, \bs{k}} \delta(K, k_1 + \cdots + k_D) = b_{D, K}.
\label{eq:Bnkb}
\end{equation}

It takes more work to demonstrate \rfw{Bnkb} (see Appendix \ref{app:Bnkb}), but doing so affirms that \rfw{Bnkfin} is consistent with its simpler manifestations. 

As a final consistency check, we note that there should be a summation condition for the total number of ways to select $k_1$ elements of type $1$, $k_2$ elements of type $2$, $\ldots$, $k_D$ elements of type $D$ to be placed amongst the available $r_1 + \cdots +r_D$ positions without regard to whether the elements are placed in their original positions. For the case of simple derangements \rfw{derang_simp}, this summation condition is 
\begin{equation}
N! = \sum_{m=0}^{N} \binom{N}{m} d_{N-m}.
\label{eq:derang_sum}
\end{equation}
\rfw{derang_sum} represents the fact that the total number of ways to order $N$ unique elements is also the number of ways to select $m$ fixed elements and derange the rest summed over all possible values of $m$. It is straightforward to check that \rfw{derang_simp} satisfies \rfw{derang_sum}. 

Towards finding an analogous summation condition for $B_{\bs{r}, \bs{k}}$, we note that $\binom{r_1}{m_1} \cdots \binom{r_D}{m_D}B_{\bs{r}-\bs{m}, \bs{k}-\bs{m}}$ is the number of ways to choose $m_i$ out of $r_i$ positions (for $i=1, \ldots, D$) to contain their original elements while the remaining $k_i-m_i$ elements are completely deranged with respect to the $r_i-m_i$ remaining original positions of type $i$. If we sum this quantity over all possible values of $m_i$, as in 
\begin{equation}
I_{\bs{r}, \bs{k}} \equiv \sum_{m_1 =0}^{k_1} \cdots \sum_{m_D=0}^{k_D} \binom{r_1}{m_1} \cdots \binom{r_D}{m_D}B_{\bs{r}-\bs{m}, \bs{k}-\bs{m}},
\label{eq:Ink0}
\end{equation}
we should obtain the number of ways to arrange (and not necessarily derange) $k_i\leq r_i$ elements of type $i$ for $i=1, \ldots, D$ across a total of $r_1 + \ldots+r_D$ lattice sites. 

Calculating this quantity another way, we note that (including filled and empty sites) we are technically trying to order a total of $r_1+\cdots +r_D$ sites: There are $k_1 +\cdots+k_D$ filled sites and $r_1-k_1 + \cdots +r_D-k_D$ empty sites. Consequently there are $(r_1 + \cdots + r_D)!$ ways to order the total collection. Given that the filled-site elements occur in multiple copies, we must correct for equivalent orderings by dividing this count by $k_j!$ for each element type. Also, since the empty sites act as an extra "type" of element, we must also divide the count by $(r_1-k_1 + \cdots + r_D-k_D)!$, the number of ways to reorder these empty sites. Thus we should find
\begin{equation}
I_{\bs{r}, \bs{k}} =\frac{(r_1 + \cdots + r_D)!}{k_1! \cdots k_D! (r_1-k_1 + \cdots + r_D-k_D)!}.
\label{eq:multi_nom}
\end{equation}
In Appendix \ref{app:multi_nom}, we show that \rfw{Ink0} produces \rfw{multi_nom}.

With our combinatorial expression found and consistency affirmed, we can now work towards building the partition function for the system.

\section{General Partition Function \label{sec:assmb_part}}

We recall that our objective is to study the equilibrium thermodynamics of the physical system depicted in \reffig{biophys_ligand_receptor}. The system is one where a fixed set of ligands can exist as bound or unbound to a collection of receptors. When a ligand is bound to a receptor, it can be bound either to an optimal receptor or to a suboptimal receptor. To study the thermodynamics of such a system, we needed to compute a combinatorial factor that counts the number of ways ligands can be bound to receptor sites where some of these bindings are suboptimal. Having computed this quantity in \refsec{part_gen_derang}, we can now use what we found to calculate the partition function. 

However, before using this derangement formalism, we will begin with minimal assumptions and write the most general expression possible for the partition function of the system. The intent in starting here is to show the intractability of the most general form of the partition function and thereby demonstrate the analytical benefits afforded by considering derangements from a pre-defined sequence. 
 
We start by defining numerical quantities in the system. Say that we have $D$ different types of ligands and a corresponding set of $D$ different types of receptors. A ligand type and a receptor type are labeled with $i$ for $i=1,\ldots, D$. The ligand of type $i$ has $n_i$ copies in the system, and the receptor of type $i$ has $r_i$ copies in the system. Each ligand can either be bound to a receptor or be unbound and free to move in the space surrounding the receptor sites. There are $N_R \equiv r_1 + \cdots + r_D$ total receptors and each of the $N_{L} \equiv n_1 + \cdots + n_D$ ligands can bind to any one of them, provided there is an available binding site.

In the most general theoretical formulation of the problem, we can represent the system microstate by a matrix ${\cal C}$ with elements ${\cal C}_{i, j}$ defined as 
\begin{equation}
{\cal C}_{i, j} =  \text{$\#$ of bindings between ligand of type $i$ and receptor of type $j$}. 
\end{equation}
If we can specify all elements of the $D\times D$ matrix  ${\cal C}$ then we have completely defined the system. 
Given our counts for the total number of ligands and receptors of each type, there are only three constraints on the elements ${\cal C}_{i, j}$: Each element must be an integer, $\sum_{j=1}^D {\cal C}_{i, j} \leq n_i$, and $\sum_{i=1}^D {\cal C}_{i, j} \leq r_j$. 

Next, we ask how we can incorporate binding parameters to represent the way energy affects the probability of a microstate. We will take $Q_{i, j}^B$ to be the single-particle partition function for a ligand of type $i$ that is bound to a receptor of type $j$. Therefore, the multi-particle partition function for all ligands of type $i$ that are bound to receptors of type $j$ is  $(Q^B_{i, j}) ^{{\cal C}_{i, j}}$. There is no factorial correction in this expression because ligands that are bound to specific receptor sites are distinguishable by virtue of the distinguishability of the receptor sites themselves. Conversely, taking $Q_{i}^F$ to be the single-particle partition function for an unbound ligand of type $i$, and given that $n_i - \sum_{j=1}^D{\cal C}_{i, j}$ is the total number of unbound ligands of type $i$, the multi-particle partition function for the ligand of type $i$ that is unbound is ${\left(Q^F_{i}\right)^{n_i - \sum_{j=1}^D{\cal C}_{i, j}}}/{(n_i - \sum_{j=1}^D{\cal C}_{i, j})!}$ where the factorial correction is because these ligands are in free space. Putting the pieces together, and including appropriate products to account for various ligand and receptor types, we find that the general partition function for this system is 
\begin{equation}
{\cal Z}_{\bs{n}, \bs{r}} = \sum_{\{{\cal C}_{i, j}\}} \prod_{i=1}^{D}\frac{\left(Q^F_{i}\right)^{n_i - \sum_{j=1}^D{\cal C}_{i, j}}}{(n_i - \sum_{j=1}^D{\cal C}_{i, j})!} \,\prod_{j=1}^{D} (Q^B_{i, j}) ^{{\cal C}_{i, j}},
\label{eq:base_part_func}
\end{equation}
where $\bs{n} \equiv (n_1, \ldots, n_D)$ and $\bs{r} = (r_1, \ldots, r_D)$. The product over $j$ is the product over all types of receptors to which ligand $i$ is bound. The product over $i$ is the product over all types of ligands, with the associated factors representing free and bound ligands. 
The external summation is a sum over all possible values of ${\cal C}_{i, j}$ according to the constraints of the physical system (i.e., ${\cal C}_{i, j}$ is an integer, $\sum_{j=1}^D {\cal C}_{i, j} \leq n_i$ and $\sum_{i=1}^D {\cal C}_{i, j} \leq r_j$). 

With \rfw{base_part_func}, we have our exact partition function for the system. However, it is unclear how to make this exact expression analytically useful. Ostensibly we need to enumerate and then sum over all possible matrices ${\cal C}$, but this is unfeasible given the number of microstates associated with even simpler contact matrices. For the products, simplifications often occur when products can be turned into summations, but the matrix nature of the factors in \rfw{base_part_func} seems to preclude this conversion. Therefore, we cannot compute analytical expressions for observables from \rfw{base_part_func} since such quantities depend on tractable calculations of the partition function. Because of these reasons, we need to make some simplifying assumptions to make progress. 

In our system, each ligand of type $i$ will bind to a receptor of type $j$ with a binding Boltzmann factor $Q_{i, j}^{B}$. The matrix nature of this expression seems to be the principal complication in our partition function since it prevents us from converting the product over $j$ into a sum in the power. Therefore, to simplify this expression we will make an assumption that reduces the dimensionality of the parameters space of this matrix. 

While still assuming that any ligand can bind to any receptor, we will also assume that every ligand of type $i$ has the same binding affinity to every type of receptor \textit{except} to a receptor of type $i$ to which the ligand of type $i$ binds with an additional energy of binding of $\Delta_i$ (with $\Delta_i \geq 0$). Mathematically, we can encode this assumption into the model by making the transformation  
\begin{equation}
Q_{i, j}^{B} \quad \longrightarrow \quad (Q^B_{i})e^{\beta\Delta_i \delta_{i, j}}, 
\label{eq:simplifying}
\end{equation}
where $\delta_{i, j}$ is the Kronecker delta, and $\beta = 1/k_BT$ for a system at temperature $T$. Essentially, \rfw{simplifying} asserts that all ligands of type $i$ that are bound to a receptor have the same single-particle partition function $Q^B_{i}$ except for the type $i$ ligands that are bound to type $i$ receptors which acquire an additional Boltzmann factor $e^{\beta \Delta_i}$. We call the latter such bindings "optimal" or "correct" bindings; all other bindings are termed "sub-optimal." 

One benefit of \rfw{simplifying} is that it reduces the number of parameters that are needed to define the system: Rather than have $N^2$ binding parameters defined by the elements of $Q_{i, j}^{B}$, we have $2N$ parameters from $(Q^B_i)$ and $\Delta_i$ together, a reduction that makes our modeling more tractable.

Regarding the physical motivation of this assumption,  \rfw{simplifying} closely approximates the ligand-receptor interactions of signaling pathways such as the Wnt-Fz pathway which exhibits specific binding for the initiation of signals (i.e., a nonzero $\Delta_i$) while still having ligands with promiscuous interactions with receptors \cite{eubelen2018molecular}. 

How does \rfw{simplifying} change our partition function? Starting from \rfw{base_part_func} and incorporating the transformation \rfw{simplifying}, we obtain 
\begin{align}
{\cal Z}_{\bs{n}, \bs{r}} 
 & = \sum_{\{{\cal C}_{i, j}\}} \prod_{i=1}^{D}\frac{\left(Q^F_{i}\right)^{n_i - \sum_{j=1}^D{\cal C}_{i, j}}}{(n_i - \sum_{j=1}^D{\cal C}_{i, j})!} \, (Q^B_{i})^{\sum_{j=1}^D {\cal C}_{i, j}}e^{\beta \Delta_i \sum_{j=1}^{D} {\cal C}_{i, j} \delta_{i, j}}
\label{eq:base_part_func2}
\end{align}
From, here we make a change of variables motivated by the quantities that appear in the expression. We define $k_i = \sum_{j=1}^D {\cal C}_{i, j}$ as the total number of bound (optimally or not) ligands of type $i$ and $m_i = \sum_{j=1}^D {\cal C}_{i, j}\delta_{j,i}$ as the total number of optimally bound ligands of type $i$.  \rfw{base_part_func2} then becomes 
\begin{align}
{\cal Z}_{\bs{n}, \bs{r}} 
 & = \sum_{\bs{k}} \sum_{\bs{m}} \Omega(\bs{k}, \bs{r}, \bs{m}) \prod_{i=1}^{D}\frac{\left(Q^F_{i}\right)^{n_i - k_i}}{(n_i - k_i)!} \, (Q^B_{i})^{k_i}e^{\beta \Delta_i m_i}
\label{eq:base_part_func3}
\end{align}
where we defined the summations as
\begin{equation}
\sum_{\bs{k}}  \equiv \prod_{i=1}^{D} \sum_{k_i = 0}^{n_i}, \qquad \sum_{\bs{m}}  \equiv \prod_{i=1}^{D} \sum_{m_i = 0}^{k_i},
\label{eq:sum_defs}
\end{equation}
and the combinatorial factor as
\begin{equation}
\Omega(\bs{k}, \bs{r}, \bs{m})  = \sum_{\{{\cal C}_{i, j}\}}\prod_{i=1}^D \delta\left(k_i, \textstyle\sum_{j=1}^D{\cal C}_{i, j}\right) \delta\left( m_i ,  \textstyle\sum_{j=1}^D{\cal C}_{i, j}\delta_{i, j}\right) 
\end{equation}
resulting from the change of summation variables. Qualitatively $\Omega(\bs{k}, \bs{r}, \bs{m})$ is the number of ways to fill $m_i$ out of $r_i$ receptor sites (for $i=1, \ldots , D$) with their optimal binding partner ligands while having the remaining $k_i -m_i$ bound ligands (again for $i=1, \ldots, D$) \textit{not} in their associated optimal site. \rfw{base_part_func2} is already an improvement over \rfw{base_part_func} since there are no longer any matrices as factors and our summations are over specific integer-valued variables (i.e., $k_i$ and $m_i$) rather than elements of a set (i.e., $\{{\cal C}_{i, j}\}$). However, it seems that in order to compute \rfw{base_part_func3}, we would need to compute $ \Omega(\bs{k}, \bs{r}, \bs{m})$ which according to \rfw{sum_defs} seems to require a summation over the elements of the aforementioned set. Fortunately this is not the case: We already have the necessary quantities to compute this factor. Given the qualitative definition of $\Omega$, we can assert that 
\begin{equation}
\Omega(\bs{k}, \bs{r}, \bs{m}) = B_{\bs{r}-\bs{m}, \bs{k}- \bs{m}}  \prod_{i=1}^{D}\binom{r_i}{m_i},
\label{eq:omeg_def}
\end{equation}
where $B_{\bs{n}, \bs{k}}$ is defined in \rfw{Bnkfin}. On the right-hand side of \rfw{omeg_def}, the factor $B_{\bs{r}-\bs{m}, \bs{k}- \bs{m}}$ represents the number of ways to select and arrange $k_i-m_i$ ligands of type $i$ (for $i=1, \ldots, D$) across a total set of $r_1 -m_1 + \ldots r_D-m_D$ receptors such that no ligand of type $i$ is bound to one of its $r_i-m_i$ optimal receptors. The factors $\binom{r_1}{m_1} \cdots \binom{n_D}{m_D}$ count the number of ways to choose $m_i$ receptors from the $r_i$ possible receptors for $i=1, \ldots, D$ to be occupied by optimal-binding partner ligands. Thus, we see that the product of $B_{\bs{r}-\bs{m}, \bs{k}- \bs{m}}$ and $\binom{r_1}{m_1} \cdots \binom{n_D}{m_D}$ is indeed the number of ways to fill $m_i$ out of $r_i$ receptor sites (for $i=1, \ldots , D$) with their optimal binding partner ligands while having the remaining $k_i -m_i$ bound ligands (again for $i=1, \ldots, D$) \textit{not} in their associated optimal sites.

With the observation that $\Omega$ can be computed from \rfw{omeg_def}, we can now obtain a simpler expression for \rfw{base_part_func3}: We have
\begin{align}
{\cal Z}_{\bs{n}, \bs{r}} = \sum_{\bs{k}} \sum_{\bs{m}} B_{\bs{r}-\bs{m}, \bs{k}- \bs{m}}  \prod_{i=1}^{D}\binom{r_i}{m_i} e^{\beta m_i\Delta_i} (Q_{i}^B)^{k_i} \frac{(Q_i^F)^{n_i-k_i}}{(n_i-k_i)!},
\label{eq:init_part}
\end{align}
We recall that in the summands of \rfw{init_part}, $k_i\leq n_i$ represents the total number of ligands of type $i$ that are bound to receptors, and $m_i \leq k_i$ represents the total number of ligands of type $i$ that are optimally bound to receptors.

For notational simplicity, we will define some additional constants. We define 
\begin{equation}
c_{\bs{n}} \equiv \prod_{i=1}^D{(Q_i^F)^{n_i}}, \qquad \delta_i \equiv e^{\beta \Delta_i}, \qquad \gamma_i \equiv \frac{Q_i^B}{Q_i^F}, 
\label{eq:coeff_def}
\end{equation}
thus giving us 
\begin{align}
{\cal Z}_{\bs{n}, \bs{r}} = c_{\bs{n}} \sum_{\bs{k}} \sum_{\bs{m}} B_{\bs{r}-\bs{m}, \,\bs{k}- \bs{m}}  \prod_{i=1}^{D}\binom{r_i}{m_i} \frac{1}{(n_i-k_i)!} \,\delta_{i}^{m_i} \gamma_i^{k_i}.
\label{eq:init_part2}
\end{align}

Given that pre-factors do not affect physical predictions in canonical partition functions, \rfw{init_part2} reveals that it is only the ratios of our single-particle partition functions that are thermodynamically relevant. This result makes sense given that only free-energy differences (i.e., logarithms of partition function ratios) should affect the physics of a system. Thus, without loss of generality, we can impose $Q_i^F=1$ for all $i$ under the assumption that the thermal dependence of each $Q_i^F$ can be absorbed into a redefinition of $\gamma_i$ and $\delta_i$ with no change in the physical implications of \rfw{init_part2}. With this imposition we have 
\begin{equation}
c_{\bs{n}} =1 \qquad \text{[free-particle partition function normalization]}
\label{eq:free_norm}
\end{equation}
Moving forward, we recognize that the partition function becomes more analytically useful to us if we can replace the discrete summation with an integral since integrals, unlike discrete summations, are more amenable to the methods of analysis. To do so we make use of the integral form of $B_{\bs{n}, \bs{k}}$ in \rfw{Bnkfin} and a few Laguerre polynomial identities. After some work (see Appendix \ref{app:gen_part_func}), we obtain
\begin{equation}
{\cal Z}_{\bs{n}, \bs{r}}  =  \frac{1}{2\pi i} \oint_{\Gamma} \frac{dz}{z} \int^{\infty}_{0} dx\, e^{z-x} \prod_{j=1}^{D} (\gamma_j (\delta_j-1))^{n_j} \left(\frac{x}{z}\right)^{r_j-n_j} L_{n_j}^{(r_j-n_j)} \left(\frac{x(z \gamma_j+1)}{z\gamma_j(1-\delta_j)} \right).
\label{eq:master_derang_part01}
\end{equation}
We can write this result in a more mathematically useful form by expressing the integrand as the exponential of a potential function:
\begin{align}
{\cal Z}_{\bs{n}, \bs{r}}(\bs{\delta}, \bs{\gamma}) =  \frac{1}{2\pi i} \oint_{\Gamma} \frac{dz}{z} \int^{\infty}_{0} dx\, \exp\big[{\cal F}_{\bs{n}, \bs{r}}(z, x;\bs{\delta}, \bs{\gamma})\big],
\label{eq:master_derang_part}
\end{align}
where $\Gamma$ is a closed contour about the origin in the complex plane and 
\begin{align}
{\cal F}_{\bs{n}, \bs{r}}(z, x; \bs{\delta}, \bs{\gamma}) & = z-x +(N_R- N_{L})\ln (x/z)\mm
 &\qquad \qquad +\sum_{j=1}^{D}\ln \left[(\gamma_j (\delta_j-1))^{n_j}L_{n_j}^{(r_j-n_j)} \left(\frac{x(z \gamma_j+1)}{z\gamma_j(1-\delta_j)} \right) \right],
\label{eq:master_FRdef}
\end{align}
with $L^{(\alpha)}_n(x)$ the $n$th generalized Laguerre polynomial, and $N_R \equiv \sum_{j=1}^{D} r_j$ and $N_{L} \equiv \sum_{j=1}^D n_j$, respectively, the total number of receptors and total number of ligands in the system. \rfw{master_derang_part} provides the starting point for our thermal equilibrium analysis. But first we will derive expressions for the order parameters written in terms of this partition function.

From \rfw{init_part2}, we can derive expressions for the two main observables of the system. The average number of bound ligands and the average number of optimally-bound ligands are, respectively,
\begin{equation}
\langle k \rangle = \sum_{j=1}^{D} \langle k_j \rangle= \sum_{j=1}^{D} \gamma_j\frac{\partial}{\partial \gamma_j} \ln {\cal Z}_{\bs{n}, \bs{r}}, \qquad \langle m \rangle = \sum_{j=1}^{D} \langle m_j\rangle= \sum_{j=1}^{D} \delta_j \frac{\partial}{\partial \delta_j} \ln {\cal Z}_{\bs{n}, \bs{r}}.
\label{eq:obs}
\end{equation}

We can use the second equation in \rfw{obs} to write an alternative expression for $\langle m \rangle$. For the function $f_n^{(\alpha)}(x; q) = (q-1)^n L_n^{(\alpha)}(x/(1-q))$, the identity $u \,\partial_u L_n^{(\alpha)}(u) = nL_{n}^{(\alpha)}(u) - (n+\alpha)L_{n-1}^{(\alpha)}(u)$ allows us to prove 
\begin{equation}
\partial_qf_n^{(\alpha)}(x; q) = (n+\alpha) f_{n-1}^{(\alpha)}(x; q). 
\label{eq:fnderiv}
\end{equation}
We can then show
\begin{align}
\frac{\partial}{\partial \delta_i}\left[ (\gamma_i (\delta_i-1))^{n_i} L_{n_i}^{(r_i-n_i)} \left(\frac{x(z \gamma_i+1)}{z\gamma_i(1-\delta_i)} \right) \right]
& = r_i \gamma_i (\gamma_i(\delta_i-1))^{n_i-1} L_{n_i-1}^{(r_i-n_i)} \left(\frac{x(z \gamma_i+1)}{z\gamma_i(1-\delta_i)} \right).
\label{eq:fnderiv_imp}
\end{align}
Thus from \rfw{obs}, we have
\begin{equation}
\langle m \rangle = \sum_{j=1}^{D} r_j \gamma_j \delta_j \frac{{\cal Z}_{\bs{n}_j, \bs{r}_j}(\bs{\delta}, \bs{\gamma})}{{\cal Z}_{\bs{n}, \bs{r}}(\bs{\delta}, \bs{\gamma})},
\label{eq:m_reduced}
\end{equation}
where $\bs{n}_{j}$ is $\bs{n}$ with 1 subtracted from the $j$th component: $\bs{n}_{j}= (n_1, \ldots, n_{j}-1, \ldots, n_D)$. The vector $\bs{r}_j$ is defined similarly. Although the term $r_i-1$ does not appear in \rfw{fnderiv_imp}, we had to introduce $\bs{r}_j$ into the expression \rfw{m_reduced} to ensure that $r_j -n_j$ in \rfw{master_derang_part01} remained unchanged when we replaced $\bs{n}$ with $\bs{n}_j$. \rfw{m_reduced} could also have been derived from \rfw{init_part2} by differentiating with respect to $\delta_j$ and making the dummy variable replacement $k_j' = k_j -1$ and $m'_j = m_j-1$. There is no simplified expression for $\langle k \rangle$ analogous to \rfw{m_reduced}. 

As is common for partition functions written as integrals, approximating the partition function by the maximum (or, in the case of complex values, stationary) value of its integrand allows us to derive more tractable expressions for the equilibrium conditions. Before we pursue these conditions, we will show how \rfw{master_derang_part} is a generalization of a result established in a previous paper. The purpose of establishing such a generalization is to extrapolate some of the physical results explored in that previous paper to this more complex case. 

\subsection{Gendered Dimer-System Assembly \label{sec:gender}}

\begin{figure}[t]
\centering
\includegraphics[width=.45\linewidth]{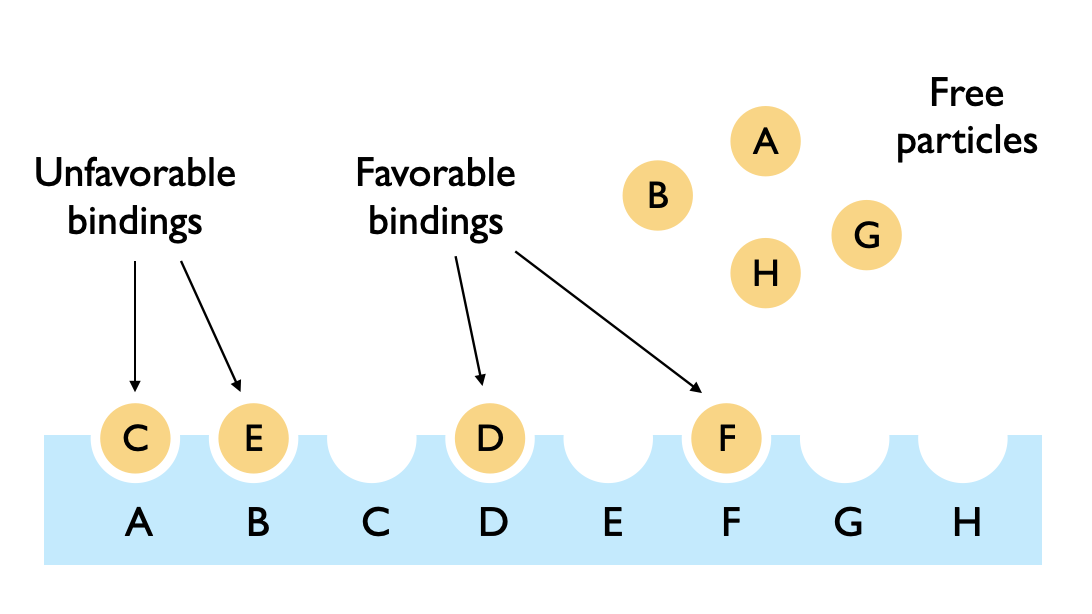}
	\caption{System for gendered dimer assembly: The general system studied in this work for the case of $n_i =1$. For this case, there is one copy of each particle species and each particle has a one-to-one correspondence with an optimal binding site. Also, the binding affinities and optimal-binding affinity advantages for each particle vary with the particle species. Consequently, this case is a slight generalization of the case of dimer assembly with fixed binding sites treated in \cite{williams2019self}. }
	\label{fig:dimer_assembly}
\end{figure}

In the appendix of \cite{williams2019self}, we analyzed a physical system termed "gendered dimer assembly." We recall that this refers to a system where there are two types of particles and where a particle of one type can only bind to a particle of the other type. By fixing the positions of all particles of one type, we were able to apply the results to the case of particles binding to a lattice of possible sites (e.g., ligands binding to receptors). We assumed each particle type had a single copy and that all particles had the same binding affinities to the lattice and the same optimal-binding affinities to their correct sites. 

In this section we use the general expression \rfw{master_derang_part} to derive a slight generalization of these past results. In \cite{williams2019self} we assumed a global binding affinity and optimal-binding affinity for all types of particles, but here we will assume that particles' binding affinities and optimal-binding affinities vary according to particle type. In effect, in \rfw{master_derang_part} we will take $r_i = n_i = 1$ for all $i$ but retain the index dependence of $\delta_i$ and $\gamma_i$. Since gendered dimer assembly (with one of the "genders" fixed in space) is a more specific case of the ligand-receptor binding system considered in this work, we should find that the equilibrium equations derived for this special case of \rfw{master_derang_part} match those found for the gendered dimer system in \cite{williams2019self}.

First, imposing the condition $r_i = n_i=1$ for all $i$ on \rfw{master_derang_part}, we find the partition function 
\begin{equation}
Z_D(\bs{\delta}, \bs{\gamma}) =  \frac{1}{2\pi i} \oint \frac{dz}{z} \int^{\infty}_{0} dx\, \exp\left[ F_{D}(z, x; \bs{\delta}, \bs{\gamma}) \right],
\label{eq:ZR_gen}
\end{equation}
where 
\begin{equation}
F_D(z, x; \bs{\delta}, \bs{\gamma} ) = z-x + \sum_{i=1}^D\ln \Big(\gamma_j(\delta_j-1+x) + \frac{x}{z} \Big).
\label{eq:FR_def}
\end{equation}
The corresponding average number of bound particles and average number of optimally bound particles for a particle of type $j$ are the same as what is given in \rfw{obs}:
\begin{equation}
\langle k_j \rangle = \gamma_j \frac{\partial}{\partial \gamma_j} \ln Z_D, \qquad \langle m_j \rangle = \delta_j \frac{\partial}{\partial \delta_j} \ln Z_D.
\label{eq:kmdef0}
\end{equation}
Applying the large $N$ integral approximation (specifically $D\gg1$ in this case) to \rfw{ZR_gen} yields 
\begin{equation}
Z_D(\bs{\delta}, \bs{\gamma}) \simeq \frac{1}{(\bar{z}^2 \det H)^{1/2}} \exp\left[F_D(\bar{z}, \bar{x}; \bs{\delta}, \bs{\gamma} ) \right] ,
\label{eq:ZR_gen_0}
\end{equation}
where $H$ is the hessian matrix with second-order derivative components $H_{\alpha, \beta} = \partial_{\alpha, \beta}F_D\big|_{z = \bar{z}, x = \bar{x}}$ ($\alpha, \beta \in \{x, z\}$), and $\bar{z}$ and $\bar{x}$ are defined by the conditions
\begin{equation}
0 = \partial_{x} F_{D}(z, x; \bs{\delta}, \bs{\gamma})\Big|_{z = \bar{z} x = \bar{x}}, \qquad 0 = \partial_{z} F_{D}(z, x; \bs{\delta}, \bs{\gamma})\Big|_{z = \bar{z}, x = \bar{x}}.
\label{eq:FR_eqbm_0}
\end{equation}
In order to compute equilibrium conditions for $\langle k_j \rangle$ and $\langle m_j \rangle$ from \rfw{ZR_gen_0} we first need the conditions for $\bar{x}$ and $\bar{z}$\footnote{When applying the saddle-point approximation or Laplace's method to an integral, one should check that the second derivative matrix has the correct stability properties to ensure the validity of the approximation. In the most general case in this work, this check requires us to compute the spectral properties of a $2\times2$ complex matrix. The difficulty in proving stability in this general scenario has led us to instead use simulations to heuristically vet the validity of the approximation}. Using \rfw{FR_def} and \rfw{FR_eqbm_0} to find these conditions, we have, from $\partial_{x} F_{D} =0$ and $\partial_{z} F_{D}=0$, respectively,
\begin{equation}
1 = \sum_{j=1}^D \frac{\bar{z}\gamma_j +1}{\bar{x} + \bar{z}\gamma_j(\delta_j -1 + \bar{x})}, \qquad \bar{z} = \bar{x} \sum_{j=1}^D \frac{1}{\bar{x} + \bar{z} \gamma_j (\delta_j-1 + \bar{x})}.
\label{eq:eqbm_zx_0}
\end{equation}
Next, computing $\langle k_j \rangle$ and $\langle m_j \rangle$, we have
\begin{align}
\langle k_j \rangle  &= \gamma_j \frac{\partial}{\partial \gamma_j} F_D = \frac{\bar{z}\gamma_j (\delta_j -1 + \bar{x})}{\bar{z}\gamma_j (\delta_j -1 + \bar{x})+\bar{x}} \label{eq:obs01_k} \\
 \langle m_j \rangle & = \delta_j \frac{\partial}{\partial \delta_j} F_D= \frac{\bar{z}\gamma_j \delta_j}{\bar{z}\gamma_j (\delta_j -1 + \bar{x})+\bar{x}} \label{eq:obs01_m}
\end{align}
where, in applying \rfw{kmdef0} to \rfw{ZR_gen_0}, we neglected the exponential pre-factor in the latter since it is subleading in the $D\gg1$ limit.

Using \rfw{eqbm_zx_0} to eliminate the $\bar{x}$ and $\bar{z}$ from \rfw{obs01_k} and \rfw{obs01_m} (see Appendix \ref{app:gendered}), we find the coupled equilibrium conditions
\begin{align}
\sum_{j=1}^D\frac{1}{\gamma_j}\Big( \langle k_j \rangle - \langle m_j \rangle(1-\delta_j^{-1})\Big) & = \Big(N- \langle k \rangle\Big)^2 \label{eq:dimer_eq1}\\
\sum_{j=1}^D\langle m_j \rangle \delta_{j}^{-1} & = \frac{ \langle k \rangle - \langle m\rangle + \sum_{j=1}^D\langle m_j\rangle\delta_j^{-1}}{N - \langle m \rangle + \sum_{j=1}^D\langle m_j\rangle\delta_j^{-1}},
\label{eq:dimer_eq2}
\end{align}
where $N \equiv \sum_{j=1}^D n_j = \sum_{j=1}^D r_j = D$. \rfw{dimer_eq1} and \rfw{dimer_eq2} define how the average number of bound and optimally-bound particles for each species $j$ vary with one another and with the parameters for binding affinity $\gamma_j$ and optimal-binding affinity $\delta_j$. For practical purposes, when trying to solve this system of equations it is necessary to first solve \rfw{eqbm_zx_0} and then insert the obtained values of $\bar{z}$ and $\bar{x}$ into  \rfw{obs01_k} and \rfw{obs01_m} to find $\langle k_j \rangle$ and $\langle m_j \rangle$. But \rfw{dimer_eq1} and \rfw{dimer_eq2} do provide an affirming pathway to more familiar results. If we take $\delta_j = \delta$ and $\gamma_j = \gamma$ for all $j$, and note that $\langle k \rangle = \sum_{j=1}^D \langle k_j \rangle$ (and similarly for $\langle m_j\rangle$), we find 
\begin{align}
\frac{1}{\gamma}\Big( \langle k \rangle - \langle m\rangle(1-\delta^{-1})\Big) = \Big(N- \langle k\rangle\Big)^2, \qquad \langle m \rangle \delta^{-1} = \frac{\langle k\rangle - \langle m\rangle(1-\delta^{-1})}{ N - \langle m \rangle (1-\delta^{-1})}.
\label{eq:dimer_eqs02}
\end{align}
The results in \rfw{dimer_eqs02} are the very same ones we found in \cite{williams2019self} for the gendered dimer system. 

By imposing the condition $\langle k \rangle = \langle m \rangle$ on the second equation in \rfw{dimer_eqs02}, we can show 
\begin{equation}
\langle k \rangle = \langle m \rangle = \frac{N-1}{1- \delta^{-1}},
\label{eq:gend_cond}
\end{equation}
suggesting that the condition $\langle k \rangle = \langle m \rangle$ only occurs when essentially all the particles are bound to their optimal binding sites. Inserting this value for $\langle k \rangle$ and $\langle m\rangle$ into the second equation of \rfw{dimer_eqs02} yields the thermal condition under which this fully optimal binding configuration occurs. We find
\begin{equation}
N-1 = \gamma \delta \frac{(1 - N\delta^{-1})^2}{1- \delta^{-1}}.
\label{eq:simp_bind_cond}
\end{equation}
In \cite{williams2019self}, we used \rfw{simp_bind_cond} to infer the existence of generally two types of binding systems with quite different relationships between $\langle k \rangle$ and $\langle m \rangle$. When $\delta\gg \gamma > 1$, \rfw{simp_bind_cond} became $ \gamma\delta \simeq N $ and we had a "search-limited" system in which optimal binding was primarily limited by the ability of particles to find their optimal binding site in the surrounding volume; when $\gamma \gg \delta > 1$, \rfw{simp_bind_cond} became $\delta \simeq N$ and we had a "combinatorics-limited" system in which optimal binding was primarily limited by the ability of particles to avoid the combinatorial sea of suboptimal contacts. 

As we increased the temperature in search-limited systems, the value of $\langle m\rangle$ remained close to the value of $\langle k \rangle$ thus indicating that such systems could have partial binding to sites but with all such bindings being optimal. Conversely, in combinatorics-limited systems, increasing the temperature led to the value of $\langle m \rangle$ being much lower than the value of $\langle k \rangle$ indicating that when particles were bound, such binding was likely suboptimal. With some heuristic arguments, we suggested that biophysical systems are more likely to be of the search-limited type, but such an inference was limited by the simplicity of our model. 

In this work, we want to extend the analysis in this simpler case to one where there are multiple particle types of various copy number and various binding and optimal-binding affinities. For this general case, the objective is to find a condition akin to \rfw{simp_bind_cond} that will allow us to distinguish various binding behaviors in the system and thus tell us if our previous combinatorics-limited and search-limited framings still apply. Due to its incorporation of multiple-copy number and type-dependent binding affinities, this more general case will be more biophysically relevant and could thus serve as a firmer basis for categorizing biophysical systems as one of the two types.

But before we consider this most general case, we consider two more specific cases to build the intuition and techniques for how binding and combinatorics affect ligand-receptor systems. 

\section{Large $N$ Limits of Special and General Cases \label{sec:cases}}

In studying the system modeled by \rfw{ZR_gen}, we will first work through two special cases that establish the intuition and methods we will later apply to the most general case. The first special case is that of $\delta_i =1$ for all $i$. This is the case where different ligand-types may have different binding affinities to the set of receptors, but all receptors are equivalent from the perspective of a single ligand-type. The second special case is that of $\gamma_i \to \infty$ corresponding to a system where ligands can only  exist as attached to a receptor and where the various microstates consist of derangements of the ligands amongst the set of receptors. With these two cases established, we will then consider the general case with no prior assumptions on the values of $\delta_i$ and $\gamma_i$.

\begin{figure*}[t]
\begin{centering}
\begin{subfigure}[t]{0.43\textwidth}
\centering
\includegraphics[width=\linewidth]{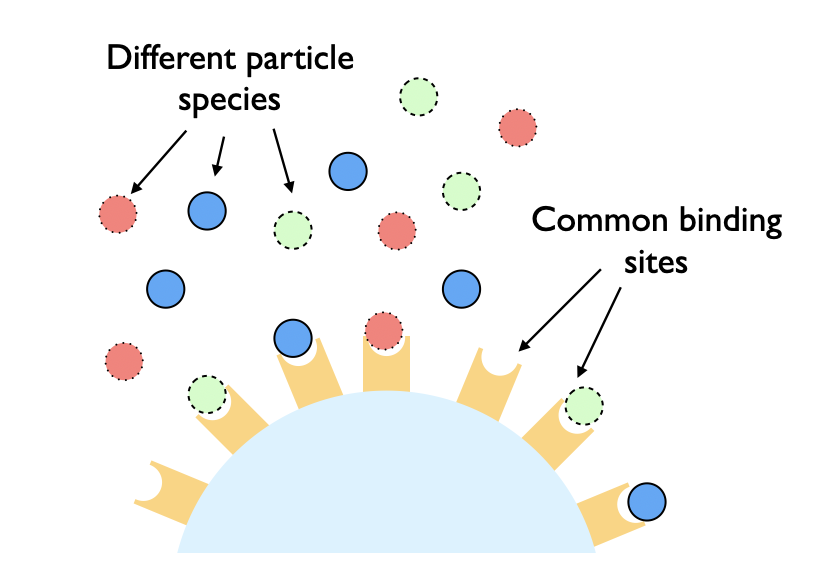}
	\caption{}
	\label{fig:binding}
\end{subfigure} \qquad
\begin{subfigure}[t]{0.43\textwidth}
\centering
\includegraphics[width=\linewidth]{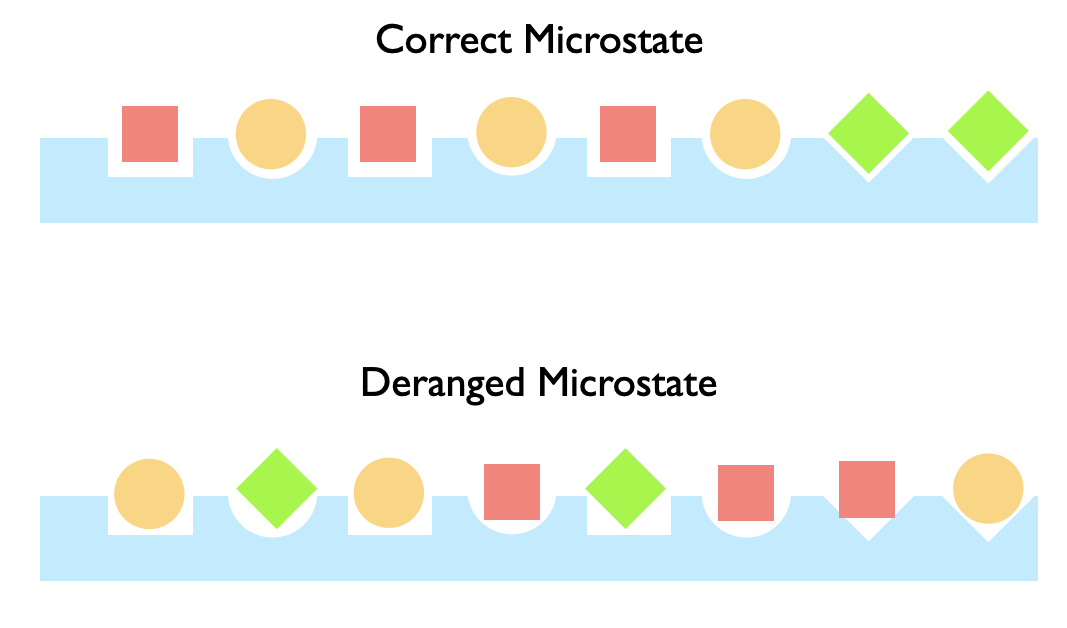}
	\caption{}
	\label{fig:deranged_micro}
\end{subfigure} 
	\caption{Two limiting cases for the general model. In (a), we have the case of $\delta_i =1$. For this case, all receptors are equivalent and a distinct ligand species (denoted with a distinct color or border in the figure) of type $i$ has a binding affinity of $\gamma_i$.  In (b), we have the case where $\gamma_i \to \infty$. For this case, all ligands are bound to the lattice, and the system microstates consist of the various ways to permute a list with repeated elements. The top figure shows the "correct microstate" (where each ligand is in its optimal binding site), and the bottom figure shows a "deranged microstate" (where each ligand is in a suboptimal binding site). The most general case of ligand-receptor binding is a combination of these two cases. }
	\label{fig:two_cases}
\end{centering}
\end{figure*}

\subsection{Simple Binding Model \label{sec:bind}}

One simplification of the most general scenario associated with  \rfw{master_derang_part}  is to have each ligand type have the same binding affinity regardless of to which receptor it binds. This simplification amounts to taking $\delta_i =1$ (or $\Delta_i =0$ by \rfw{coeff_def}). Phrased differently, this condition implies that, for a single type, the optimally and suboptimally bound ligand partition functions are equal, and thus there is no thermal advantage for a ligand to be bound to any particular receptor; from the perspective a single type of ligand, all receptors are thermodynamically identical. However, the ligands of different types are not thermodynamically identical to each other: Since $\gamma_i$ is not presumed to the be the same for all $i$, each ligand type has a different binding affinity to an arbitrary receptor. Thus, our system contains many distinguishable ligands in the presence of many identical receptors. 

The thermal implications of this condition can be found by computing the partition function. Taking $\delta_i \to 1$ in \rfw{master_derang_part01},
we can define the new partition function 
\begin{equation}
W_{\bs{n}, \bs{r}}(\bs{\gamma}) \equiv \lim_{\{\delta_i \to 1\}} {\cal Z_{\bs{n}, \bs{r}}(\bs{\delta}, \bs{\gamma})}.
\label{eq:Wdef}
\end{equation}
Using the limit identity $
\lim_{\lambda\to 0}\, \lambda^{n} L_{n}^{(\alpha)}\left(\frac{x}{\lambda}\right) = (-1)^n {x^n}/{n!}$, the definition \rfw{Wdef} ultimately gives us
\begin{equation}
W_{\bs{n}, \bs{r}}(\bs{\gamma}) =  \frac{N_R!}{\prod_j n_j!} \frac{1}{2\pi i}  \oint_{\Gamma} \frac{dz}{z^{N_R+1}} \,e^{z}\prod_{j=1}^{D}\left(1+\gamma_j z \right)^{n_j},
\label{eq:Wdef0}
\end{equation}
where we used the definition $N_R \equiv \sum_{j=1}^D r_j$. The factor ${N_R!}/{\prod_j n_j!}$ is an important normalization that ensures that we have the correct counting of microstates. For example, the $\gamma_j\gg1$ (i.e., mostly bound-ligands) limit should yield (for $\sum_j r_j \geq \sum_j n_j$) a partition function that is proportional to the number of ways to arrange $n_i$ identical objects of type $i$ for $i = 1, \ldots, D$ amongst $r_1+ \cdots + r_D$ sites. 

We could have obtained \rfw{Wdef0} without use of the limiting case by starting from the expression
\begin{equation}
W_{\bs{n}, \bs{r}}(\bs{\gamma})  = \sum_{\bs{k}} I_{\boldsymbol{r},\bs{k}} \prod_{j=1}^D \frac{1}{(n_j-k_j)!}\gamma_j^{k_j} = \frac{N_R!}{\prod_{j} n_j!}\sum_{\bs{k}} \frac{1}{(N_R-\sum_jk_j)!} \prod_{j=1}^D \binom{n_j}{k_j} \gamma_j^{k_j},
\label{eq:W0}
\end{equation}
where the sum for each $k_j$ runs from $0$ to $n_j$. In the summand of the first equality of \rfw{W0}, $I_{\bs{r}, \bs{k}}$ (defined in \rfw{multi_nom}) is the number of ways to arrange $k_i$ bound ligands of type $i$, for $i=1, \ldots, D$ amongst $r_1 + \cdots + r_D$ receptor sites; the quantity $1/(n_j-k_j)!$ is the the free-particle partition function for type $j$ ligands (there is no additional factor in the numerator due to \rfw{free_norm}); and $\gamma_j^{k_j}$ is the bound-particle partition function for all ligands of type $j$ that are bound to receptors. By using the contour integral definition of the inverse factorial, we can show that \rfw{W0} is equivalent to \rfw{Wdef0}.

Using \rfw{obs} and \rfw{W0}, we find that the average number of bound ligands in the system can be written as 
\begin{equation}
\langle k \rangle = N_R  \sum_{i=1}^D\gamma_i \frac{W_{\bs{n}_i, \bs{r}_i}(\bs{\gamma}) }{W_{\bs{n}, \bs{r}}(\bs{\gamma}) },
\label{eq:ord_wn}
\end{equation}
where $\bs{n}_j$ is $\bs{n}$ with 1 subtracted from the $j$th component: $\bs{n}_{j}= (n_1, \ldots, n_{j}-1, \ldots, n_D)$.  The vector $\bs{r}_j$ is defined similarly. \rfw{ord_wn} provides us with a reliable means for computing the order parameter presuming we have a reliable means for computing the partition function. For our use case, we will use the large $N$ saddle-point approximation as the basis for this latter computation.  

First, defining 
\begin{equation}
{\cal A}_{\bs{n}, \bs{r}}(z;\bs{\gamma})\equiv z - N_R \ln z+ \sum_{j=1}^{D}n_j \ln\left(1 + \gamma_jz \right) + \ln  \frac{N_R!}{\prod_j n_j!} ,
\label{eq:Adef}
\end{equation}
and then applying the saddle point approximation to $W_{\bs{n}, \bs{r}}$ defined in \rfw{Wdef0}, we obtain the approximate partition function
\begin{align}
W_{\boldsymbol{n}, \bs{r}}(\bs{\gamma})  & =   \frac{1}{2\pi i} \oint_{\Gamma} \frac{dz}{z} \exp\big[{\cal A}_{\bs{n}, \bs{r}}(z;\bs{\gamma})\big]
\simeq
\frac{1}{\left( 2\pi \bar{z}^2 A''_{\bs{n}, \bs{r}}(\bar{z}; \bs{\gamma})\right)^{1/2}}\exp \left[A_{\bs{n}, \bs{r}}(\bar{z}; \bs{\gamma})\right] , 
\label{eq:approx_wn}
\end{align}
where $\bar{z}$ is defined by the constraint $0 = \partial_zA_{\bs{n}, \bs{r}}(z; \bs{\gamma})|_{z = \bar{z}} \equiv A_{\bs{n}}'(z; \bs{\gamma})|_{z = \bar{z}}$. \rfw{approx_wn} is an approximation, but we henceforth use an equality symbol for notational sparsity. Using \rfw{Adef} to compute $\bar{z}$ given its constraint definition, we find the condition
\begin{equation}
0 = \bar{z} - N_R+ \sum_{j=1}^D \frac{n_j \gamma_j\bar{z}}{1+ \gamma_j \bar{z}}
\label{eq:z0_soln}
\end{equation}
Next, from \rfw{obs} and \rfw{Wdef}, we can identify the average number of bound ligands of type $i$, $\langle k_i \rangle$, as 
\begin{equation}
\langle k_i \rangle =   \gamma_i\frac{\partial}{\partial \gamma_i} \ln W_{\bs{n}, \bs{r}}.
\end{equation}
From this definition and \rfw{approx_wn} and \rfw{Adef}, we thus find 
\begin{equation}
\langle k_i \rangle =\frac{n_i \gamma_i \bar{z}}{1+\gamma_i \bar{z}},
\label{eq:ki}
\end{equation}
where we dropped sub-leading terms. This expression could also have been derived by starting from \rfw{ord_wn} and using \rfw{approx_wn} (and dropping the exponential pre-factor) with the fact that 
\begin{equation}
A_{\bs{n}_i, \bs{r}_i}(\bar{z}; \bs{\gamma})- A_{\bs{n}, \bs{r}}(\bar{z} ; \bs{\gamma})= \ln \bar{z} - \ln (1 + \gamma_i \bar{z}) + \ln \frac{n_i}{N_R}.
\end{equation}

\rfw{ki} and \rfw{z0_soln} provide a means for approximately determining the number of bound ligands of a specific type. Given the list of binding affinities $\bs{\gamma} = (\gamma_1, \ldots, \gamma_D)$ for our ligands, we can numerically solve \rfw{z0_soln} and insert the result into \rfw{ki}. However, there is a different perspective on these results that connects them to the way binding is typically represented in chemistry. We can use our system of equations to solve for $\bar{z}$ in terms of $\langle k_j \rangle$ in two ways: We have 
\begin{equation}
\bar{z} = N_R- \langle k \rangle, \qquad \bar{z}\left( N_L -N_R + \bar{z}\right) = \sum_{j=1}^D \langle k_{j}\rangle \gamma_{j}^{-1},
\label{eq:barz}
\end{equation}
where we used $N_L \equiv \sum_{j=1}^D n_j$. The first equation is found by substituting \rfw{ki} into \rfw{z0_soln}. The second equation is found by dividing \rfw{ki} by $\gamma_i \bar{z}$, noting that $n_j\gamma_j\bar{z}/(1+ \gamma_j \bar{z}) = n_j - n_j/(1+ \gamma_j \bar{z}) = n_j - \langle k_j \rangle/\gamma_j \bar{z}$ and inserting this expression into \rfw{z0_soln}. Eliminating $\bar{z}$ from \rfw{barz}, we then have 
\begin{equation}
\left(N_R-\langle k \rangle \right)\left(N_L-\langle k \rangle \right)  = \sum_{j=1}^D \langle k_j \rangle \gamma_{j}^{-1}
\label{eq:ligand_eq}
\end{equation}
which is the law of mass action for a system with $N_R$ receptors and $N_L$ ligands  where the ligand of type $i$ has a binding affinity $\gamma_i$ to any receptor. We could have anticipated this result: When $\delta_i =1$, the combinatorics of various binding configurations becomes irrelevant since all re-orderings of the same set of bound ligands are thermally equivalent, and thus the equilibrium properties should be governable by averages constrained by the law of mass action alone. 

The order of our approximation does not allow for $\langle k \rangle = N_R$ or $\langle k \rangle = N_L$, which are the cases where all receptors are occupied or where all ligands are bound, respectively. However, it does allow for $\langle k \rangle =N_R-1$ or $\langle k \rangle = N_L -1$ assuming $N_R\leq N_L$ and $N_L \leq N_R$, respectively. These are states where essentially all receptors are bound with a ligand or all ligands are attached to a receptor. To move forward, we will assume we have a system where there are more receptor sites than ligands, i.e.,  $N_L \leq N_R$; the alternative case can be easily analyzed as well. Our almost-completely bound state is then $\langle k \rangle = N_L-1$ and with our definition of $\bar{z}$ in \rfw{barz}, we have $\bar{z} = N_R- N_L +1$. Taking $N_L\gg1$ (as is the case in our large $N$ approximation), the thermal condition that defines this almost-completely bound state is therefore
\begin{equation}
N_R- N_L +1 = \sum_{j=1}^{D} n_j \gamma_{j}^{-1} +O(\gamma_{j}^{-2}).
\label{eq:gamma_condRL}
\end{equation}
To make this problem easier to analyze, we will constrain ourselves to work within the case where $n_j = r_j$ for all $j$, namely matched population of ligands and receptors. This case is also the one we will explore in the simulation comparison of this result. For this case, \rfw{gamma_condRL} becomes 
\begin{equation}
1 = \sum_{j=1}^{D} n_j \gamma_{j}^{-1} +O(\gamma_{j}^{-2}). 
\label{eq:gamma_cond}
\end{equation}
Given a temperature-dependence for $\gamma_j$, we can numerically solve \rfw{gamma_cond} for the temperature at which essentially all of the ligands are bound to receptor sites. This case of $n_j = r_j$ for all $j$ is also the case we will explore in the simulation comparison of the results in this section. 

\subsection{Derangement-Only Model \label{sec:perm}}

Another simplification of the most general scenario associated with \rfw{master_derang_part} is to consider it as purely a combinatorial one in which the only available microstates are those consisting of permutations of ligand positions amongst the receptor sites. This can only occur if the bound-ligand partition function is infinitely larger that the corresponding unbound-ligand partition function. Quantitatively, by \rfw{coeff_def}, this amounts to taking $\gamma_i \to \infty$. In such a case, all ligands are bound to a receptor (if $\sum_{j} n_j < \sum_j r_j$) or all receptors are occupied (if $\sum_j r_j < \sum_{j} n_j$), and thermal fluctuations only lead the ligands to switching receptors. 

For simplicity going forward, we will assume $r_j \geq n_j$ for all $j$. This corresponds to the situation where there are more receptor sites than ligands. Taking the partition function \rfw{master_derang_part}  to the $\gamma_i \to \infty$ limit and dividing out the thermodynamic pre-factor $ \prod_{i=1}^D \gamma_i^{n_i}$ representing the bound-ligand partition functions, we can define
\begin{equation}
X_{\bs{n}, \bs{r}}(\bs{\delta}) \equiv \lim_{\{\gamma_i \to \infty\}} \frac{1}{ \prod_{i=1}^D \gamma_i^{n_i}}{\cal Z}_{\bs{n}, \bs{r}}(\bs{\delta}, \bs{\gamma}),
\label{eq:gen_derang}
\end{equation}
Computing the limit gives us
\begin{equation}
X_{\bs{n}, \bs{r}}(\bs{\delta}) = \frac{1}{(N_R-N_L)!} \int^{\infty}_{0} dx\, e^{-x} x^{N_R-N_L}\prod_{i=1}^{D} (\delta_i-1)^{n_i} L_{n_i}^{(r_i-n_i)} \left(\frac{x}{1-\delta_i} \right),
\label{eq:gen_derang0}
\end{equation}
where $L_n(x)$ is the $n$th Laguerre polynomial. In defining \rfw{gen_derang}, we used the fact that the $\{\gamma_i\}$ are thermodynamically irrelevant for a system consisting only of bound ligands and thus their factors can be divided out of the partition function.

We could have derived \rfw{gen_derang} without a limiting case by recognizing that the various microstates of this system are "partial derangements" of a list. Specifically, we could have written $X_{\bs{n}, \bs{r}}$ as a summation over these derangements:
\begin{equation}
X_{\boldsymbol{n}, \bs{r}}(\bs{\delta})  = \sum_{\bs{m}} B_{\bs{r}-\bs{m}, \bs{n}-\bs{m}} \prod_{j=1}^D \binom{r_j}{m_j}\delta_j^{m_j}, 
\label{eq:Xnsimp}
\end{equation}
where the summation for each $m_j$ runs from $0$ to $n_j$, and $B_{\bs{n}, \bs{k}}$ is defined in \rfw{Bnkfin}. Deriving \rfw{gen_derang0} from \rfw{Xnsimp} requires  the identity \rfw{sumLaga} derived in Appendix \ref{app:multi_nom}.

Also, \rfw{gen_derang} is an "elements with repeats" generalization of the permutation glass considered in \cite{williams2018permutation}. If we take $r_i = n_i=1$ for all $i$ in the product in \rfw{gen_derang}, we find 
\begin{equation}
X_{\bs{n}, \bs{r}}(\bs{\delta})\Big|_{r_i = n_i =1} = \int^{\infty}_{0} dx\, e^{-x} \prod_{i=1}^{N} \big(\delta_i -1+x\big),
\label{eq:derang_part_simp}
\end{equation}
which, with $\delta_i = e^{\beta \Delta_i}$, is identical to the partition function derived in \cite{williams2018permutation}\footnote{In the original paper, we defined our microstates in terms of energy \textit{penalties} rather than energy benefits so the partition function here differs from the original by a multiplicative constant.}. In that work, we derived necessary but not sufficient conditions for the system to settle into the "completely correct" microstate, which in our case corresponds to all ligands being bound in their optimal receptors. Here we attempt to derive analogous conditions for this more general case.

Using \rfw{obs}, \rfw{gen_derang}, \rfw{gen_derang0}, and the identity \rfw{fnderiv}, we find that the average number of ligands bound to their optimal receptors is
\begin{equation}
\langle m \rangle  = \sum_{j=1}^{D} r_{j}\delta_j \frac{X_{\bs{n}_j, \bs{r}_j}(\bs{\delta}) }{X_{\bs{n}, \bs{r}}(\bs{\delta}) } ,
\label{eq:ord_zn}
\end{equation}
where $\bs{n}_{j}$ is $\bs{n}$ with 1 subtracted from the $j$th component (i.e., $\bs{n}_{j}= (n_1, \ldots, n_{j}-1, \ldots, n_D)$) and $\bs{r}_j$ is defined similarly. \rfw{ord_zn} tells us  that if we have a consistent means for computing the partition function $X_{\bs{n}, \bs{r}}(\bs{\delta})$, we can calculate the order parameter with little extra work. For this system, the consistent means we have for computing the partition function is the large $N$ approximation. 

To implement this approximation, we first define
\begin{equation}
F_{\bs{n}, \bs{r}}(x; \bs{\delta}) \equiv x - (N_R-N_L) \ln x - \sum_{j=1}^D\ln \left[\left(\delta_j-1\right)^{n_j}\, L_{n_j}^{(r_j-n_j)} \left( \frac{x}{1-\delta_j} \right)\right] + \ln [(N_R-N_L)!]\,.
\label{eq:Fdef}
\end{equation}
Then applying Laplace's method to $X_{\bs{n}, \bs{r}}$ defined in \rfw{gen_derang0}, we have the approximation
\begin{align}
X_{\bs{n}, \bs{r}}(\bs{\delta})  & = \int^{\infty}_{0} dx\, \exp\left[ - F_{\bs{n}, \bs{r}}(x; \{\delta_i\})\right]
 \simeq   
\left(\frac{2\pi}{F''_{\bs{n}, \bs{r}}(\bar{x}; \bs{\delta})}\right)^{1/2} \exp \left[-F_{\bs{n}, \bs{r}}(\bar{x}; \bs{\delta})\right] 
\label{eq:approx_zn}
\end{align}
where $\bar{x}$ is defined by the constraint $0 = \partial_xF_{\bs{n}}(x; \{\delta_i\})|_{x = \bar{x}} \equiv F_{\bs{n}}'(x; \{\delta_i\})|_{x = \bar{x}}$. \rfw{approx_zn} is an approximation, but we henceforth use an equality symbol for notational sparsity.  Applying the constraint $0 = F_{\bs{n}}'(x; \{\delta_i\})|_{x = \bar{x}}$ to \rfw{Fdef} and using the recursive Laguerre identity  $u \,\partial_u L_n^{(\alpha)}(u) = nL_{n}^{(\alpha)}(u) - (n+\alpha)L_{n-1}^{(\alpha)}(u)$, we find that $\bar{x}$ can be computed from
\begin{equation}
\bar{x} = \sum_{j=1}^{D} r_j\Biggl(1- \frac{\displaystyle L_{n_j-1}^{(\omega_j)} \left( \bar{\sigma}_{j}\right)}{\displaystyle  L_{n_j}^{(\omega_j)} \left( \bar{\sigma}_{j} \right)}\Biggr); \qquad \bar{\sigma}_{j} \equiv \frac{\bar{x}}{1-\delta_j},
\label{eq:barx}
\end{equation}
where we defined $\omega_j \equiv r_j -n_j$. Using \rfw{ord_zn} with \rfw{approx_zn} (and the fact that we are in the $N\gg1$ limit), we see that the average number of optimal bindings is 
\begin{align}
\langle m\rangle & =   \sum_{j=1}^D \frac{r_j \delta_j}{\delta_j-1} \frac{\displaystyle L_{n_j-1}^{(\omega_j)} \left( \bar{\sigma}_{j}\right)}{\displaystyle  L_{n_j}^{(\omega_j)}  \left( \bar{\sigma}_{j} \right)}.
\label{eq:jtox}
\end{align}

With \rfw{jtox}, we can determine whether a microstate consisting of all ligands bound to their optimal receptors can be achieved in this system. Finding the condition that makes such a microstate possible would require us to look at the low temperature behavior of the system, but let's momentarily go in the opposite direction. 

What is the behavior of \rfw{jtox} when $T$ goes to $\infty$? First, as $T\to \infty$, $\delta_i$  goes to $1$. Given the definition of the Laguerre polynomial, we can derive the limit $\lim_{\lambda\to 0}\, \lambda^{m} L^{(\alpha)}_{m}\left({x}/{\lambda}\right) = (-1)^m {x^m}/{m!}.$  Using this limit, we find for $T\to \infty$ that $\langle m \rangle \to \sum_{j=1}^D n_jr_j /\bar{x}$ and $\bar{x} \to \sum_{j=1}^D r_j$. Therefore, 
\begin{equation}
\lim_{T \to \infty} \langle m \rangle  = \frac{\sum_{j=1}^{D} n_jr_j}{\sum_{j=1}^Dr_j}
\label{eq:mlargeT}
\end{equation}
There are three things to note about this result: First, the fact that it is non-zero; second, the vanishing scaling with $r_j$ as $r_j\to \infty$; third, the linear scaling with $n_j$.

With regard to the first fact, a naive entropic argument might lead us to think that there would be \textit{no} optimal bindings at infinite temperature (i.e., a physical regime where the energy advantage of optimal bindings is irrelevant) since the macrostate of completely deranged ligands ($\langle m \rangle \simeq 0$) would supposedly have the largest number of microstates and thus the largest entropy. However, \rfw{mlargeT} suggests that the macrostate for completely deranged bindings is in fact not entropically favored at infinite temperature, and thus that such a macrostate takes up less configuration space than seemingly more ordered and constrained macrostates. Indeed, when you have various types of ligands each of which occurs in large numbers, then it becomes more constraining to require no ligand to be in its optimal binding site than it is to have some ligands be optimally bound. 

With regard to the second fact, we see that since $r_j$ appears in both the denominator and numerator of \rfw{mlargeT}, the infinite temperature limit of $\langle m \rangle$ remains finite as $r_j \to \infty$. This means that the number of optimal bindings that result from random assortment does not change as the number of available receptor sites increases. Increasing the number of receptors doesn't make such optimal binding more likely if the number of ligands remains constant.

Conversely, we see that \rfw{mlargeT} does increase with increasing $n_j$ meaning that increasing the number of ligands in the system does increase the number of optimal bindings that can occur. This makes sense because with more ligands in the system there is a greater chance that one of those existing ligands will make their way to an optimal lattice site. 

Pharmacologically, these latter two results imply that if one could only make one change in order increase the odds of thermally random optimal ligand-receptor binding, one should increase the concentration of the ligands in the system rather than increasing the number of receptor sites available to them. It is important to note, however, that recalling our initial assumption, the limiting behavior shown in \rfw{mlargeT} is valid only for the case where $r_j \geq n_j$. 

We can obtain an alternative interpretation of the meaning \rfw{mlargeT} by writing it in terms of a covariance. Defining the bar-average of a parameter $O_j$ as $\overline{O} = \sum_{j=1}^D O_j/D$ and the covariance $\text{cov}(r, n) = \overline{r\, n} - \overline{r} \overline{n}$ we have
\begin{equation}
\lim_{T \to \infty} \langle m \rangle = \bar{n} \left(1 + \frac{\text{cov}(r, n)}{\bar{r}\,\bar{n}}\right).
\label{eq:mlargeT0}
\end{equation}
The quantity $\bar{n}$ is the average particle-number across all types of ligands and $\text{cov}(r, n)$ represents how much $r_j$ and $n_j$ vary together. \rfw{mlargeT0} shows that the more that $r_j$ and $n_j$ vary together, the larger the lower limit on the number of optimally bound ligands. With more concurrent variability in the number of ligands and receptors of each type, it becomes more likely that at least some ligands, just from random assorting, will be bound to their optimal receptors. 

Now, we consider the opposite temperature limit under the frame of a specific  question: At what temperature are all of the ligands bound to their optimal receptors? For analytical simplicity, we will subsequently work within the case $n_j = r_j$ for all $j$, namely matched population of ligands and receptors. Thus all ligands are bound to their optimal receptors when $\langle m \rangle = N_R = N_L$. This case allows us to study the combinatorial properties of this model more simply without having to consider the various ways subsets of receptors are occupied. 

To answer this question, we first use \rfw{barx} and \rfw{jtox} to obtain the identity
\begin{equation}
\bar{x} = N - \sum_{j=1}^D \langle m_j \rangle (1- \delta_{j}^{-1}), 
\label{eq:xmrel}
\end{equation}
where we defined $N \equiv \sum_j^D r_j = \sum_j^Dn_j$ for this case,  and the $\langle m_j \rangle$s are the elements of the sum in \rfw{jtox}. When all ligands are bound to their optimal receptors, we have $\langle m_j \rangle = n_j = r_j$. Thus, at this desired critical temperature, we have the condition
\begin{equation}
\bar{x} =  \sum_{j=1}^Dn_j\delta_{j}^{-1}.
\label{eq:xTc}
\end{equation}
To move forward, we will make two assumptions whose consistency we will check at the end of the calculation. First, we assume that the desired temperature is sufficiently low that $\delta_i\gg1$ for all $i$. Second, we assume that $\delta_i \gg \bar{x}$. With these assumptions, we find $L^{(0)}_{n_j-1}(\bar{\sigma}_j)/L^{(0)}_{n_j}(\bar{\sigma}_j) = 1 - \bar{x} \delta_{j}^{-1}  + O(\delta_{j}^{-2})$.  Expanding \rfw{jtox} to first order in $\delta_{j}^{-1}$ yields
\begin{equation}
\langle m \rangle = N +  (1-\bar{x}) \sum_{j=1}^{D}  n_j \delta_{j}^{-1} + O(\delta^{-2}),
\label{eq:jtox2}
\end{equation}
where $O(\delta^{-2}) \equiv \sum_{j=1}^{D} O(\delta_j^{-2})$ and $O(\delta_j^{-2})$ represents terms of order $\delta_j^{-2}$. At the temperature at which all ligands are optimally bound, we have $\langle m \rangle = N$. Thus, \rfw{jtox2} implies $\bar{x} =1+O(\delta_{j}^{-2})$ and by \rfw{xTc} we obtain the final condition
\begin{equation}
1 = \sum_{j=1}^{D} n_j \delta_{j}^{-1}+ O(\delta^{-2}). 
\label{eq:temp_def}
\end{equation}
Given the temperature dependence for $\delta_j$, we can numerically solve \rfw{temp_def} for the temperature at which all of the ligands are optimally bound to receptor sites. We will do so in \refsec{perm_sim} when we simulate this system. To check consistency with our two initial assumptions (i.e., $\delta_j \gg1$ and $\delta_j \gg \bar{x}$), we note that, for the large particle-number limit, \rfw{temp_def} implies $\delta_j >n_j \gg1 = \bar{x} + O(\delta_j^{-2})$, as we assumed.

\subsection{General Case \label{sec:general}}

Having explored various limiting cases, we are now ready for the full case. In this section, our objective is two fold: First, determine the equations for both the average number of bound and optimally-bound ligands of each type; second, use these equations to determine the thermal conditions that define the system settling into the microstate in which each ligand is bound to its optimal receptor (i.e., the fully optimally bound). To get to either objective, we first need to approximate the partition function and compute the standard observables (\rfw{obs}) according to this approximation. We will apply methods similar to those applied to the limiting cases to analyze this general case.

Applying the saddle point approximation to \rfw{master_derang_part}, we have
\begin{align}
{\cal Z}_{\bs{n}, \bs{r}}(\bs{\delta}, \bs{\gamma}) & =   \frac{1}{2\pi i} \oint_{\Gamma} \frac{dz}{z} \int^{\infty}_{0} dx\, \exp\big[{\cal F}_{\bs{n}, \bs{r}}(z, x;\bs{\delta}, \bs{\gamma})\big]
\simeq \frac{1}{({\bar{z}}^2 \det H)^{1/2}}  \exp\big[{\cal F}_{\bs{n}, \bs{r}}(\bar{z}, \bar{x}; \bs{\delta}, \bs{\gamma})\big]
\label{eq:master_derang_part0}
\end{align}
where 
\begin{align}
{\cal F}_{\bs{n}, \bs{r}}(\bar{z}, \bar{x};\bs{\delta}, \bs{\gamma}) &  \equiv \bar{z}-\bar{x} + (N_R - N_L) \ln (x/z)+ \sum_{j=1}^{D}\ln \left[(\gamma_j (\delta_j-1))^{n_j} L_{n_j}^{(r_j-n_j)} \left(\bar{\phi}_j \right) \right]; \label{eq:master_FRdef}\\[.5em]
 \quad \bar{\phi}_{j}&\equiv \frac{\bar{x}}{1-\delta_j}\left(1+ \frac{1}{\bar{z} \gamma_j}\right).
\end{align}
The quantity $H$ is the complex hessian matrix of ${\cal F}_{\bs{n}}$
\begin{equation}
H \equiv \partial_{\alpha,\beta} {\cal F}_{\bs{n}, \bs{r}}(z, x;\bs{\delta}, \bs{\gamma})\Big|_{x, z = \bar{x}, \bar{z}}, \qquad 
\end{equation}
where the variables $\alpha$ and $\beta$ can be $x$ or $z$. The critical points $\bar{z}$ and $\bar{x}$ are defined by the conditions
\begin{equation}
0 = \partial_z {\cal F}_{\bs{n}, \bs{r}}(z, x;\bs{\delta}, \bs{\gamma})\Big|_{x, z = \bar{x}, \bar{z}}, \qquad 0 = \partial_x {\cal F}_{\bs{n}, \bs{r}}(z, x;\bs{\delta}, \bs{\gamma})\Big|_{x, z = \bar{x}, \bar{z}}.
\end{equation}
Applying the critical point conditions to \rfw{master_FRdef}, we find
\begin{equation}
\bar{z} = N_R-N_L + \sum_{j=1}^{D} \frac{1}{\bar{z} \gamma_j +1} \left(n_j- r_j\frac{\displaystyle L_{n_j-1}^{(\omega_j)} \left( \bar{\phi}_{j}\right)}{\displaystyle  L_{n_j}^{(\omega_j)} \left( \bar{\phi}_{j} \right)}\right), \qquad \bar{x} = \sum_{j=1}^{D} n_j\left(1- \frac{\displaystyle L_{n_j-1}^{(\omega_j)} \left( \bar{\phi}_{j}\right)}{\displaystyle  L_{n_j}^{(\omega_j)} \left( \bar{\phi}_{j} \right)}\right),
\label{eq:xandzbar}
\end{equation}
where we defined $\omega_j \equiv r_j -n_j$. The associated values of $\langle k\rangle$ and $\langle m\rangle$ can then be found by applying \rfw{obs} to the approximated partition function \rfw{master_derang_part0} while ignoring the subleading pre-factor. Doing so, we obtain
\begin{equation}
\langle k_j \rangle = \frac{1}{\bar{z}\gamma_j +1} \left(n_j\bar{z}\gamma_j +r_j\frac{\displaystyle L_{n_j-1}^{(\omega_j)} \left( \bar{\phi}_{j}\right)}{\displaystyle  L_{n_j}^{(\omega_j)} \left( \bar{\phi}_{j} \right)}\right) , \qquad \langle m_j \rangle  =  \frac{r_j\delta_j}{\delta_j-1} \frac{\displaystyle L_{n_j-1}^{(\omega_j)} \left( \bar{\phi}_{j}\right)}{\displaystyle  L_{n_j}^{(\omega_j)} \left( \bar{\phi}_{j} \right)}.
\label{eq:kandmdef}
\end{equation}
With \rfw{kandmdef}, we can use the solutions for $\bar{x}$ and $\bar{z}$ determined from \rfw{xandzbar} to find the average number of bound and correctly bound ligands of each type, thus fulfilling the first objective. 

For the second objective of determining the thermal conditions for fully optimal ligand-receptor binding, we first use \rfw{xandzbar} and \rfw{kandmdef} together to obtain two equations relating the four quantities:
\begin{equation}
\bar{z} = N_R - \langle k \rangle , \qquad \bar{x} = N_R - \sum_{j=1}^D \langle m_j \rangle (1-\delta_j^{-1}),
\label{eq:four_eqns}
\end{equation}
where we took $\langle k \rangle = \sum_{j} \langle k_j \rangle$. In \refsec{gender}, we showed that in the gendered dimer assembly system, the correct assembly condition yielded the result $\langle k \rangle = \langle m \rangle = (N-1)/(1-\delta^{-1}) = N -1 + O(\delta^{-1})$. With this result, we were then able to find the thermal condition that defined the fully optimized state. We want to do something similar for the more general case considered in this section.

We will work within the case $n_j = r_j$ for all $j$, namely matched population of ligands and receptors. This case is the most amenable to an analysis of the competing influences of disorder and binding since we would not need to consider the multiple ways to select subsets of receptors or ligands for binding. 

We start by defining the state of fully optimal ligand-receptor binding in a way analogous to the definition in  \rfw{gend_cond}: We assert that the system is in the state where all ligands are optimally bound to receptors when $\langle k \rangle$ and $\langle m \rangle$ satisfy
\begin{equation}
\langle k \rangle = \langle m \rangle = N-1 + O(\delta^{-1}),
\label{eq:gen_bind_cond}
\end{equation}
where we defined $N \equiv N_R = N_L$, $O(\delta^{-1}) \equiv \sum_{j=1}^{D} O(\delta_j^{-1})$ and $O(\delta_j^{-1})$ represents terms of order $\delta_j^{-1}$. To find the thermal condition that defines fully optimal binding, we need to find the thermal condition that is consistent with \rfw{gen_bind_cond}. Applying \rfw{gen_bind_cond} to \rfw{four_eqns} and \rfw{kandmdef}, we find 
\begin{equation}
1 = \sum_{j=1}^{D} n_j \delta_j^{-1} \left(1 + \gamma_{j}^{-1}\right) + O(\delta^{-2}), 
\label{eq:master_therm}
\end{equation}
where $O(\delta^{-2}) \equiv \sum_{j=1}^D O(\delta_j^{-2})$ (see Appendix \ref{app:therm_cond}). 
\rfw{master_therm} is the general thermal condition that defines fully optimal binding. Given the temperature dependences of $\delta_j$ and $\gamma_j$, we can use \rfw{master_therm} to determine the temperature at which all ligands of all types settle into their optimal receptors. 

Practically, if we wanted to compute the number of bound and optimally bound ligands of type $j$ (for the case where $r_j =n_j$ for all $j$), we would use the first and second equations in \rfw{kandmdef}, respectively, assuming our system is in the $n_j\gg1$ regime. Furthermore, if our system satisfied $\delta_j \gg 1$ at low temperature, then we could use \rfw{master_therm} to compute the thermal conditions under which the system achieves fully optimal binding. We perform these calculations for a concrete example in the next section. 

\section{Simulation Comparison \label{sec:simulation}}

\begin{figure}[t]
\centering
\includegraphics[width=.75\linewidth]{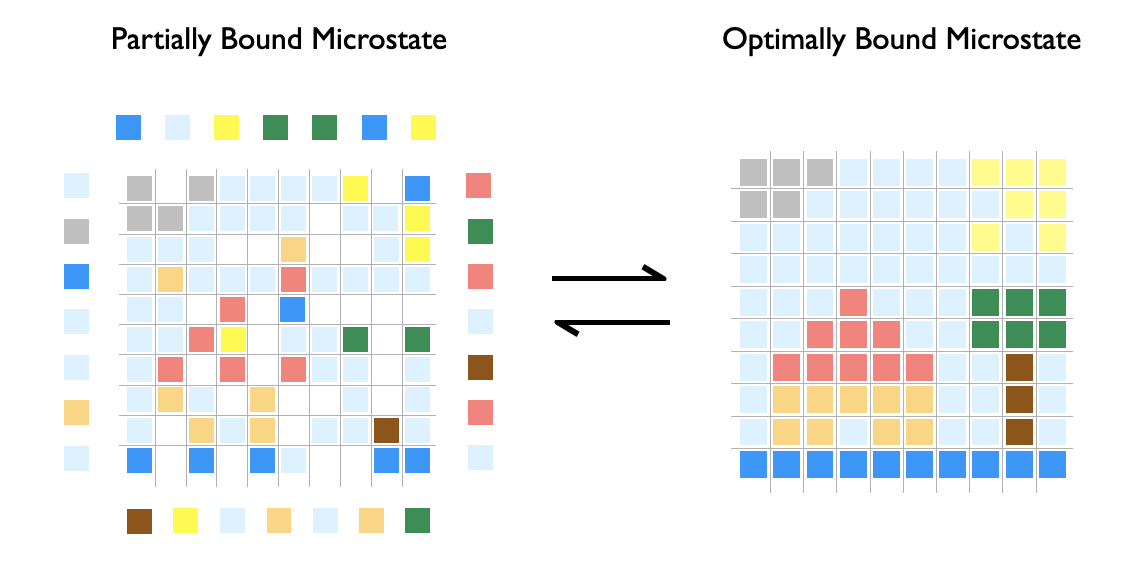}
	\caption{Depiction of particle binding on a grid. The system is characterized by different particles (i.e., the colored-squares) each of which occurs in multiples copies and is associated with a set of "optimally bound positions" on the grid. In the figure, the optimal positions for a colored square are the positions that would lead it to reproduce the microstate on the right. On the left, some squares are on the grid (i.e., bound) and other squares are off the grid (i.e., free). In the language of the model, we say that particle species $i$ has a binding affinity of $\gamma_i$ to the grid and an additional binding affinity factor of $\delta_i$ when it is bound to its optimal site. }
	\label{fig:assembly_microstate}
\end{figure}

Consider a two-dimensional grid of square lattice sites each of which can be filled with various colored squares. The colored squares represent the ligands of the system with a specific color defining a ligand type, and the lattice sites are the receptors. Each color-type binds optimally to a particular collection of lattice sites. For pictorial convenience we can arrange the optimal lattice sites for each particle type such that a figure is created. This way it is obvious whether our system is in the fully optimally bound configuration. We depict this system in \reffig{assembly_microstate}.

As a clarifying point, the model we developed for ligands binding to receptors applies equally well to a one-dimensional chain as to an $n$-dimensional grid as long as both are finite. This is because coordinates on a finite grid can be mapped one-to-one to a finite list such that a fixed collection of objects exploring various positions in the multi-dimensional grid is equivalent to those objects being placed in various orderings in a list. 

 In what follows, we use this graphical lattice model to present simulation results for the two limiting cases and the general case discussed in \refsec{cases}. For this grid-assembly system, the number of ligands and the corresponding number of optimal receptors for each type are equal, and thus we study our system for the case where $n_i = r_i$.

\subsection{Simple Binding Model Simulation $(\delta_i =1)$ \label{sec:bind_sim}}

In this section, we affirm the theoretical results in \refsec{bind} by simulating a simplification of the grid system in \reffig{assembly_microstate} at various temperatures.
\begin{figure*}[t]
\begin{subfigure}[t]{0.35\textwidth}
\includegraphics[width=\linewidth]{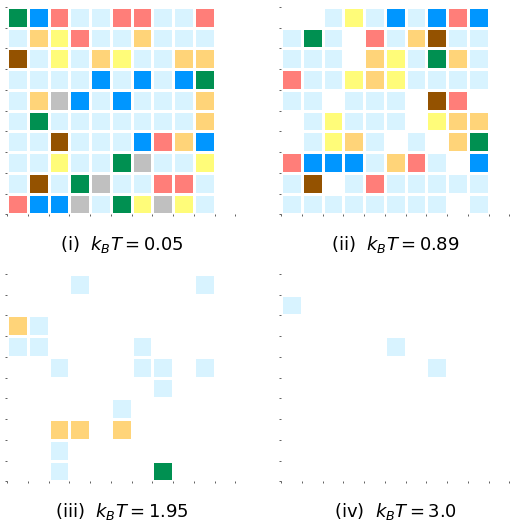}
	\caption{}
	\label{fig:grid_binding}
\end{subfigure} $\qquad$
\begin{subfigure}[t]{0.5\textwidth}
\centering
\includegraphics[width=\linewidth]{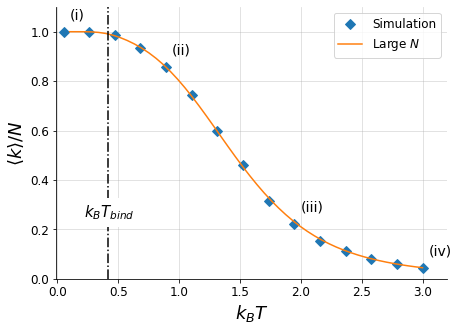}
	\caption{}
	\label{fig:binding} 
\end{subfigure}
	\caption{Grid snapshots and theory vs. simulation for simple ligand-receptor binding from \refsec{bind}: We defined $\gamma_j = (\beta E_V)^{3/2}e^{\beta E_j}$ where $E_V = 10^{-3}$ and $E_j$ was sampled from a normal distribution ${\cal N}(\mu_E, \sigma_E^2)$ with $\mu_E = 6.0$ and $\sigma_E = 2.0$. All energy parameters were taken to be dimensionless. In (a) we see snapshots of the simulated system at various equilibrium temperatures.  Particles not bound to the grid (i.e., free particles) are not shown. In (b) we plot a theory vs. simulation comparison for $\langle k \rangle$ as a function of temperature, and we mark the binding temperature $k_BT_{\text{bind}}$ computed from \rfw{gamma_cond1}. The "Large $N$" values of $\langle k \rangle$ were computed from \rfw{ki} summed over $i$. The "Simulation" values of $\langle k \rangle$ were computed from the results of a Metropolis Hastings algorithm where each point is the average of the result of five simulations. We see that \rfw{gamma_cond1} indeed identifies the temperature at which the system is nearly completely bound and that \rfw{ki} accurately models the simulation. (See \textit{Supplementary Code} in \refsec{code} for link to code repository used to produce this figure.)}
	\label{fig:ligand_bind}
\end{figure*}
The simplification is to assume particles have no "optimal" position on the grid (i.e., $\delta_j=1$) and thus a single particle has the same binding affinity to every site on the grid. The system was simulated using the Metropolis Hastings algorithm in which the microstates transitioned into one another dependent on free energy differences of the form $\beta E = \sum_{j} k_j \ln \gamma_j$, where $k_j$ is the number of bound particles of type $j$ in the microstate and $\gamma_j$ is the associated binding affinity. We allowed for two types of microstate transitions: particle binding to the grid and particle dissociation from the grid. As is typical for Metropolis Hastings algorithms, to fully determine the transition probabilities we also had to incorporate the difference in probabilities of selecting the particles for the forward- and reverse-transitions between microstates (See Appendix \ref{app:simulation} for a description of a more general simulation system and \textit{Supplementary Code} in \refsec{code} for associated code).

To incorporate an explicit temperature dependence into the system, we set $\gamma_j = (\beta E_V)^{3/2} e^{\beta E_j}$ where $\beta = 1/k_BT$. The quantity $E_V$ represents the volume-based energy of a free particle in the system (e.g., $E_V \equiv h^2/2\pi mV^{2/3}$ for an ideal gas particle of mass $m$) and thus $(\beta E_V)^{3/2}$ represents the ratio between the kinetic partition functions for free and bound particles. The quantity $E_j$ represents the binding energy of a particle of type $j$. For simplicity, we did not assume a $j$ dependence for $E_V$. 

Taking $k_BT_{\text{bind}} = \beta_{\text{bind}}^{-1}$ to be the critical temperature at which the complete (or, more precisely, "almost-complete") binding state is achieved, \rfw{gamma_cond} thus became
\begin{equation}
1 = (\beta E_V)^{-3/2}\sum_{j=1}^{D} n_j  e^{-\beta_{\text{bind}} E_j} +O(e^{-2\beta_{\text{bind}} E_j}), 
\label{eq:gamma_cond1}
\end{equation}

Solving \rfw{gamma_cond1} for $k_BT_\text{bind}$ gives us the temperature at which the thermal advantage of each particle binding to the grid (at any site) is large enough to overcome the entropic disadvantage of the particle existing statically in the grid rather than freely in the volume. In a sense, \rfw{gamma_cond1} defines the thermal condition under which all particles are able to search for and successfully find the grid in the space they occupy. This "searching" is encoded by the product of the $V$ and $n_j$ factors in the equation: As $V$ (defined in $E_V$) and $n_j$ increase, the volume in which a particle must search increases, and the number of particles doing the searching increases, respectively. Both increases make it more difficult for the system to settle into a state in which all particles are bound: Increasing volume increases the space in which particles must search for the grid; increasing the number of particles increases the number of units that need to conduct this search successfully. Thus increasing either of these values makes achieving the complete binding state more difficult, unless the temperature is lowered sufficiently so that the binding energy is strong enough to overcome the entropic disadvantage of having the particles exist freely. It is only below $k_BT_{\text{bind}}$ that the searching entropy succumbs to the energy advantage and the system settles into its full binding state. We employ this spatial search metaphor to distinguish this system from one grounded in a combinatorial search of possible states. We discuss this latter system in the next section.

To simulate the system, we chose numerical values for all parameters. For simplicity, we took all energy parameters in the system to be dimensionless.  The values of $E_j$ defining $\gamma_j$ were sampled from a Gaussian distribution ${\cal N}(\mu_E, \sigma_E^2)$ with mean $\mu_E = 6.0$ and variance $\sigma_E = 2.0$. The value of $E_V$ was set to $E_V = 10^{-3}$. The values of $n_j$ were determined directly from \reffig{assembly_microstate}: Inspecting the count of squares for each of the $D=8$ colors and taking each color to be a particle type, we have  $\bs{n} = \bs{r} =  (9, 9, 10, 5, 7, 6, 3, 51)$. 

In \reffig{grid_binding}, we show the simulated grid at various equilibrium temperatures. The particles are colored squares where particle-type is distinguished by color. Particles not bound to the grid are not shown. The values of $k_BT$ are dimensionless because we are taking the energy parameters of the system to be dimensionless. In the (i) image  of \reffig{grid_binding}, we see that all particles are bound although they are not in their "correct positions" as defined by the fully optimally bound state in \reffig{assembly_microstate}. This is of course because, with $\delta_i =1$, there is no thermal advantage to being in such entropically limited positions. As the temperature increases, fewer particles occupy the grid which confirms the basic intuition that the system should "melt" at higher temperatures. 

In \reffig{binding}, we plot the theoretical temperature-dependence of $\langle k \rangle = \sum_{j=1}^{D} k_j$ against the simulated temperature-dependence. We mark the points in the curve that are associated with the grid depictions in \reffig{grid_binding}. The temperature computed from \rfw{gamma_cond1} is denoted as $k_BT_{\text{bind}}$. We see excellent agreement between the simulation results and the theoretical results. Moreover, the predicted temperature computed from \rfw{gamma_cond1} accords with the results of the simulation. Inspecting (ii) in \reffig{grid_binding}, as we expect, above the critical temperature computed from \rfw{gamma_cond1}, the grid is no longer completely bound. 

\subsection{Derangement-Only Model Simulation $(\gamma_i \to \infty)$ \label{sec:perm_sim}}

In this section, we affirm the theoretical results in \refsec{perm} by simulating a simplification of the grid system in \reffig{assembly_microstate} at various temperatures. The simplification is to consider the system for the case in which all particles remained on the grid (i.e., $\gamma_j \to \infty$) and where state transitions are confined to particles exchanging positions within one another. The system was simulated using the Metropolis Hastings algorithm where microstates transitioned into one another contingent on free energy differences of the form $\beta E = \sum_j m_j \ln \delta_j$, where $m_j$ is the number of optimally bound particles of type $j$ in the microstate and $\delta_j$ is the additional binding affinity factor for optimal binding. We allowed for only one type of transition: single-step permutations of particle positions (See Appendix \ref{app:simulation} for a description of a more general simulation system and  \textit{Supplementary Code} in \refsec{code} for associated code).

To incorporate temperature into the system, we returned to our original expression for $\delta_j$ in \rfw{obs}: $\delta_j = e^{\beta \Delta_j}$. We recall that $\Delta_j$ is the energy-advantage an optimal binding has over any other binding for a ligand of type $j$. The associated critical temperature at which all particles were optimally bound was defined as $k_BT_{\text{derang}}$. Taking $k_BT_{\text{derang}} = \beta_{\text{derang}}^{-1}$, \rfw{temp_def} yields
\begin{equation}
1 = \sum_{j=1}^{D} n_j e^{-\beta_\text{derang}  \Delta_j} + O(e^{-2\beta_{\text{derang}} \Delta}).
\label{eq:temp_def0}
\end{equation}

Solving \rfw{temp_def0} for $k_BT_\text{derang}$ gives us the temperature at which the thermal advantage of each particle settling into its optimal site is large enough to overcome the entropic disadvantage of choosing that site in the space of all other combinatorial possibilities. The influence of combinatorics is encoded by $\bs{n} = (n_1, n_2, \ldots, n_D)$: As $n_j$ increases, the number of possible combinatorial states in the system increases and thus it becomes more difficult for a ligand to thermally select the optimal site in a sea of suboptimal ones, unless the temperature is lowered to diminish how much the combinatorial entropy influences the free energy. It is only below $k_BT_{\text{derang}}$ that combinatorial entropy succumbs to the energy-advantage of optimal sites, and the system settles into its fully optimal state.

To simulate the system, we chose numerical values for all parameters.  The values of $\Delta_j$ were sampled from a Gaussian distribution ${\cal N}(\mu_{\Delta}, \sigma_{\Delta}^2)$ with mean $\mu_{\Delta} = 4.0$  and variance $\sigma_{\Delta} = 2.0$; energy parameters were taken to be dimensionless. The values of $n_j$ were determined directly from \reffig{assembly_microstate}: Taking each particle to represent a particle type, we have $\bs{n} = \bs{r} =(9, 9, 10, 5, 7, 6, 3, 51)$.

\begin{figure*}[t]
\begin{centering}
\begin{subfigure}[t]{0.35\textwidth}
\centering
\includegraphics[width=\linewidth]{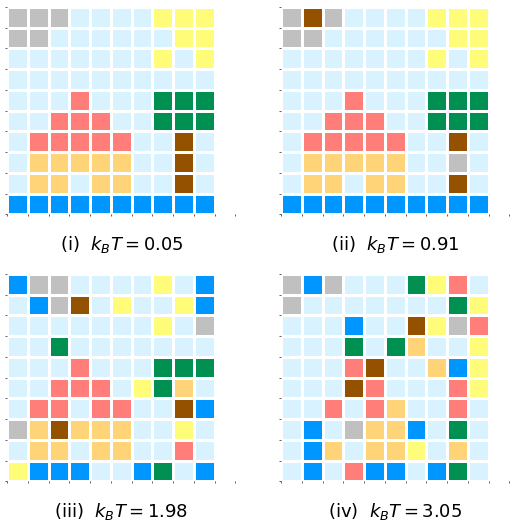}
	\caption{}
	\label{fig:grid_derang}
\end{subfigure} $\qquad$
\begin{subfigure}[t]{0.5\textwidth}
\centering
\includegraphics[width=\linewidth]{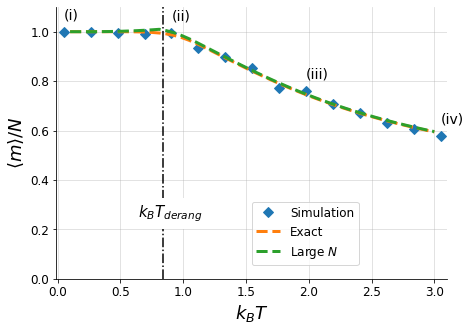}
	\caption{}
	\label{fig:derangements}
\end{subfigure}
	\caption{Grid snapshots and theory vs. simulation for derangement model \refsec{perm}: We used $\delta_j = e^{\beta \Delta_j}$ where $\Delta_j$ was sampled from a normal distribution ${\cal N}(\mu_\Delta, \sigma_\Delta^2)$ with $\mu_\Delta = 4.0 $ and $\sigma_\Delta = 2.0$. All energy parameters were taken to be dimensionless. In (a) we see snapshots of the simulated system at various equilibrium temperatures. As temperature increases, the system becomes more "deranged," meaning particles are less likely to assume their optimal binding sites.  In (b) we plot a theory vs. simulation comparison for $\langle m \rangle$ as a function of temperature, and we mark the binding temperature $k_BT_{\text{derang}}$ computed from \rfw{temp_def0}. The "Exact" values of $\langle m\rangle$ were computed from   \rfw{gen_derang0} and  \rfw{ord_zn}. The "Large $N$" values of $\langle m \rangle$ were computed from \rfw{jtox}. The "Simulation" values of $\langle m \rangle$ were computed from the results of a Metropolis Hastings algorithm where each point is the average of the result of five simulations. We see that \rfw{temp_def0} indeed identifies the temperature at which the system is in its fully optimally bound state and that both \rfw{ord_zn} and  \rfw{jtox} accurately model the simulation. (See \textit{Supplementary Code} in \ref{sec:code} for link to code repository used to produce this figure.)}
	\label{fig:derang_plot}
\end{centering}
\end{figure*}

In \reffig{grid_derang}, we show the simulated grid at various equilibrium temperatures. The particles are colored squares where particle-type is distinguished by color. The values of $k_BT$ are dimensionless because we are taking the energy parameters of the system to be dimensionless. In the (i) image  of \reffig{grid_derang}, we see that all particles are bound in their "correct positions" as defined by the fully optimally bound state in \reffig{assembly_microstate}. As the temperature increases, the particles become increasingly "deranged" from their correct positions, though we note that even at high temperatures some particles (in particular the ones with large $n_j$) do maintain many of their correct positions. This latter result is consistent with the discussion following \rfw{mlargeT}. 

In \reffig{derangements}, we plot the theoretical temperature-dependence of $\langle m \rangle = \sum_{j=1}^{D} \langle m_j\rangle $ against the simulated temperature-dependence. In particular we compare the simulations to the "Exact" theoretical prediction defined in \rfw{ord_zn}, and the "Large $N$" theoretical prediction defined in \rfw{jtox}. We mark the points in the curve that are associated with the grid depictions in \reffig{grid_derang}. The temperature computed from \rfw{temp_def0} is denoted as $k_BT_{\text{derang}}$. We see excellent agreement between the simulation results and the theoretical results. Moreover, the predicted temperature computed from \rfw{temp_def0} accords with the results of the simulation. Inspecting (ii) in \reffig{grid_binding}, as we expect, beyond the critical temperature the grid begins to show deranged particle states. 

Having explored the two limiting cases of the general model, we are now prepared for the fully general case. We will proceed as we did in these two example sections: Starting with a theoretical analysis stemming from an approximation and then finally simulating our results. The objective is to obtain a condition similar to \rfw{gamma_cond1} and \rfw{temp_def0} that implicitly defines the temperature at which the fully optimally bound state is achieved. 

\subsection{General Case Simulation \label{sec:gen_sim}}

In this section, we simulate the system outlined in \refsec{general} for the lattice grid depicted in \reffig{assembly_microstate}.

To incorporate temperature into the system we defined $\delta_j = e^{\beta \Delta_j}$ and $\gamma_j = (\beta E_V)^{3/2} e^{\beta E_j}$, where $\Delta_j$ is the energy advantage for the particle of type $j$ binding to its optimal lattice site. The quantity $E_V$ represents the volume-based energy of a free particle in the system (e.g., $E_V \equiv h^2/2\pi mV^{2/3}$ for an ideal gas particle of mass $m$) and thus $(\beta E_V)^{3/2}$ represents the ratio between the kinetic partition functions for free and bound particles. The quantity $E_j$ is the "base-binding energy" of the particle of type $j$ to any position on the lattice. For example, if each particle of type $j$ had a non-zero binding affinity only when bound to its optimal site, we would have $E_j =0$ and $\Delta_j>0$ for all $j$. With these thermal dependences, and taking $k_B T_{\text{crit}} = \beta_{\text{crit}}^{-1}$ to be the critical temperature at which the system achieves the fully optimally bound state, \rfw{master_therm} becomes
\begin{equation}
1 = \sum_{j=1}^{D}n_j e^{-\beta_{\text{crit}} \Delta_j}\left( 1+ (\beta_{\text{crit}} E_V)^{-3/2} e^{-\beta_{\text{crit}} E_j}\right) + O(e^{-2\beta_{\text{crit}} \Delta_j}). 
\label{eq:master_therm_subs}
\end{equation}
Comparing \rfw{master_therm_subs} to \rfw{gamma_cond1} and \rfw{temp_def0}, it appears that the first is a combination of the latter two. Moreover, given our "search" and "combinatorics" interpretation of \rfw{gamma_cond1} and \rfw{temp_def0} respectively, it appears that \rfw{master_therm_subs} embodies aspects of both limits contingent on various relative values of the parameters. In the next section, we will explore these relationships further.

To simulate the system, we chose numerical values for all parameters.  The values of $\Delta_j$ were sampled from a Gaussian distribution ${\cal N}(\mu_{\Delta}, \sigma_{\Delta}^2)$ with mean $\mu_{\Delta} = 3.0$  and variance $\sigma_{\Delta} = 1.0$.  The values of $E_j$ were sampled from a Gaussian distribution ${\cal N}(\mu_E, \sigma_E^2)$ with mean $\mu_E = 12.0$ and variance $\sigma_E = 4.0$. The value of $E_V$ was set to $E_V = 10^{-3}$. The values of $n_j$ were determined directly from \reffig{assembly_microstate}: Specifically, taking each color to represent a particle type, we had $\bs{n} = \bs{r} = (9, 9, 10, 5, 7, 6, 3, 51) $

\begin{figure*}[t]
\begin{centering}
\begin{subfigure}[t]{0.35\textwidth}
\centering
\includegraphics[width=\linewidth]{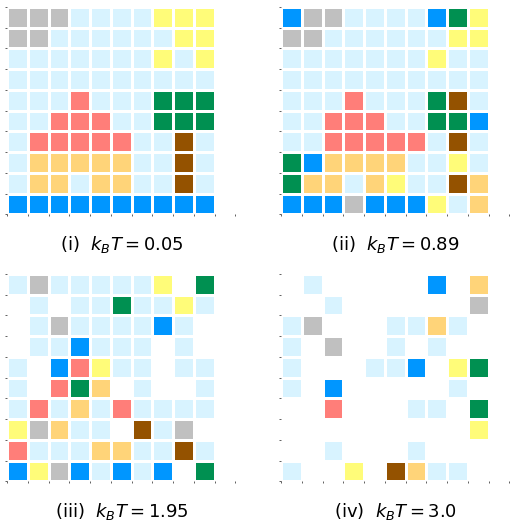}
	\caption{}
	\label{fig:grid_assembly3}
\end{subfigure} $\qquad$
\begin{subfigure}[t]{0.5\textwidth}
\centering
\includegraphics[width=\linewidth]{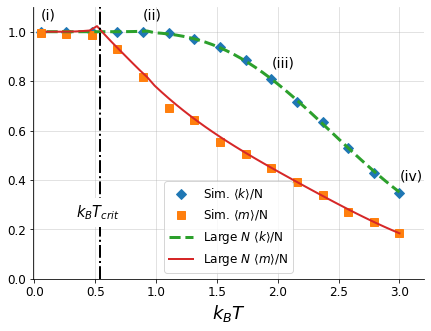}
	\caption{}
	\label{fig:thermal_plot3}
\end{subfigure}
	\caption{Grid snapshots and theory vs. simulation for general ligand-receptor binding from \refsec{general}: We used $\delta_j = e^{\beta \Delta_j}$ and $\gamma_j = (\beta E_V)^{3/2}e^{\beta E_j}$ where $\Delta_j$ was sampled from a normal distribution ${\cal N}(\mu_\Delta, \sigma_\Delta^2)$ with $\mu_\Delta = 3.0$ and $\sigma_\Delta =1.0 $, $E_V = 10^{-3}$, and $E_j$ was sampled from a normal distribution ${\cal N}(\mu_E, \sigma_E^2)$ with $\mu_E = 14.0$ and $\sigma_E = 2.0$. All energy parameters were taken to be dimensionless. In (a), we have snapshots of the simulated system at various temperatures. Particles not bound to the grid are not shown. In (b) we plot a theory vs. simulation comparison for both $\langle k \rangle$ and $\langle m \rangle$ as functions of temperature. The critical temperature computed from \rfw{master_therm_subs} is denoted by $k_BT_{\text{crit}}$. The theoretical values of $\langle k \rangle$ and $\langle m \rangle$ were computed from \rfw{kandmdef} and summed over $j$. The "Simulation" values of $\langle k \rangle$ and $\langle m \rangle$ were computed from the results of a Metropolis Hastings algorithm where each point is the average of the result of five simulations.  We see that \rfw{master_therm_subs} identifies the temperature at which the system is in the fully optimally bound state and that the theory curves well match the simulation results. Thus the validity of the Large-$N$ solutions are affirmed in this case. Finally, we note that the $\langle m \rangle$ curve falls away faster than the $\langle k \rangle$ curve which is a consequence of the specific parameter choices we made. In \refsec{system_type} we explore how different choices could lead to different behaviors for $\langle m \rangle$ relative to $\langle k \rangle$. (See \textit{Supplementary Code} in \ref{sec:code} for link to code repository used to produce this figure.)}
	\label{fig:gen_grid_assembly}
\end{centering}
\end{figure*}

In \reffig{gen_grid_assembly}, we display the results of simulated thermodynamics for our grid assembly system for this general case. For such equilibrium simulations, we can choose whatever state transitions we like as long as detailed balance is satisfied, that is, as long as the ratios between the forward and reverse transitions are equivalent to the ratios of the Boltzmann factors between the final and initial states \cite{krauth2006statistical}. Therefore when simulating the equilibrium behavior of the system, we chose state transitions which led to an an efficient-in-time exploration of the state space even if such transitions were unphysical. We included three state-transitions: particle binding (i.e, ligand association to a receptor), particle unbinding (i.e., ligand dissociation from a receptor), and binding permutation (i.e., bound ligands switching receptor sites). The binding permutation transition does not occur in real biomolecular systems, but it was useful for our simulations since it ensured that the system did not remain trapped in non-equilibrated states at low temperature. Theoretically with particle binding and unbinding alone, the system should always find its true equilibrium \textit{eventually}, but consistently finding such an equilibrium on the finite time scales of realistic simulations is difficult. Thus the principal effect of including the binding permutation transition is to reduce the time needed to simulate these systems.

In \reffig{grid_assembly3}, we show the simulated grid at various equilibrium temperatures.  In the (i) image  of \reffig{grid_assembly3}, we see that all particles are bound in their "optimal positions" as defined by the fully optimally bound state in \reffig{assembly_microstate}. As the temperature increases, fewer particles occupy the grid \textit{and} fewer of the particles which occupy the grid are in their optimal binding sites which is consistent with our intuition that the system should lose both binding and combinatorial order as the temperature increases. 

In \reffig{thermal_plot3}, we plot the theoretical temperature-dependences of $\langle k \rangle = \sum_{j=1}^{D} \langle k_j\rangle$ and $\langle m \rangle = \sum_{j=1}^D \langle m_j \rangle$ against their simulated temperature-dependences. We mark the points in the curve that are associated with the grid depictions in \reffig{grid_assembly3}. The temperature computed from \rfw{master_therm_subs} (i.e., the temperature at which fully optimal binding is achieved) is denoted as $k_BT_{\text{crit}}$. We see excellent agreement between the simulation results and the theoretical results. Moreover, the computed critical temperature is in accordance with the results of the simulations. Inspecting (ii) in \reffig{grid_assembly3} (which is at a temperature above the critical temperature) the system is no longer in its fully optimally bound configuration, as we should expect.

The $\langle m \rangle$ and $\langle k \rangle$ curves depicted in \reffig{thermal_plot3} represent only one type of relationship between the temperature dependences of total binding and optimal binding. In this case, we see that as we heat the system above the critical temperature, the average number of optimally bound particles falls more quickly than does the average number of total bound particles. Therefore, slightly above the critical temperature we have a grid-image such as that depicted in (ii) of \reffig{grid_assembly3}: particles are mostly bound but not all of them are in their optimal stites.   

This thermal relationship between $\langle k \rangle$ and $\langle m \rangle$ was pre-determined by our parameter choices for $\Delta_j$, $E_j$, and $E_V$ as was the temperature computed from \rfw{master_therm_subs}. In the discussion following \rfw{simp_bind_cond} for the gendered dimer assembly model, we noted how such parameter choices could lead us to categorize the extremes of that assembly system as one of two types. In the next section, we will attempt to do something similar with this general ligand-receptor system.  

\section{Search and Combinatorics Limited Systems \label{sec:system_type}}

In \cite{williams2019self}, we found that systems of dimer assembly could often be characterized as either search-limited or combinatorics-limited contingent on the relationships between the binding parameters. Given that the system we are currently studying is a generalization of a version of gendered dimer assembly (as shown in \refsec{gender}), we can naturally ask if the ligand-receptor binding system exhibits similar divisions. 

We again narrow our analysis to the case of matched populations of ligands and receptors. Without this assumption, we would have to analyze separately the cases where $r_j < n_j$ and $n_j < r_j$ with little additional benefit in conceptual understanding. Consider the theoretical condition defining the microstate where all ligands are optimally bound to a receptor site (rewritten here from \rfw{master_therm}): 
\begin{equation}
1 = \sum_{j=1}^{D} n_j \delta_j^{-1} \left(1 + \gamma_{j}^{-1}\right) + O(\delta^{-2}),
\label{eq:master_therm_repeat}
\end{equation}
To obtain \rfw{master_therm_repeat}, we took $\delta_j\gg1$. Therefore, all of our system-limits will necessarily be defined with the base assumption that the optimal ligand-receptor binding affinity is large. But even with this base assumption, we can still find two important limits that give us different approximations for the critical temperature.

Assume first that $\gamma_j \gg 1$ for all $j$ at this critical temperature. Then $1/\gamma_j$ is sub-dominant in the second term in the parentheses of \rfw{master_therm_repeat}, and we have the approximation
\begin{equation}
1 =\sum_{j=1}^{D} n_j \delta_{j}^{-1} + O(\gamma^{-1})+ O(\delta^{-2}). \qquad \text{[Combinatorics Limiting Condition]}
\label{eq:comb_limit}
\end{equation}
Recalling that $\gamma_j$ is the base-binding affinity of a type $j$ ligand to any receptor, we know that if $\gamma_j \gg1$ for all $j$, then all ligands have a sufficiently strong binding to \textit{any} receptor that ligand-receptor binding will occur even at low temperature. Thus, whether the system consists entirely of \textit{optimal} bindings is primarily determined by whether $\delta_j$ is strong enough to bias such optimal bindings over their combinatorial disadvantage. We call \rfw{comb_limit} a "combinatorics-limiting condition" (akin to \rfw{temp_def}) for fully optimal binding since the achievement of fully optimal binding is limited by the influence of the combinatorics on the system.

On the other hand, assume that $\gamma_j \ll 1$ (with $\gamma_j \delta_j \gg1$) for all $j$ at the critical temperature. Then $1/\gamma_j$ is dominant in the second term in the parentheses for \rfw{master_therm} and we have the approximation
\begin{equation}
1 = \sum_{j=1}^{D} n_j \delta_{j}^{-1}\gamma_{j}^{-1} +O(\gamma^{0})+ O(\delta^{-2}) .  \qquad \text{[Search Limiting Condition]}
\label{eq:search_limit}
\end{equation}
In this case, the product $\delta_j\gamma_j$ represents the net-binding affinity for an unbound ligand to not merely bind to any receptor but to specifically bind to its optimal receptor. A small value of $\gamma_j$ means that unbound ligands are not attracted to suboptimal receptors, and thus without the additional optimal-binding affinity factor $\delta_j$, ligands would generally not bind at all. In particular, since $\gamma_j\ll1$ but $\delta_j\gamma_j\gg1$ the optimal binding affinity is already strong enough to overcome the combinatorial disadvantage of such bindings because no other bindings besides the optimal ones are thermally favored. Thus achieving the fully optimally bound state is not limited by the influence of combinatorics. Instead, we call \rfw{search_limit} a "search-limiting condition" (akin to \rfw{gamma_cond}) to highlight the fact that for this case, achieving the fully optimally bound state is primarily limited by ligands' abilities to search for their optimal receptor sites in the volume they occupy. 

Given temperature dependences for $\delta_j$ and $\gamma_j$, we can compute the temperatures corresponding to the approximations \rfw{comb_limit} and \rfw{search_limit}. We term these temperatures, respectively, $T_\text{comb}$ and $T_\text{search}$. Using the temperature dependences which yielded \rfw{master_therm_subs}, we find \rfw{comb_limit} and \rfw{search_limit} become, respectively, 
\begin{align}
1 = \sum_{j=1}^{D} n_j e^{-\beta_{\text{comb}}\Delta_j}, \qquad
1= (\beta_{\text{search}} E_V)^{-3/2}\sum_{j=1}^{D} n_j e^{-\beta_{\text{search}}(\Delta_j+ E_j)} \label{eq:therm_limits}
\end{align}
where $k_BT_{\text{search}} = \beta_{\text{search}}^{-1}$ and similarly for $k_{B}T_{\text{comb}}$, and we dropped sub-leading terms for notational simplicity. We can use the equations in \rfw{therm_limits} to establish upper bounds on the true critical temperature $T_{\text{crit}}$: From how the temperature parameter appears in each equation and comparing each to \rfw{master_therm_subs}, it is straightforward to show
\begin{equation}
T_{\text{crit}} <T_{\text{comb}}, \,T_{\text{search}}.
\label{eq:temp_inequality}
\end{equation}

Now why is it important to define these limiting cases at all and then compute temperatures from them? Because the relative values of the temperatures are associated with qualitatively different behaviors for the $\langle k \rangle$ and $\langle m \rangle$ curves, and thus computing these temperatures can immediately provide us with a sense of how the difference $\langle k\rangle - \langle m \rangle$ varies with temperature.

\begin{figure*}[t]
\begin{centering}
\begin{subfigure}[t]{0.325\textwidth}
\centering
\includegraphics[width=\linewidth]{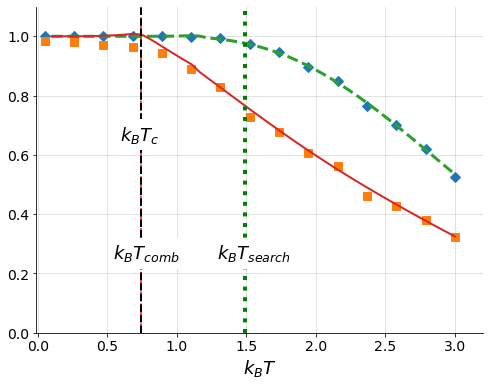}
	\caption{}
	\label{fig:comb_limit}
\end{subfigure}\hfill
\begin{subfigure}[t]{0.325\textwidth}
\centering
\includegraphics[width=\linewidth]{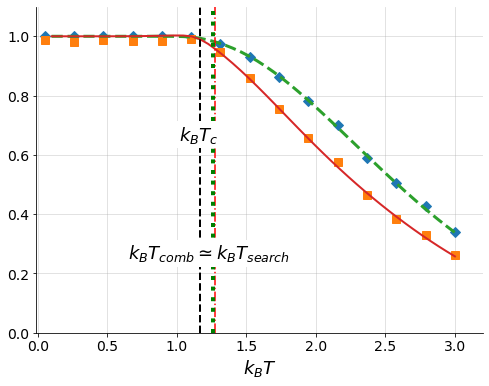}
	\caption{}
	\label{fig:indet_limit}
\end{subfigure}\hfill
\begin{subfigure}[t]{0.325\textwidth}
\centering
\includegraphics[width=\linewidth]{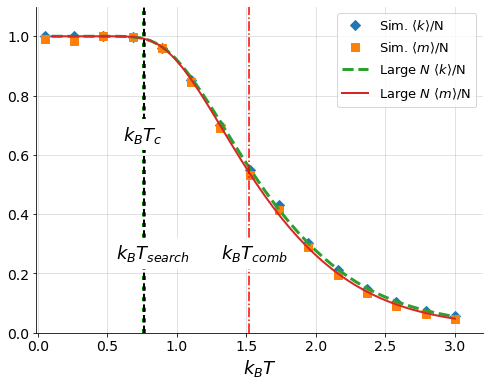}
	\caption{}
	\label{fig:search_limit} 
\end{subfigure}
	\caption{Search-limited, indeterminate, and combinatorics-limited systems: We took $\delta_j = e^{\beta \Delta_j}$ and $\gamma_j = (\beta E_V)^{3/2}e^{\beta E_j}$ where $\Delta_j$ was sampled from a normal distribution ${\cal N}(\mu_\Delta, \sigma_\Delta^2)$, $E_V = 10^{-3}$ and $E_j$ was sampled from a normal distribution ${\cal N}(\mu_E, \sigma_E^2)$. In (a) with $(\mu_\Delta, \sigma_\Delta) =  (4.75, 2.0)$ and $(\mu_E, \sigma_E) =  (16.0, 3.0)$, we have a combinatorics-limited system; In (b) with $(\mu_\Delta, \sigma_\Delta) =(6.75, 2.0)$ and $(\mu_E, \sigma_E) = (10.75, 3.0)$, we have a system of indeterminate or in-between type; In (c) with $(\mu_\Delta, \sigma_\Delta) = (7.7501, 2.0)$ and $(\mu_E, \sigma_E) = (3.0,1.0)$, we have a search-limited system. In the combinatorics-limited system there is a significant difference between $\langle k \rangle$ and $\langle m \rangle$ above the critical temperature, indicating that such systems can have a significant fraction of suboptimally bound ligands. In the search-limited system there is little difference between $\langle k \rangle$ and $\langle m\rangle$, indicating that even when ligands in such systems are partially bound, the ligands are primarily bound optimally. (See \textit{Supplementary Code} in \ref{sec:code} for link to code repository used to produce this figure.)}
	\label{fig:system_type_plot}
\end{centering}
\end{figure*}

We show this in \reffig{system_type_plot}. In each figure, we plot simulation vs theory curves for $\langle k \rangle$ and $\langle m \rangle$ (akin to that displayed in \reffig{thermal_plot3}), for various parameter distributions of $\gamma_i$ and $\delta_i$. We see that depending on these distributions, $k_BT_\text{comb}$ and $k_BT_\text{search}$ have different relative values and these relative values can in turn be used to infer properties of the relationship between $\langle k \rangle$ and $\langle m\rangle$. Specifically, if $T_{\text{comb}} < T_{\text{search}}$, then above the critical temperature, the difference $\langle k\rangle - \langle m \rangle$ grows quickly indicating that a significant fraction of ligands can be bound suboptimally to the set of receptors (\reffig{comb_limit}). Conversely if $T_{\text{comb}} > T_{\text{search}}$, then above the critical temperature, the difference $\langle k\rangle - \langle m \rangle$ is small, indicating that even when the system consists of only partially bound ligands, most of these ligands are attached to their optimal receptor sites (\reffig{search_limit}). 

We can also use these temperatures to approximately define when a system is either search-limited or combinatorics-limited. According to the arguments used to derive \rfw{therm_limits} (and as is affirmed by the results in \reffig{system_type_plot}), a system is combinatorics-limited when 
\begin{equation}
\text{Combinatorics-limited system:} \quad T_{\text{crit}} \simeq T_{\text{comb}}
\end{equation}
and a system is search-limited when
\begin{equation}
\text{Search-limited system:} \quad T_{\text{crit}} \simeq T_{\text{search}}.
\end{equation}

Thus, similar to what was found for dimer system self-assembly \cite{williams2019self}, we have found that we can categorize the ligand-receptor system as constrained by two extremes: A search-limited extreme and a combinatorics-limited extreme. Given the results of the special cases considered in \refsec{bind} and \refsec{perm}, we could also rename the search-limited condition and combinatorics-limited condition as "binding-limited" and "derangement-limited" conditions, respectively; the binding-limited condition defining whether particles can go from free-space to attaching to their preferred site on the grid, and the derangement-limited condition defining whether the particles attach to their preferred site relative to other sites. 

\subsection{Hints at Non-Equilibrium Behavior \label{sec:hints}}

In simulating the general system in \refsec{gen_sim} and \refsec{system_type}, we included the non-physical "binding permutation transition" in order to efficiently explore the state space. Including such a transition to model equilibrium behavior is allowed given the constraints of detailed balance (i.e., any transition is permitted as long as forward and backward transition ratios equate to Boltzmann factor ratios), but if we want to understand realistic non-equilibrium behavior---such as how a system of initially free ligands binds over time to a collection of receptors---we can only use the physical transitions of binding and unbinding. Limiting our transition choices in this way reveals another difference between combinatorics and search-limited systems: When only using physically realistic state transitions, combinatorics-limited systems were more likely to get trapped in non-equilibrated metastable states than were search-limited systems. In particular, when we only allowed for particle binding and unbinding transitions in systems where $\gamma_j \gg 1$ and the temperature satisfied $T< T_{\text{crit}}$ (thus indicating the system should satisfy $\langle k \rangle = \langle m \rangle \simeq N-1$), the simulated system did not always find the "true equilibrium" of fully optimal binding even if analytical predictions suggested it should. 

\begin{figure*}[t]
\begin{centering}
\begin{subfigure}[t]{0.325\textwidth}
\centering
\includegraphics[width=\linewidth]{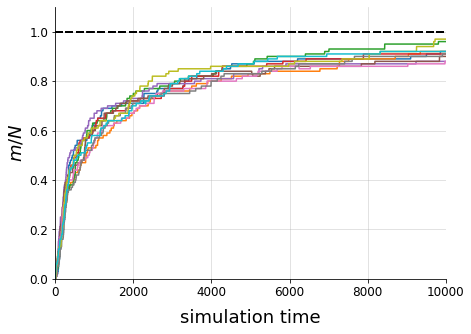}
	\caption{}
	\label{fig:hint_comb_limit}
\end{subfigure}\hfill
\begin{subfigure}[t]{0.325\textwidth}
\centering
\includegraphics[width=\linewidth]{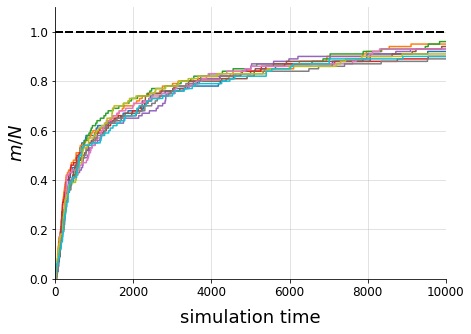}
	\caption{}
	\label{fig:hint_indet_limit}
\end{subfigure}\hfill
\begin{subfigure}[t]{0.325\textwidth}
\centering
\includegraphics[width=\linewidth]{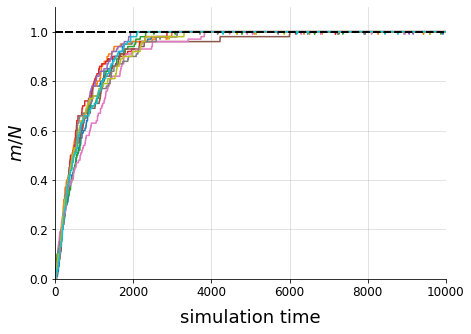}
	\caption{}
	\label{fig:hint_search_limit} 
\end{subfigure}
\begin{subfigure}[t]{0.325\textwidth}
\centering
\includegraphics[width=\linewidth]{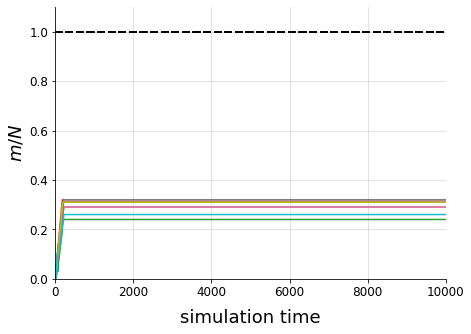}
	\caption{}
	\label{fig:hint2_comb_limit}
\end{subfigure}\hfill
\begin{subfigure}[t]{0.325\textwidth}
\centering
\includegraphics[width=\linewidth]{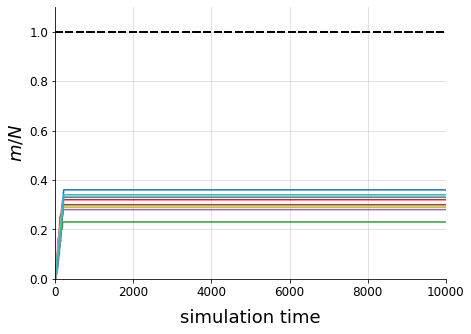}
	\caption{}
	\label{fig:hint2_indet_limit}
\end{subfigure}\hfill
\begin{subfigure}[t]{0.325\textwidth}
\centering
\includegraphics[width=\linewidth]{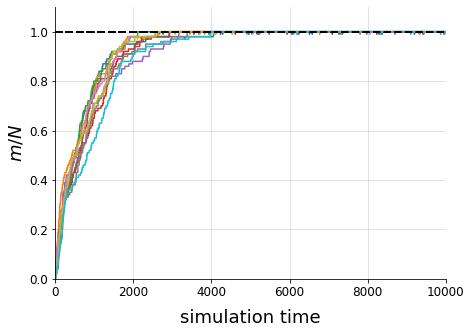}
	\caption{}
	\label{fig:hint2_search_limit} 
\end{subfigure}
	\caption{Simulations in time of the $k_BT = 0.5$ values of $m/N$ for the systems in \reffig{system_type_plot} allowing for unphysical (top-row) and physical (bottom-row) types of transitions. Each simulation began with all particles unbound from the grid. The black dashed line at $m/N =1$ represents the expected equilibrium value that all simulations should approach (as inferred from the plots in \reffig{system_type_plot}). For (a), (b), and (c), we used all the same state transitions used to simulate \reffig{system_type_plot}, namely particle binding, particle unbinding, and particle permutation. For (d), (e), and (f), we only used the physically relevant state transitions of particle binding and unbinding. Plots (a) and (d) have the combinatorics-limited parameters used for \reffig{comb_limit}. Plots (b) and (e) have the indeterminate system parameters used for \reffig{indet_limit}. Plots (c) and (f) have the search-limited parameters used for \reffig{search_limit}. We see that while all system types approach their expected equilibrium values when using the unphysical (but efficient-in-time) particle permutation transition, only the search-limited system reproduces the equilibrium estimate on the finite time scale of simulations when only particle binding and unbinding are allowed transitions. This suggests that, for optimal ligand-receptor binding, only search-limited systems are able to avoid kinetic traps in real non-equilibrium situations. (See \textit{Supplementary Code} in \ref{sec:code} for link to code repository used to produce this figure.)}
	\label{fig:hints_neqbm}
\end{centering}
\end{figure*}

In \reffig{hints_neqbm}, we again simulated the systems whose equilibrium properties are depicted in \reffig{system_type_plot}, except in this case we only tracked the evolution of the number of optimally bound particles, denoted $m$, over the course of the simulation. In each case, the system started from a state consisting entirely of free particles, and then it evolved according to its transition properties. The black dashed line at $m/N =1$ represents the approximate predicted equilibrium value of $\langle m \rangle/N$ for the associated system type at $k_BT = 0.5$ (as obtained from the plots in \reffig{system_type_plot}). The colored continuous lines are various simulations of the system at the given temperature. For \reffig{hint_comb_limit}, \reffig{hint_indet_limit}, and \reffig{hint_search_limit}, we used the same state-transitions used to simulate \reffig{system_type_plot}: Particle binding, particle unbinding, and particle permutation. For \reffig{hint2_comb_limit}, \reffig{hint2_indet_limit}, and \reffig{hint2_search_limit}, we only used the physically relevant state transitions of particle binding and unbinding.  Plots \reffig{hint_comb_limit} and \reffig{hint2_comb_limit} have the combinatorics-limited parameters used for \reffig{comb_limit}. Plots \reffig{hint_indet_limit} and \reffig{hint2_indet_limit} have the indeterminate system parameters used for \reffig{indet_limit}. Plots \reffig{hint_search_limit} and \reffig{hint2_search_limit} have the search-limited parameters used for \reffig{search_limit}. We see that while all system types approach their "correct" equilibrium value when using the unphysical (but efficient-in-time) particle-permutation transition, only the search-limited system reproduces the equilibrium result when only particle binding and unbinding are allowed transition steps. 

The physical explanation for this behavior is simple. If we begin in a state of free ligands in a system satisfying $\gamma_j \gg1$, then the ligands have sufficiently strong binding to all receptors to bind to any one of them and not necessarily to their optimal receptors. Once such ligands are bound, it is unlikely they will dissociate from these receptors and then bind to their true optimal receptor because $\gamma_j$ is so large. This situation exists in distinction to that of a search-limited system where $\gamma_j \ll 1$ and thus where dissociation from suboptimal receptors is thermodynamically feasible and ligands only bind strongly (and largely irreversibly) to their optimal binding sites.

Therefore, combinatorics-limited and search-limited are not merely convenient equilibrium distinctions between systems. They also typify distinctions in realistic non-equilibrium behavior. Theoretically, for infinite time, all systems should reach their true equilibrium regardless of whatever transitions they manifest. However, achieving such a true equilibrium in finite time is made difficult when kinetic traps exist in the system. In contrast to combinatorics-limited systems, search-limited systems have virtually no kinetic-traps since strong binding only occurs when receptors bind to their optimal receptor sites which are associated with the true equilibrium. Thus, in cases where ligands need to specifically bind to certain receptors in finite time, the systems will necessarily need to be search-limited to ensure for rapid optimal binding. 

This "prediction" is necessarily a soft one because we have yet to define at a parameter-level the distinction between the two principal system types. In \cite{williams2019self} we derived necessary but insufficient conditions typifying the distinction between combinatorics and search-limited systems for dimer assembly, but, in the current work, the index-dependence of our biophysical parameters make the analogous such conditions difficult to derive. 

\section{Biophysical Implications \label{sec:implications}}

The theoretical investigations of the previous sections were motivated by the biophysics of ligand-receptor binding. Now we will consider whether our results can help us better understand this starting point. 

 For the system of ligand-receptor binding depicted in \reffig{biophys_ligand_receptor}, the results \rfw{kandmdef} afford us the ability to predict $\langle k_j \rangle$ and $\langle m_j \rangle$ for various ligand species given $\gamma_j$, $\delta_j$, $n_j$, and $r_j$. With \rfw{master_therm_repeat} and temperature dependences for the binding affinities, we can determine the temperature at which a system (that has a matched population of ligands and receptors) settles into the fully optimally bound configuration. Such a prediction would first require finding values for $\gamma_j$, $\delta_j$, and $n_j$. Given that $\gamma_j$ and $\delta_j$ are "effective" model parameters representing, respectively, a ligand of type $j$'s binding affinity to a non-optimal site and the binding affinity advantage to an optimal site, these values would have to be approximated from available data on ligand-receptor interactions. A simple approach to this approximation would amount to taking $\gamma_j$ to be the average of the binding affinities for a particular ligand $j$ to all receptors (besides the optimal one) in a system and $\delta_j$ to be the additional binding affinity factor to the ligand's optimal receptor (i.e., $\gamma_j \delta_j$ would be ligand $j$'s \textit{absolute} binding affinity to this optimal receptor). However, ligand-receptor affinities, though useful theoretical quantities for modeling, are notoriously difficult to calculate in practice \cite{jarmoskaite2020measure} so obtaining accurate predictions of $k_BT_{\text{crit}}$ might be similarly challenging. Moreover, the assumption of matched ligand and receptor populations, though useful for obtaining intuitive analytical results, is clearly not a general one.

Still, we can use the theoretical properties of \rfw{master_therm_repeat} to make qualitative statements about the properties of optimally bound systems. First, we recognize that the terms within the sum must each be less than unity. Therefore as one of the factors of a single term increases, the other factors must decrease in order to keep the total product less than unity. For a system with a large value of $n_j$ for ligand $j$, this means         we need a correspondingly large $\delta_j$ in order for the thermal constraint condition to be soluble. Conceptually, this implies that the more copies we have of a ligand-type, then the more strongly that ligand must bind to its optimal receptor in order for the entire system's fully optimally bound configuration to be achievable.

This result is also true for the \textit{total} number of ligands in the system:  The larger the number of total ligands in the system, the greater the average optimal binding affinities of the ligands must be in order for the system to settle into its fully optimally bound configuration at constant temperature. We can see this by applying Jensen's inequality to the first equation in \rfw{therm_limits} and using \rfw{temp_inequality}. Doing so we obtain
\begin{equation}
k_B T_{\text{crit}} < \frac{1}{\ln N}\sum_{j=1}^D \frac{n_j}{N} \Delta_j,
\label{eq:therm_fin}
\end{equation}
where we defined $N \equiv \sum_{j=1}^D n_j$. \rfw{therm_fin} sets an upper limit on the the temperature at which all ligands settle into their optimal receptors. Since $n_j/N$ is the fraction of elements of type $j$ and $\Delta_j$ is the associated binding energy benefit for being in an optimal site, \rfw{therm_fin} shows that the critical temperature is bounded above by the weighted average of energy benefits, $\overline{\Delta} \equiv \sum_{j=1}^D n_j \Delta_j/N$. Moreover, if we want this bound to remain the same as we increase $N$, the average optimal binding energy must increase in tandem. Thus the more ligands we have in the system the greater we expect the average binding affinity to be, presuming the energetically optimal binding configuration is a desired state in the system.

One could likely have guessed a result of the form in \rfw{therm_fin}. Namely, it makes sense that the critical temperature should be of the same order as an average binding energy in the system. However, what is perhaps surprising is that the limiting temperature scales as $1/\ln N$: As the total number of ligands in the system increases, the temperature at which the low energy system is accessible decreases in tandem but does so logarithmically. Though seemingly novel, this scaling appears to be archetypal for combinatorial statistical physics systems where state spaces are combinations and permutations of a set \cite{williams2018permutation, williams2019self}.

We could also invert this inequality and use it to establish an upper-limit on the number of ligands in the system. Assuming we know $k_BT_{\text{crit}}$, we can impose an upper limit on $N$ as 
\begin{equation}
N < e^{\beta_{\text{crit}} \bar{\Delta}}.
\label{eq:Nineq}
\end{equation}
As $\bar{\Delta}$ increases, so too does the limit on $N$. Thus systems with larger average optimal binding energies can also admit more ligands and still achieve the fully optimally bound configuration at nearly the same temperatures. The main biophysical implication is that cells with more proteins should also have larger average binding affinities for those proteins. For example, human cells and prokaryotic cells, which differ in numbers of proteins by a few order of magnitude \cite{milo2013total}, should also differ in the average optimal binding energies for such proteins. In particular if the compared collection of cells are typically found in similar thermal environments (i.e., exist at the same $k_BT$) and the particle number was stringently bound by \rfw{Nineq}, we would expect the relative average values of the binding energies for human and prokaryotic ligands should be the same as the order of magnitude that differentiates their relative number of proteins. 

However, the limiting effect of \rfw{Nineq} is only relevant if the associated system is specifically combinatorics-limited, i.e., if \rfw{comb_limit} well approximates the general thermal condition \rfw{master_therm_repeat}. The results in \refsec{hints} suggest that real biophysical systems (where optimal ligand-receptor binding is functionally important) are likely search-limited rather than combinatorics-limited since it is only in the former that fully optimal binding can be achieved through physical transitions on finite time scales. We recall that search-limited systems are those for which the energy advantage for optimal binding is sufficiently high that the only limiting factor to ligands finding their optimal site is their ability to "search" for these sites in the constituent volume. More formally, search-limited systems are roughly defined as those for which $T_{\text{search}}$ (computed from \rfw{search_limit}) provides a good approximation for $T_{\text{crit}}$ (computed from \rfw{master_therm_repeat})

Given that real ligand-receptor systems are likely search-limited, they would also likely easily satisfy the combinatorics condition \rfw{comb_limit} at their typical temperature. Therefore, the derived inequality \rfw{Nineq} would not be a strong limit on the number of particles in systems where optimal binding is functionally required. Instead, a stronger limit would be found by using the search-limiting condition \rfw{search_limit} to find a bound on particle number. Finding such a bound first requires us to determine a form for $\gamma_j$ that is a convex function of temperature, and such a $\gamma_j$ depends on our exact model of how free ligands and bound ligands exist in the system of interest. But as a toy-case, we can also use our grid-assembly expression for $\gamma_j$ to get a sense of the general form of the limit\footnote{We note that the expression for $\delta_j = e^{\beta \Delta_j}$ is also associated with the toy-model, but since $\delta_j$ is defined as a ratio of binding affinities, a purely exponential representation of $\delta_j$ is arguably general. Taking $\gamma_j = (\beta E_V)^{3/2} e^{-\beta E_j}$ was primarily for analytic convenience and real systems would have a different thermal dependence for $\gamma_j$.}. Using the grid-assembly expression for $\gamma_j$ amounts to using the second equation in \rfw{therm_limits} to define the critical temperature. The Jensen-inequality based derivation that shows how this equation limits $N$ is similar to that which leads to \rfw{Nineq}. Ultimately, we find 
\begin{equation}
N < (\beta_{\text{crit}} E_V)^{3/2} e^{\beta_{\text{crit}}(\bar{\Delta} + \bar{E})},
\label{eq:Nineq1}
\end{equation}
where $\bar{E} \equiv \sum_{j=1}^D n_jE_j/N$. Thus the implication is the same as that which follows \rfw{Nineq}: Ligand-receptor systems which have a large number of ligands and where optimal binding is physically important, should also have large average binding energies for those ligands.

\rfw{master_therm_repeat} gives the thermal condition under which fully optimal ligand-receptor binding can be achieved, but there is not a universal reason for why such a bound configuration would be \textit{functionally} optimal in all biomolecular systems. Perhaps the microstate where all ligands are in their optimal receptor sites is too rigid to be biophysically useful, and thus it is preferred if the system exists in a partially-bound state where most (but not all) of the bound ligands are in their optimal configuration. Thus the value in this formalism might not exist in its ability to predict the temperature at which the system is in a fully bound state, but rather in its power to predict and categorize binding properties above this temperature. For this value, the system-type distinction is the major biophysical utility of this formalism: Ligand-receptor systems are inherently combinatorial, but it seems that if nature were to evolve such systems so as to avoid the kinetic traps of suboptimal bindings, it would engineer ligands to have sufficiently high binding affinities to their optimal-receptors as to easily achieve the combinatorics-limit \rfw{comb_limit} at \textit{in vivo} temperatures. 

\section{Discussion \label{sec:extend}}

We began this work with the goal of using combinatorics to model how the competition between distinct ligands affects the ligands' equilibrium binding properties to receptors. Before developing a physical model of such binding, we needed to solve a modified version of a well known problem on derangements. After solving this problem and using it to compute a general partition function, we were able to derive implicit formulas for the average number of bound ligands and the average number of optimally bound ligands for each type (\rfw{kandmdef}). Narrowing our focus to the case of matched receptor and ligand populations (i.e., $r_j =n_j$ for all $j$), we were able to derive the condition under which all ligands are bound to their optimal receptors (\rfw{master_therm_repeat}). 

The observables defined in \rfw{kandmdef} allowed us to compute the equilibrium binding properties of our ligand-receptor system at any temperature (provided we have thermal dependences for $\gamma_j$ and $\delta_j$ for all $j$). But the model also provided softer biophysical predictions. \rfw{Nineq} and \rfw{Nineq1} can be seen as such predictions, labeled as "soft" because rather than explicitly predicting that some observable has a value, they predict that, in order for a system condition to be satisfied, an observable cannot exceed a certain value. The main implication of these inequalities is that the larger the binding energy advantage for optimal contacts, the more particles the system can have and still be capable of achieving the fully optimally bound state. Thus, systems with more ligands should also have larger optimal binding energies supposing such bindings have functional importance. Note that \rfw{Nineq1} was derived for the simple model where free-ligands were taken to be point particles, but an analogous inequality could be obtained for more physically informed values of $\gamma_j$.   

For the task of quantitatively modeling optimal ligand-receptor binding, there are many limitations to the introduced model. 

First, we are only partly modeling the combinatorics of our system, a combinatorics that can be characterized by derangements from a pre-defined "correct" or optimal configuration. A more general and flexible model would not \textit{a priori} define such a configuration and would instead have an interaction matrix defining how various ligand types interact with various receptor types, and would \textit{then} use this matrix to determine the optimal matchings between ligands and receptors. However, it is not clear how such a more general formalism would be soluble. We started \refsec{assmb_part} by representing our problem in terms of a general interaction matrix (in \rfw{base_part_func}), but then had to make the simplifying assumption of a pre-defined optimal configuration in order to put the partition function in an analytically tractable form. If we wanted to maintain analytical tractability while pursuing a more general model, we would likely need a similar simplifying assumption. 

Second, to derive our various optimal binding conditions (\rfw{gamma_cond}, \rfw{temp_def}, and \rfw{master_therm}), we assumed that the system contained the same number of receptor sites as ligands for each type (i.e., $n_j = r_j$ for all $j$). This assumption was convenient for the framework of derangements, but does not match a realistic biophysical scenario for which the number of ligands and receptors are not necessarily equal. Thus, a reasonable extension to the Biophysical Implications \refsec{implications} would be to find the analytical conditions that define optimal binding for the case of unmatched receptor and ligand populations of each type. However, the general form of the observables in this system (given in \rfw{kandmdef}) do not make the assumption of $n_j =r_j$ for all $j$, and thus can still be used to numerically compute the number of bound and unbound-particles of various types as a function of temperature.

Third, we did not discuss how the spatial organization of receptors can affect the propensity of ligands to bind to them. Instead we assumed that all receptors were on equal footing in terms of spatial accessibility, and all binding variances could be encoded into the collection of parameters $\boldsymbol{\gamma} = (\gamma_1, \ldots, \gamma_D)$ and $\boldsymbol{\delta} = (\delta_1,  \ldots, \delta_D)$. How spatial organization affects binding could likely be better modeled through a matrix-based interaction scheme where spatially occluded receptors have reduced binding affinities, but, as previously mentioned, such an interaction scheme is less analytically tractable than that assumed in the paper.

The first and the third of these realistic additions would complicate the combinatorial simplicity of what we have presented here, so much so that an entirely new model would likely have to be erected in its place.  As in all modeling, building such a new model would require us to balance the increase in relevance from an incorporation of more realistic properties with a concomitant loss in transparency from the new model's relative insolubility.

Finally, the model we developed does not answer the implicit question posed in the introduction. In \refsec{ligand_receptor}, we first motivated our initial steps towards the general model presented in \refsec{assmb_part} by noting that many ligand-receptor systems exhibit many-to-many interactions in which each type of ligand can bind to many receptors and vice versa. This fact led us to build a model in which even when ligands interacted optimally with a few receptors, these ligands still had the potential to interact with sub-optimal receptors. However, the results of \refsec{hints} suggest that if optimal ligand-receptor binding (where each ligand binds to its energetically optimal receptor) is functionally important in the system, then that binding has to be highly specific to combat its combinatorial disadvantage. So although our model was motivated by multi-specific ligand binding, the model's results primarily pertain to highly-specific ligand binding and the thermal conditions needed to achieve such binding. Therefore left over from the introduction is the question of how does one quantitatively model the ligand-receptor binding underlying the combinatorial codes that make multi-specific systems so necessary in pharmacology and real biomolecular environments. Building such a model would likely require us to make use of an interaction matrix since it is only through such a matrix that one can encode different sets of ligands interacting optimally with different sets of receptors.

As an ending remark, we note that although \refsec{perm} explored a limiting case to our more general model, the model introduced there can also stand alone as a probabilistic model of derangements. For a state space of permutations of a list with repeated elements (e.g., all of the ways to permute the characters and spaces in a sentence), we assume that there is a single permutation where the elements are said to be in their "correct order." Let the parameter $w_i$ be proportional to the probability that an element of type $i$ is in its correct position. Then, the partition function we previously computed in \rfw{gen_derang0} becomes the weighted sum of states
\begin{equation}
X_{\boldsymbol{n}}(\bs{w})  = \int^{\infty}_{0} dx\, e^{-x} \, \prod_{k=1}^{D}\left(w_k-1\right)^{n_k}\, L^{(0)}_{n_k} \left( \frac{x}{1-w_k} \right),
\label{eq:Inr}
\end{equation}
where we took $n_j = r_j$ since we are considering permutations of a list where there are matched populations of correct positions and elements. \rfw{Inr} is essential to the computation of two quantities: First, the probability that $\ell_i$ elements of type $i$, for $i=1, \ldots, D$, are in their correct positions; Second, the average number of elements that are in their correct positions. Respectively, these quantities are
\begin{equation}
P_{\boldsymbol{\ell}} = \frac{G_{\boldsymbol{n-\ell}}}{X_{\boldsymbol{n}}(\bs{w}) }   \prod_{k=1}^D \binom{n_k}{\ell_k} w_{k}^{\ell_k}, \qquad \langle \ell_{\text{tot}} \rangle  = \sum_{j=1}^{D} n_{j} w_j \frac{X_{\boldsymbol{n}_{j}}(\bs{w}) }{X_{\boldsymbol{n}}(\bs{w})},
\label{eq:In_res}
\end{equation}
where $G_{\bs{n}}$  (defined in \rfw{gillis_even} and related to $B_{\bs{r}, \bs{k}}$ through $G_{\bs{n}} =B_{\bs{n}, \bs{n}}$) is the generalized derangement formula, and $\boldsymbol{n}_{j}$ is $\boldsymbol{n}$ with 1 subtracted from the $j$th component: $\boldsymbol{n}_{j}= (n_1, \ldots, n_{j}-1, \ldots, n_D)$. Aside from deriving them directly, we can obtain the expressions in \rfw{In_res} by translating the Boltzmann factor expressions in \refsec{perm} into the language of probability weights by taking $\delta_j \to w_j$. Analogous to \rfw{therm_fin}, using Jensen's inequality, we find that a necessary, but not sufficient, condition for the average $\langle \ell_{\text{tot}} \rangle$ to be equal to $N$ (i.e., for all elements to be in their correct order) is 
\begin{equation}
1 \leq \sum_{j=1}^D \frac{n_j \ln w_j}{N \ln N}.
\label{eq:gen_therm_fin}
\end{equation}
This probabilistic model of derangements would be relevant in a computing context where one is interested in the probability of various derangements that differ from a given sequence by a fixed number of elements.

\section{Conclusion}

We introduced a derangement model of ligand-receptor interactions. The model advances the subject of ligand-receptor modeling in two directions: First, it concretely frames the question of how a finite collection of ligands can compete for a finite collection of receptors. Such a question is important because ligands in real biomolecular systems always exist in crowded environments with other types of ligands, and, with the multi-specificity of ligand-receptor binding, such crowded environments ordinarily exhibit ligands struggling to bind to their optimal sites in a sea of suboptimal ones. Second, to better organize the counting of the system's microstates, the model introduces a combinatorial problem (and its solution) as the foundational framework for such systems. This approach to building biomolecular models by beginning with combinatorial questions appears to be generalizable to similar contexts since biomolecular systems often contain finite numbers of particles where particles can only interact in precise ways that can often be defined by some combinatorial set.

\section{Acknowledgements}
The author thanks Michael Brenner and Rostam Razban for helpful comments on the work. 

\section{Author's Statements}
\begin{itemize}
\item \tbf{Financial Support:} This research received no specific grant from any funding agency, commercial or nonprofit sectors.
\item \tbf{Conflict of Interests Statement:} The author has no conflicts of interest to disclose.
\item \tbf{Ethics Statement:} This research did not require ethical approval.
\end{itemize}

\section{Supplementary Code \label{sec:code}}

The code used to generate all figures is found in the repository \\ \href{https://github.com/mowillia/LigandReceptor}{https://github.com/mowillia/LigandReceptor}.

\appendix

\section{Simulations \label{app:simulation}}

To simulate our system and create the plots in \reffig{binding}, \reffig{derangements},  \reffig{thermal_plot3}, and \reffig{system_type_plot} we implemented a Metropolis Hastings algorithm \cite{krauth2006statistical}. The code for recreating these figures is linked to in \refsec{code}. Here we review the salient parts of the implementation. 

For these simulations, we needed to define a microstate,  the probability of transitions between microstates, and the types of transitions between microstates.

\subsection{Microstate Definition}

A microstate of our system was defined by two lists: one representing the collection of unbound particles, and the other representing particles bound to their various binding sites. The particles themselves were denoted by unique strings and came in multiple copies according to the system parameters. For example, a system with $D=3$ types of particles with $n_1 = 2$, $n_2 = 3$, and $n_3 = 1$ could have a microstate defined by $\texttt{unbound\_particles} = [A_2,  A_2, A_3 ]$ and $\texttt{bound\_particles} = [A_1, -, A_2, -, A_1, - ]$ where "$-$" in the bound list stands for an empty binding site. 

Since the number of optimally bound particles was an important observable for the system, we also needed to define the optimal binding configuration for the microstates. Such an optimal configuration was chosen at the start of the simulation and was defined as a microstate with no unbound particles and all the bound particles in a particular order. For example, using the previous example, we might define the optimal binding configuration as $ \texttt{optimal\_bound\_config} = [A_1, A_1, A_2, A_2, A_2, A_3]$, in which case the number of optimally bound particles of each type in $\texttt{bound\_particles} = [A_1, -, A_2, -, A_1, - ]$ is $m_1=1$, $m_2 =1$, and $m_3 = 0$. The number of bound particles of each type is $k_1 = 2$, $k_2 = 1$, and $k_3 = 0$. We note that the order of the elements in \texttt{unbound\_particles} is not physically important, but, since the number of optimally bound particles is an important observable, the order of the elements in \texttt{bound\_particles} is physically important.

For these simulations, the temperature-normalized  energy of a microstate with $k_i$ bound particles of type $i$ and $m_i$ optimally bound particles of type $i$ was defined as
\begin{equation}
\beta E(\bs{k}, \bs{m}) = \sum_{i=1}^D(m_i \ln \delta_i + k_i \ln \gamma_i),
\label{eq:sim_en}
\end{equation}
where $\bs{k} = (k_1, k_2, \ldots, ,k_D)$, $\bs{m} = (m_1, m_2, \ldots, m_D)$, $\gamma_i$ is the binding affinity, and and $\delta_i$ is the optimal-binding affinity. 

\subsection{Transition Probability}

For transitioning between microstates, we allowed for three different transition types: Particle binding to a site; particle unbinding from a site; permutation of two particles in two different binding sites. Particle binding and unbinding both occur in real physical systems, but permutation of particle positions is unphysical. This latter transition type was included to ensure an efficient-in-time sampling of the state space. For simulations of \textit{equilibrium} systems it is valid to include physically unrealistic transition types as long as the associated transition probabilities obey detailed balance. 

At each time step, we first randomly selected one of the three transition types with equal probability for each type, then randomly selected the final proposed microstate given the initial microstate, and finally computed the probability that said proposal was accepted. By the Metropolis Hastings algorithm \cite{krauth2006statistical}, the probability that the transition is accepted is given by 
\begin{equation}
\text{acceptance prob}(\text{init} \to \text{fin}) = \min \left\{1, e^{- \beta (E_{\text{fin}}-E_{\text{init}})}\frac{\pi(\text{fin} \to \text{init})}{\pi(\text{init} \to \text{fin})} \right\},
\label{eq:sim_prob}
\end{equation}
where $\beta$ is inverse temperature, $E_{\text{init}}$ is the energy of the initial microstate state, and $E_{\text{fin}}$ is the energy of the final microstate, with energy defined in \rfw{sim_en}. The quantity $\pi(\text{init} \to \text{fin})$ is the probability of randomly proposing the final microstate state given the initial microstate state and $\pi(\text{fin} \to \text{init})$ is defined similarly. The ratio $\pi(\text{fin} \to \text{init})/\pi(\text{init} \to \text{fin})$ varied for each transition type. 

\subsection{Transition Types and Examples}
Below we give examples of the three types of transitions along with the value of the ratio $\pi(\text{fin} \to \text{init})/\pi(\text{init} \to \text{fin})$ in each case. In the following, $N_f$ and $N_b$ represent the number of free particles and the number of bound particles, respectively, before the transition. 

\begin{itemize}
\item \tbf{Particle Binding to Site:} One particle was randomly chosen from the \texttt{unbound\_particles} list and placed in a randomly chosen empty site in the \texttt{bound\_particles} list. $\pi(\text{fin} \to \text{init})/\pi(\text{init} \to \text{fin})= (N_b+1)^{-1}/(N_f^{-1}\times N_f^{-1}) ={N_f^2}/{(N_b+1)}$. \\[.5em]
\textit{Example:} \\$\texttt{unbound\_particles} = [A_2,  A_2, A_3 ]$ and $\texttt{bound\_particles} = [A_1, -, A_2, -, A_1, - ]$ \\[.5em] (transitions to) $\\[.5em]  \texttt{unbound\_particles} = [A_2,  A_3 ]$ and $\texttt{bound\_particles} = [A_1, A_2, A_2, -, A_1, - ]$; \\[.5em] Transition weight: $\pi(\text{fin} \to \text{init})/\pi(\text{init} \to \text{fin}) =9/4 $
\item \tbf{Particle Unbinding from Site:} One particle was randomly chosen from the \texttt{bound\_particles} list and placed in the \texttt{unbound\_particles} list. $\pi(\text{fin} \to \text{init})/\pi(\text{init} \to \text{fin})= ((N_f+1)^{-1}\times (N_f+1)^{-1})/(N_b^{-1})= {N_b}/{(N_f+1)^2}$. \\[.5em]
\textit{Example:} \\$\texttt{unbound\_particles} = [A_2,  A_2, A_3 ]$ and $\texttt{bound\_particles} = [A_1, -, A_2, -, A_1, - ]$ \\[.5em]  (transitions to) $\\[.5em] \texttt{unbound\_particles} = [A_2,  A_2, A_3, A_2 ]$ and $\texttt{bound\_particles} = [A_1, -, - , -, A_1, - ]$; \\[.5em] Transition weight: $\pi(\text{fin} \to \text{init})/\pi(\text{init} \to \text{fin}) = 3/16 $
\item \tbf{Particle Permutation:} Two particles were randomly selected from the \texttt{bound\_particles} list, and their positions in the list were switched. $\pi(\text{fin} \to \text{init})/\pi(\text{init} \to \text{fin})= 1.$ \\[.5em]
\textit{Example:} \\$\texttt{unbound\_particles} = [A_2,  A_2, A_3 ]$ and $\texttt{bound\_particles} = [A_1, -, A_2, -, A_1, - ] $ \\[.5em]  (transitions to) $\\[.5em] \texttt{unbound\_particles} = [A_2,  A_2, A_3 ]$ and $\texttt{bound\_particles} = [A_2, -, A_1, -, A_1, - ]$; \\[.5em] Transition weight: $\pi(\text{fin} \to \text{init})/\pi(\text{init} \to \text{fin}) =1 $
\end{itemize}

For impossible transitions (e.g., particle binding when there are no free particles) the probability for accepting the transition was set to zero. At each temperature, the simulation was run for anywhere from 10,000 to 30,000 time steps depending on observed convergence, and the final $2\%$ of the time steps were used to compute
ensemble averages of $\langle k \rangle$ and $\langle m \rangle$. These simulations were repeated five times, and each point in \reffig{binding}, \reffig{derangements},  \reffig{thermal_plot3}, and \reffig{system_type_plot} represents the average $\langle k \rangle$ and $\langle m \rangle$ over these five runs.

\section{Consistency Checks for $B_{\bs{r}, \bs{k}}$}

In this section, we affirm the various consistency checks for $B_{\bs{r}, \bs{k}}$. The first check ensures that $B_{\bs{r}, \bs{k}}$ has the proper limiting case when $\bs{r}$ has components with value $1$. The second check ensures that $B_{\bs{r}-\bs{m}, \bs{k}-\bs{m}}$ has the correct normalization when summed over all possible values of $\bs{m}$. 

\subsection{Checking \rfw{Bnkb} \label{app:Bnkb}}

We want to check that 
\begin{equation}
\sum_{k_1=0}^{1} \ldots \sum_{k_D=0}^{1} B_{\bs{r}_0, \bs{k}} \delta(K, k_1 + \cdots + k_D) = \sum_{j=0}^D (-1)^{j}\binom{D}{j} \binom{D-j}{K-j}^2(K-j)!,
\label{eq:Bnk_init}
\end{equation}
where $\bs{r}_0 = (1, \ldots, 1)$. We start from \rfw{Bnk} (reproduced here for convenience):
\begin{align}
B_{\bs{r}, \bs{k}} & = \sum_{j_1 =0}^{k_1} \cdots \sum_{j_D =0}^{k_D} \binom{r_1}{j_1} \cdots \binom{r_D}{j_D} (-1)^{j_1 + \cdots + j_D}\binom{r_1-j_1 + \cdots + r_D-j_D}{k_1-j_1 + \cdots + k_D-j_D}\mm
 & \qquad \qquad \times \frac{(k_1-j_1 + \cdots + k_D-j_D)!}{(k_1-j_1)! \cdots (k_D-j_D)!}.
\label{eq:Bnk0}
\end{align}
Given $j_i \leq k_i \leq r_i$, if $r_i=1$ then we must have $(k_i-j_i)! = 1$. We also have $\binom{r_i}{j_i} = \binom{1}{j_i} =1$. Thus \rfw{Bnk} becomes
\begin{align}
B_{\bs{r}_0, \bs{k}} & = \sum_{j_1 =0}^{k_1} \cdots \sum_{j_D =0}^{k_D}  (-1)^{j_1 + \cdots + j_D}\binom{D-j_1 - \cdots -j_D}{K-j_1 - \cdots -j_D}\mm
 & \qquad \qquad \times {(K-j_1 - \cdots -j_D)!}\label{eq:Bnk2},
\end{align}
where we defined $K \equiv \sum_{i=1}^D k_i$. With the identity,
\begin{equation}
1 = \sum_{J=0}^{K}\delta(J, j_1 + \cdots + j_D),
\end{equation}
we can introduce a Kronecker delta and obtain
\begin{align}
B_{\bs{r}_0, \bs{k}} & =\sum_{J=0}^{K} \sum_{j_1 =0}^{k_1} \cdots \sum_{j_D =0}^{k_D}  (-1)^{j_1 + \cdots + j_D}\binom{D-j_1 - \cdots -j_D}{K-j_1 - \cdots -j_D}\mm
 & \qquad \qquad \times {(K-j_1 - \cdots -j_D)!}\,\delta(J, j_1 + \cdots + j_D)\mm
  & = \sum_{J=0}^{K} (-1)^{J} \binom{D-J}{K-J} (K-J)!  \sum_{j_1 =0}^{k_1} \cdots \sum_{j_D =0}^{k_D} \delta(J, j_1 + \cdots + j_D).
  \label{eq:Bnk3}
\end{align}
Isolating the summation over the Kronecker delta yields
\begin{align}
\sum_{j_1 =0}^{k_1} \cdots \sum_{j_D =0}^{k_D} \delta(J, j_1 + \cdots + j_D) 
& = \frac{1}{2\pi i} \oint \frac{dz}{z^{J+1}}\prod_{i=1}^D \sum_{j_i =0}^{k_i} z^{j_i} \mm
 & = \frac{1}{2\pi i} \oint \frac{dz}{z^{J+1}}(1 + z)^{k_1 + \cdots + k_D}\mm
  & = \binom{K}{J}.
\end{align}
Thus, we obtain 
\begin{align}
B_{\bs{r}_0, \bs{k}} & =\sum_{J=0}^{K} (-1)^{J} \binom{D-J}{K-J} (K-J)! \binom{K}{J}.
  \label{eq:Bnk4}
\end{align}
Performing the final summation in \rfw{Bnk_init}, we obtain
\begin{align}
\sum_{k_1=0}^{1} \ldots \sum_{k_D=0}^{1} B_{\bs{r}_0, \bs{k}} \delta(K, k_1 + \cdots + k_D)  & =\sum_{J=0}^{K} (-1)^{J} \binom{D-J}{K-J} (K-J)! \binom{K}{J} \mm
 & \qquad \qquad \times \sum_{k_1=0}^{1} \ldots \sum_{k_D=0}^{1} \delta(K, k_1 + \cdots + k_D) \mm
& = \sum_{J=0}^{K} (-1)^{J} \binom{D-J}{K-J} (K-J)! \binom{K}{J} \binom{D}{K},
\end{align}
and with the identity 
\begin{equation}
\binom{K}{J} \binom{D}{K} = \binom{D}{J} \binom{D-J}{K-J}, 
\end{equation}
we have 
\begin{equation}
\sum_{k_1=0}^{1} \ldots \sum_{k_D=0}^{1} B_{\bs{r}_0, \bs{k}} \delta(K, k_1 + \cdots + k_D) = \sum_{J=0}^{K} (-1)^{J} \binom{D-J}{K-J}^2 (K-J)! \binom{D}{J},\quad \checkmark
\end{equation}
as expected. 

Why does this result make sense? When $\bs{r} = (1, \ldots, 1)\equiv \bs{r}_0$, the vector $\bs{k}$ can only have elements of $1$ or $0$. Thus $B_{\bs{r}_0, \bs{k}}$ represents the number of ways to completely derange a \textit{particular} collection of objects, defined by $\bs{k}$, out of $D$ unique objects. In order to find $b_{D,K}$, the number of ways to completely derange $K$ objects, we need to count and sum the number of derangements for \textit{all} collections of objects. Therefore, to obtain $b_{D,K}$ we need to sum $B_{\bs{r}_0, \bs{k}}$ over all possible values of $\bs{k}$ such that $\sum_j k_j = K$.

\subsection{Checking equivalence between \rfw{Ink0} and \rfw{multi_nom} \label{app:multi_nom}}
We want to find a reduced form for
\begin{equation}
I_{\bs{r}, \bs{k}} = \sum_{m_1 =0}^{k_1} \cdots \sum_{m_D=0}^{k_D} \binom{r_1}{m_1} \cdots \binom{r_D}{m_D}B_{\bs{r}-\bs{m}, \bs{k}-\bs{m}}.
\label{eq:Ink}
\end{equation}
The expression for $B_{\bs{r}, \bs{k}}$ is 
\begin{equation}
B_{\bs{r}, \bs{k}} = \frac{1}{(\sum_i\alpha_i)!}\int^{\infty}_{0} dx\, e^{-x}  \prod_{i=1}^D (-1)^{k_i} x^{\alpha_i} L_{k_i}^{(\alpha_i)}(x),
\label{eq:Bnkfinapp}
\end{equation}
Noting that $\alpha_i =  r_i - k_i= (r_i -m_i) - (k_i - m_i)$, from \rfw{Ink} and \rfw{Bnkfinapp}, we find 
\begin{equation}
I_{\bs{r}, \bs{k}} = \frac{1}{(\sum_i\alpha_i)!}\int^{\infty}_{0} dx\, e^{-x}\,x^{\sum_i\alpha_i} \prod_{i=1}^D \sum_{m_i=0}^{k_i} \binom{r_i}{m_i} (-1)^{k_i-m_i} L_{k_i-m_i}^{(\alpha_i)}(x).
\end{equation}
Next, we make the change of variables 
\begin{equation}
q_i \equiv k_i - m_i.
\end{equation}
We then have 
\begin{equation}
I_{\bs{r}, \bs{k}} = \frac{1}{(\sum_i\alpha_i)!}\int^{\infty}_{0} dx\, e^{-x} x^{\sum_i\alpha_i} \prod_{i=1}^D \sum_{q_i=0}^{k_i} \binom{k_i + \alpha_i}{k_i - q_i} (-1)^{q_i} L_{q_i}^{(\alpha_i)}(x).
\label{eq:Inkint}
\end{equation}
To simplify this result we need to find an identity for 
\begin{equation}
\sum_{q=0}^{k} \binom{k + \alpha}{k - q} U^{q} L_{q}^{(\alpha)}(x).
\end{equation}
Expanding the generalized Laguerre polynomial $L^{(\alpha)}_q(x)$ according to its definition, we obtain 
\begin{align}
\sum_{q=0}^{k} \binom{k + \alpha}{k - q} U^{q} L_{q}^{(\alpha)}(x) & = \sum_{q=0}^{k}  \binom{k + \alpha}{k - q} U^{q} \sum_{i = 0}^{q} \binom{q+\alpha}{q-i} \frac{(-1)^{i}}{i!} x^{i}. 
\label{eq:Lagalpha}
\end{align}
More identity wrangling gives us 
\begin{equation}
\binom{k + \alpha}{k - q} \binom{q+\alpha}{q-i} = \binom{k+\alpha}{k-i} \binom{k-i}{q-i}, 
\end{equation}
and thus \rfw{Lagalpha} becomes 
\begin{align}
\sum_{q=0}^{k} \binom{k + \alpha}{k - q} U^{q} L_{q}^{(\alpha)}(x) 
 &= \sum_{i=0}^{k} \binom{k+\alpha}{k-i}\frac{(-1)^{i}}{i!} x^{i}  \sum_{q = i}^{k}  U^{q} \binom{k-i}{q-i}\mm
  &= \sum_{i=0}^{k} \binom{k+\alpha}{k-i}\frac{(-1)^{i}}{i!} x^{i} U^{i}  \sum_{q = i}^{k}  U^{q-i} \binom{k-i}{q-i}\mm
    &= \sum_{i=0}^{k} \binom{k+\alpha}{k-i}\frac{(-1)^{i}}{i!} x^{i} U^{i}  (1+U)^{k-i},
    \label{eq:sumLaga0}
\end{align}
which, by the definition of the Laguerre polynomial yields
\begin{equation}
\sum_{q=0}^{k} \binom{k + \alpha}{k - q} U^{q} L_{q}^{(\alpha)}(x)  = (1+U)^{k} L_{k}^{(\alpha)}\left( \frac{xU}{1+U}\right).
\label{eq:sumLaga}
\end{equation}
Also, from the definition of the Laguerre polynomial, we can show 
\begin{equation}
\lim_{\lambda \to 0} \lambda^{k} L_{k}^{(\alpha)}\left(\frac{x}{\lambda}\right) = (-1)^{k} \frac{x^k}{k!}.
\label{eq:limLaga}
\end{equation}
Therefore, with \rfw{sumLaga}  and \rfw{limLaga}, we have 
\begin{equation}
\sum_{q=0}^{k} \binom{k + \alpha}{k - q} (-1)^{q} L_{q}^{(\alpha)}(x) = \lim_{U\to -1} (1+U)^{k} L_{k}^{(\alpha)}\left( \frac{xU}{1+U}\right) = \frac{x^{k}}{k!}. 
\end{equation}
Inserting this result into \rfw{Inkint} gives us
\begin{align}
I_{\bs{r}, \bs{k}} &  =  \frac{1}{(\sum_i\alpha_i)! }\int^{\infty}_{0} dx\, e^{-x} x^{\sum_i\alpha_i} \frac{x^{k_1 + \cdots + k_D}}{k_1! \cdots k_D!}\mm
 & = \frac{1}{(\alpha_1 + \cdots + \alpha_D)! k_1! \cdots k_D!}\int^{\infty}_{0} dx\, e^{-x} x^{r_1 + \cdots + r_D},
\end{align}
which, with the Gamma function definition, yields
\begin{equation}
I_{\bs{r}, \bs{k}}  = \frac{(r_1 + \cdots + r_D)!}{(r_1-k_1 + \cdots + r_D - k_D)! k_1! \cdots k_D!}.
\end{equation}

Why does this result make sense? From one perspective, the quantity $\binom{r_1}{m_1} \cdots \binom{r_D}{m_D}B_{\bs{r}-\bs{m}, \bs{k}-\bs{m}}$ is the number of ways to choose $m_j$ out of $r_j$ positions to contain their correct elements while the remaining $k_j-m_j$ elements  are completely deranged with respect to their $r_j-m_j$ remaining correct positions, for $j=1, \ldots, D$. If we sum over all possible values of $m_j$ we should obtain the number of ways to \textit{arrange} (i.e., not only derange) $k_j \leq r_j$ objects for $j=1, \ldots, D$ across a total of $r_1 + \ldots+r_D$ lattice sites. 

On the other hand, if we are trying to calculate the number of ways to arrange $k_j$ objects of type $j$ for $j=1, \ldots, D$ across $r_1 + \cdots + r_D$ lattice sites, we can use the language of multinomials. Say that the objects represent "filled" sites on the lattice and the spaces between objects are "empty" sites. Then there are $k_1+\cdots + k_D$ filled sites (of which $k_i$ are identical for each $i$) and $r_1-k_1 + \cdots + r_D-k_D$ empty sites. Finding the number of ways to arrange the objects amongst the $r_1+\cdots + r_D$ sites is equivalent to finding the total number of ways to order this collection of filled and empty sites while correcting for equivalent orderings due to reordering the positions of the same type of site. Including both filled and empty sites there is a total of $r_1 + \cdots + r_D$ sites amongst which we have $r_1-k_1 + \cdots+r_D-k_D$ "copies" of empty sites, $k_1$ copies of filled sites of type 1, $k_2$ copies of filled sites of type 2, $\ldots$ and $k_D$ copies of filled sites of type $D$. Counting the number of ways to order the $r_1 + \cdots + r_D$ sites and correcting for the equivalent reorderings arising from the multiple copies of various types of sites leads to the multionomial
\begin{equation}
\frac{(r_1 + \cdots + r_D)!}{(r_1-k_1 + \cdots + r_D-k_D)!k_1! \cdots k_D! },
\end{equation}
which we have shown is equivalent to the result written in terms of a summation over $B_{\bs{r-m}, \bs{k-m}}$.

\section{Deriving General Partition Function \label{app:gen_part_func}}

In this section, we derive \rfw{master_derang_part01} from \rfw{init_part2}.  First we note from the definition of $B_{\bs{r}, \bs{k}}$ in \rfw{Bnkfin} that 
\begin{equation}
B_{\bs{r}-\bs{m}, \bs{k}-\bs{m}} = \frac{1}{(\sum_j \alpha_j)!}\int^{\infty}_{0} dx\, e^{-x} x^{\sum_j \alpha_j} \prod_{j=1}^D (-1)^{k_j-m_j} L_{k_j-m_j}^{(\alpha_j)}(x),
\label{eq:Bnkfin0}
\end{equation}
where $\alpha_j = r_j - k_j$. Thus the summation over $\bs{m}$ in \rfw{init_part2} becomes 
\begin{align}
\sum_{\bs{m}}B_{\bs{r}-\bs{m}, \bs{k}-\bs{m}}\prod_{j=1}^{D}\binom{r_j}{m_j}\delta_j^{m_j} & = \frac{1}{(\sum_j \alpha_j)!}\int^{\infty}_{0} dx\, e^{-x} x^{\sum_j \alpha_j}\mm
& \qquad \times\prod_{j=1}^{D}\sum_{m_j=0}^{k_j}\binom{r_j}{m_j}\delta_j^{m_j}  (-1)^{k_j-m_j} L_{k_j-m_j}^{(\alpha_j)}(x),\mm
& = \frac{1}{(\sum_j \alpha_j)!}\int^{\infty}_{0} dx\, e^{-x} x^{\sum_j \alpha_j}\mm
& \qquad \times\prod_{j=1}^{D}\sum_{q_j=0}^{k_j}\binom{k_j + \alpha_j}{k_j-q_j}\delta_j^{k_j-q_j}  (-1)^{q_j} L_{q_j}^{(\alpha_j)}(x),
\end{align}
where we changed variables from $m_j$ to $q_j = k_j-m_j$ in the final line. Using the following Laguerre polynomial identity (derived in \rfw{sumLaga0})
\begin{equation}
\sum_{q=0}^{k} \binom{k + \alpha}{k - q} U^{q} L_{q}^{(\alpha)}(X)  = (1+U)^{k} L_{k}^{(\alpha)}\left( \frac{XU}{1+U}\right),
\end{equation}
we then have 
\begin{equation}
\sum_{m_j=0}^{k_j}\binom{k_j + \alpha_j}{k_j-q_j}\delta_j^{k_j-q_j}  (-1)^{q_j} L_{q_j}^{(\alpha_j)}(x) = (\delta_j-1)^{k_j}L_{k_j}^{(\alpha_j)} \left(\frac{x}{1-\delta_j}\right).
\end{equation}
Therefore, the summation over $\bs{m}$ becomes
\begin{align}
\sum_{\bs{m}}B_{\bs{r}-\bs{m}, \bs{k}-\bs{m}}\prod_{j=1}^{D}\binom{r_j}{m_j}\delta_{j}^{m_j}  & = \frac{1}{(\sum_j \alpha_j)!}\int^{\infty}_{0} dx\, e^{-x} \prod_{j=1}^{D}x^{\alpha_j}(\delta_j-1)^{k_j}L_{k_j}^{(\alpha_j)} \left(\frac{x}{1-\delta_j}\right)\mm
 & \equiv C_{\bs{r}, \bs{k}}
\end{align}
For the summation over $\bs{k}$, we use the contour integral identity for inverse factorial 
\begin{equation}
\frac{1}{(\sum_j \alpha_j)!} = \frac{1}{2\pi i} \oint_{\Gamma} \frac{dz}{z^{\sum_j \alpha_j+1}}e^{z},
\label{eq:Mexp}
\end{equation}
where $\Gamma$ is a closed contour about the origin, to obtain
\begin{align}
  \sum_{\bs{k}} C_{\bs{r}, \bs{k}}  & \prod_{j=1}^{D} \frac{1}{(n_j-k_j)!} \,\gamma_j^{k_j} = \frac{1}{2\pi i} \oint_\Gamma \frac{dz}{z} e^z\int^{\infty}_{0} dx\, e^{-x} \prod_{j=1}^D\sum_{k_j = 0}^{n_j}\left(\frac{x}{z} \right)^{\alpha_j}\frac{\gamma_j^{k_j}(\delta_j-1)^{k_j}}{(n_j-k_j)!}L_{k_j}^{(\alpha_j)} \left(\frac{x}{1-\delta_j}\right)\mm
   & = \frac{1}{2\pi i} \oint_{\Gamma} \frac{dz}{z} e^z\int^{\infty}_{0} dx\, e^{-x} \prod_{j=1}^{D} \left(\frac{x}{z}\right)^{r_j}\sum_{k_j=0}^{n_j}\frac{1}{(n_j-k_j)!} \left(\frac{z\gamma_j}{x}(\delta_j-1)\right)^{k_j}\mm 
   & \qquad \qquad \qquad \qquad \qquad \times L_{k_j}^{(r_j-k_j)} \left(\frac{x}{1-\delta_j}\right).
\end{align}
Using the identity
\begin{equation}
\sum_{k=0}^n \frac{Y^{k}}{(n-k)!} L_{k}^{(r-k)}(X) =Y^{n}\, L_{n}^{(r-n)}(X-Y^{-1}),
\label{eq:arg_summ}
\end{equation}
proved in Appendix \ref{app:arg_summ}, we find that the final partition function is 
\begin{align}
{\cal Z}_{\bs{n}, \bs{r}} & =  \frac{1}{2\pi i} \oint_{\Gamma} \frac{dz}{z} \int^{\infty}_{0} dx\, e^{z-x} \prod_{j=1}^{D} (\gamma_j (\delta_j-1))^{n_j} \left(\frac{x}{z}\right)^{r_j-n_j} L_{n_j}^{(r_j-n_j)} \left(\frac{x(z \gamma_j+1)}{z\gamma_j(1-\delta_j)} \right)\mm
& =  \frac{1}{2\pi i} \oint_{\Gamma} \frac{dz}{z} \int^{\infty}_{0} dx\, e^{z-x} \left( \frac{x}{z} \right)^{N_R- N_L}\prod_{j=1}^{D} (\gamma_j (\delta_j-1))^{n_j} L_{n_j}^{(r_j-n_j)} \left(\frac{x(z \gamma_j+1)}{z\gamma_j(1-\delta_j)} \right),
\label{eq:master_derang_part_app}
\end{align}
where we defined $N_R \equiv \sum_{j=1}^{D} r_j$ and $N_{L} \equiv \sum_{j=1}^D n_j$ as the total number of receptors and total number of ligands, respectively, in the system. In deriving \rfw{master_derang_part_app}, we have made no assumptions about the relative value of $r_j$ and $n_j$ for each $j$ and therefore the result is valid for both $r_j< n_j$ and $r_j \geq n_j$. 

\section{Laguerre Polynomial Argument Summation Identity \rfw{arg_summ} \label{app:arg_summ}}
In this section, we prove the identity \rfw{arg_summ}. Starting from the definition of the generalized Laguerre polynomial \rfw{gen_laguerre}, we have
\begin{align}
\sum_{k=0}^n \frac{Y^{k}}{(n-k)!} L_{k}^{(r-k)}(X) & = \sum_{k=0}^n \frac{Y^{k}}{(n-k)!} \sum_{j=0}^k \binom{k + r - k}{k-j} \frac{(-1)^j}{j!} X^j\mm
& = \sum_{k=0}^n \frac{Y^{k}}{(n-k)!} \sum_{\ell=0}^k \binom{r}{\ell} \frac{(-1)^{k-\ell}}{(k-\ell)!} X^{k-\ell}\mm
 & = \sum_{\ell=0}^n \sum_{k=\ell}^{n} Y^{k} (-1)^{k-\ell}X^{k-\ell} \binom{r}{\ell} \frac{1}{(n-\ell)!} \binom{n-\ell}{k-\ell} 
\end{align}
where we changed summation variables ($\ell = k-j$) in the second line and switched the order of summations in the third line. Separating the summations and then using the binomial theorem, we obtain
\begin{align}
\sum_{k=0}^n \frac{Y^{k}}{(n-k)!} L_{k}^{(r-k)}(X) 
 & = \sum_{\ell=0}^n \binom{r}{\ell} \frac{1}{(n-\ell)!}Y^{\ell} \sum_{k=\ell}^{n}  (-1)^{k-\ell}(XY)^{k-\ell}  \binom{n-\ell}{k-\ell} \mm
 & = \sum_{\ell=0}^n \binom{r}{\ell} \frac{1}{(n-\ell)!}Y^{\ell}(1-XY)^{n-\ell} \mm
  & = Y^{n}\sum_{\ell=0}^n \binom{r}{\ell} \frac{(-1)^{n-\ell}}{(n-\ell)!}(X-Y^{-1})^{n-\ell} \mm
    & = Y^{n}\sum_{j=0}^n \binom{n+r-n}{n-j} \frac{(-1)^{j}}{j!}(X-Y^{-1})^{j} \mm
    & = Y^{n}\,L^{(r-n)}_n(X-Y^{-1}) 
\end{align}
where we changed variables to $j= n-\ell$ in the fourth line and used the definition \rfw{gen_laguerre} in the final line. 

\section{Gendered Dimer Assembly Equilibrium Conditions \label{app:gendered}}
Here we derive the equilibrium conditions \rfw{dimer_eq1} and \rfw{dimer_eq2} from the large $N$ approximation conditions  \rfw{eqbm_zx_0} and the observable definitions \rfw{obs01_k} and \rfw{obs01_m}. We start with the conditions defining $\bar{x}$ and $\bar{z}$:
\begin{equation}
1 = \sum_{j=1}^D \frac{\bar{z}\gamma_j +1}{\bar{x} + \bar{z}\gamma_j(\delta_j -1 + \bar{x})}, \qquad \bar{z} = \bar{x} \sum_{j=1}^D \frac{1}{\bar{x} + \bar{z} \gamma_j (\delta_j-1 + \bar{x})}.
\label{eq:eqbm_zx}
\end{equation}
The expressions for $\langle k_j \rangle$ and $\langle m_j \rangle$ are 
\begin{align}
\langle k_j \rangle  &= \frac{\bar{z}\gamma_j (\delta_j -1 + \bar{x})}{\bar{z}\gamma_j (\delta_j -1 + \bar{x})+\bar{x}} \label{eq:obs1_k} \\
 \langle m_j \rangle & = \frac{\bar{z}\gamma_j \delta_j}{\bar{z}\gamma_j (\delta_j -1 + \bar{x})+\bar{x}}. \label{eq:obs1_m}
\end{align}
Next, we seek to eliminate the $\bar{z}$ and $\bar{x}$ dependence from these observables. 
From \rfw{eqbm_zx}, \rfw{obs1_k}, and \rfw{obs1_m}, we can show
\begin{equation}
\bar{z} = N - \langle k \rangle, \qquad
\bar{x} = N - \sum_{j=1}^D \langle m_j \rangle(1- \delta_j^{-1}) ,
\label{eq:zxbar}
\end{equation}
where we used $\langle k \rangle = \sum_{j=1}^D\langle k_j \rangle$, and the total particle number $N$ is equal to the number of particle types $D$ when there is one copy per particle.
With \rfw{obs1_k} and \rfw{obs1_m}, we can also show
\begin{equation}
\langle k_j \rangle - \langle m_j\rangle(1-\delta_j^{-1} \rangle = \frac{\bar{z}\gamma_j \bar{x}}{\bar{z} \gamma_j (\delta_j-1 + \bar{x}) + \bar{x}}, 
\label{eq:kjminusmj}
\end{equation}
and with the second equation in \rfw{eqbm_zx} and the first equation in \rfw{zxbar}, we obtain
\begin{equation}
\sum_{j=1}^D\frac{1}{\gamma_j}\Big( \langle k_j \rangle - \langle m_j \rangle(1-\delta_j^{-1})\Big) = \Big(N- \langle k \rangle\Big)^2,
\label{eq:eqbm_k1}
\end{equation}
which is our first equilibrium condition. 

From both equations in \rfw{eqbm_zx}, we have
\begin{equation}
\sum_{j=1}^D \frac{\bar{z}\gamma_j}{\bar{x} + \bar{z}\gamma_j(\delta_j -1 + \bar{x})} = 1 - \sum_{j=1}^D \frac{1}{\bar{x} + \bar{z}\gamma_j(\delta_j -1 + \bar{x})} = 1 - \bar{z}/\bar{x}.
\label{eq:zbar_id}
\end{equation}
Using \rfw{zbar_id} in \rfw{kjminusmj} and \rfw{obs1_m}, we find, respectively
\begin{align}
 \sum_{j=1}^D \Big( \langle k_j \rangle - \langle m_j\rangle(1-\delta_j^{-1} \rangle\Big)   & = \bar{x}-\bar{z} \label{eq:k_m_diff2}\\
\sum_{j=1}^D \langle m_j\rangle\delta_{j}^{-1} &= 1- \bar{z}/\bar{x}
\label{eq:k_m_diff3}.
\end{align}
With \rfw{k_m_diff2} and \rfw{k_m_diff3} and the second equation in \rfw{zxbar}, we then obtain 
\begin{align}
\sum_{j=1}^D\langle m_j \rangle \delta_{j}^{-1} = \frac{\langle k \rangle - \langle m \rangle + \sum_{j=1}^{D} \langle m_j \rangle \delta_{j}^{-1}}{N- \langle m \rangle + \sum_{j=1}^{D} \langle m_j \rangle \delta_{j}^{-1}}, \label{eq:eqbm_m1}
\end{align}
which is our second equilibrium condition.

\section{Derivation of \rfw{master_therm} \label{app:therm_cond}}
In this section, we derive \rfw{master_therm}, the general thermal condition for the fully optimally bound state. This state is defined as 
\begin{equation}
\langle k \rangle = \langle m \rangle = N-1 + O(\delta^{-1}).
\label{eq:corr_ass_cond}
\end{equation}
where $N \equiv \sum_{j=1}^Dn_j = \sum_{j=1}^Dr_j$. To derive the desired thermal condition, we will need some of our previous results. In particular, we need the average number of optimally bound ligands of type $j$
\begin{equation}
\langle m_j \rangle  =  \frac{n_j\delta_j}{\delta_j-1} \frac{\displaystyle L_{n_j-1} \left( \bar{\phi}_{j}\right)}{\displaystyle  L_{n_j} \left( \bar{\phi}_{j} \right)},
\label{eq:mj}
\end{equation}
and the equations relating $\bar{z}$ and $\bar{x}$ to our observables:
\begin{equation}
\bar{z} = N - \langle k \rangle , \qquad \bar{x} = N - \sum_{i=1}^D \langle m_i \rangle (1-\delta_i^{-1}).
\label{eq:four_eqns0}
\end{equation}
From \rfw{corr_ass_cond}, we can infer that for the fully optimally bound state we have
\begin{equation}
\langle m _j \rangle = n_j - O(D^{-1}) + O(\delta_j^{-1}),
\label{eq:m_approx}
\end{equation}
where we are explicitly referencing the $O(D^{-1})$ terms in order to satisfy $\sum_{j=1}^D O(D^{-1}) =1$ required of \rfw{corr_ass_cond}. Using \rfw{m_approx} in the second equation in \rfw{four_eqns0} yields
\begin{equation}
\bar{x} = 1 + \sum_{j=1}^{D} n_j \delta^{-1}_{j} + O(\delta^{-1}),
\label{eq:xbardelta}
\end{equation}
where we subsumed terms of order $O(D^{-1}\delta^{-1})$ into $O(\delta^{-1})$.

For the fully optimally bound configuration, ligands need to greatly favor their optimal receptors in order to bind only to such receptors. Thus for the fully optimally bound state, we can assume that the optimal-binding affinity for each particle-type is much greater than 1: $\delta_j \gg 1$. From this assumption we can take $\bar{\phi}_j$ defined as
\begin{equation}
\bar{\phi}_j = \frac{\bar{x}}{1-\delta_j} \left( 1+ \frac{1}{\bar{z} \gamma_j}\right)
\end{equation}
to satisfy $|\bar{\phi}_j| \ll 1$. We will check this latter assumption at the end of the calculation, but moving forward with it, we find 
\begin{equation}
\frac{\displaystyle L_{n_j-1} \left( \bar{\phi}_{j}\right)}{\displaystyle  L_{n_j} \left( \bar{\phi}_{j} \right)} = 1 + \bar{\phi}_j + O (\bar{\phi}_j^2) = 1 - \bar{x}\delta_j^{-1} \left(1 + \frac{1}{\bar{z} \gamma_j}\right) + O(\delta_j^{-2}).
\label{eq:lag_ratio_approx}
\end{equation}
 Inserting \rfw{lag_ratio_approx} into \rfw{mj} and summing over $j$, we obtain 
\begin{align}
\langle m \rangle & = \sum_{j=1}^D\frac{n_j}{1- \delta_{j}^{-1}} \frac{\displaystyle L_{n_j-1} \left( \bar{\phi}_{j}\right)}{\displaystyle  L_{n_j} \left( \bar{\phi}_{j} \right)}\mm
 & = N + \sum_{j=1}^{D} n_j \delta_j^{-1} - \bar{x}\sum_{j=1}^Dn_j \delta_j^{-1} \left(1 + \frac{1}{\bar{z} \gamma_j}\right)+ O(\delta^{-2}).
\end{align}
Using \rfw{xbardelta}, we can write this result as 
\begin{align}
\langle m \rangle & = N -1 + \bar{x}- \bar{x}\sum_{j=1}^Dn_j \delta_j^{-1} \left(1 + \frac{1}{\bar{z} \gamma_j}\right)+ O(\delta^{-1}).
\label{eq:mNminus1}
\end{align}
Therefore, we see that \rfw{mNminus1} reproduces \rfw{corr_ass_cond} if 
\begin{equation}
1 = \sum_{j=1}^Dn_j \delta_j^{-1} \left(1 + \frac{1}{\bar{z} \gamma_j}\right).
\end{equation}
With $\langle k \rangle  = N-1 + O(\delta^{-1})$ and $\bar{z} = N-\langle k \rangle$, we find $\bar{z} = 1 + O(\delta^{-1})$, thus giving us the final thermal condition 
\begin{equation}
1 = \sum_{j=1}^D{n_j} \delta_j^{-1}\left(1 + \frac{1}{ \gamma_j}\right)+ O(\delta^{-2}).
\label{eq:final_therm_res}
\end{equation}
Now we will check the assumption of $|\bar{\phi}_j|\ll1$ for consistency: From \rfw{final_therm_res}, we can infer that $\sum_{j=1}^Dn_j\delta_j^{-1} < 1$. Therefore $\bar{x}$ defined in \rfw{xbardelta} is $O(1)$. Also from \rfw{final_therm_res}, we can infer that, in the large number limit (i.e., $n_j\gg1$), we must have $\delta_j^{-1}(1+ 1/\gamma_{j})\ll 1$. Thus, we see that $\bar{\phi}_j = \bar{x}(1+{1}/{\bar{z} \gamma_j})/(1-\delta_j) \simeq - \bar{x}\delta_j^{-1}(1+ 1/\gamma_{j}) + O(\delta_j^{-2}) $ can indeed be taken to satisfy $|\bar{\phi}_j| \ll 1$.


\newcommand{\etalchar}[1]{$^{#1}$}
\newcommand{\noopsort}[1]{} \newcommand{\printfirst}[2]{#1}
  \newcommand{\singleletter}[1]{#1} \newcommand{\switchargs}[2]{#2#1}

\end{document}